\documentclass[aps,pra, two column,10pt,superscriptaddress,nofootinbib, notitlepage]{revtex4-2}

\usepackage{pgfplots}
\pgfplotsset{compat=1.18}
\newcommand\identity{1\kern-0.25cm\text{l}}
\usepackage{etex}
\usepackage{physics}
\usepackage{amsmath,amssymb,amsthm,amstext,mathrsfs}
\usepackage[colorlinks=true,citecolor=blue,urlcolor=blue]{hyperref}
\usepackage[T1]{fontenc}
\usepackage{comment}
\usepackage{eufrak}
\usepackage{enumerate}
\usepackage{times,txfonts}
\usepackage{braket}
\usepackage{svg}
\usepackage{color}

\usepackage{url}

\usepackage{amsmath,blkarray}
\usepackage{mathtools}
\usepackage{latexsym}
\usepackage{tabularx, booktabs}
\usepackage{epstopdf}
\usepackage{lipsum}
\usepackage{float}
\usepackage{amsfonts}
\usepackage{subcaption}
\usepackage{color,soul}

\newcommand{\be}{\begin{equation}}
	\newcommand{\ee}{\end{equation}}
\newcommand{\ba}{\begin{eqnarray}}
	\newcommand{\ea}{\end{eqnarray}}

\definecolor{ss}{RGB}{250,80,220}
\begin{document}
\title{Genuine Multipartite Nonlocality for  Arbitrary Input: Maximal Randomness Generation and Robust Self-Testing}
\author{Rajdeep Paul}
\email{rajdeeppaul22@gmail.com}
\affiliation{Department of Physics, Indian Institute of Technology Hyderabad, Telangana 502284, India }
\author{Ranendu Adhikary}
\email{ronjumath@gmail.com}
\affiliation{Department of Physics and Center for Quantum Frontiers of Research and Technology (QFort),
National Cheng Kung University, Tainan 701, Taiwan}
\author{Alok Kumar Pan }
\email{akp@phy.iith.ac.in}
\affiliation{Department of Physics, Indian Institute of Technology Hyderabad, Telangana 502284, India }
\begin{abstract}
{Bell nonlocality provides the foundation for device-independent (DI) certification of quantum devices. We introduce a Bell inequality capable of identifying genuine multipartite nonlocality (GMNL) in an arbitrary $m$-partite scenario with an arbitrary odd number of measurements per party. Since the multi-setting nature of this inequality precludes the use of Jordan’s Lemma, we construct an analytical sum-of-squares (SOS) decomposition to obtain the optimal quantum violation without assuming any bound on the Hilbert space dimension. This, in turn, enables self-testing of the shared entangled state and the corresponding measurement observables, up to local isometries, whose existence we confirm using a swap-based certification scheme. In addition, we show that our framework enables the extraction of maximal global DI randomness ($m$ bits) at the optimal quantum violation, thereby exceeding previous limitations in the GMNL regime. Finally, we demonstrate that the architecture of our inequality yields improved robustness to noise as 
the number of measurement settings grows, ensuring experimental feasibility.}

\end{abstract}

\maketitle

\section{Introduction} Quantum nonlocality—a phenomenon that challenges our classical worldview—reveals correlations between space-like separated parties that cannot be explained by any local causal model. 
First articulated in Bell’s 1964 theorem~\cite{bell1964einstein,bell1966problem}, nonlocality emerged from debates about the completeness of quantum theory and has since evolved into a foundational pillar of quantum information science and technology. Beyond its foundational significance, the DI character of nonlocal correlations has turned them into a powerful resource for emerging quantum technologies, including quantum key distribution~\cite{ekert1991quantum,scarani2009security} and randomness generation~\cite{Nieto-Silleras_2018,liu2018device,wooltorton2022tight}.

Bell nonlocality is most commonly studied in bipartite scenarios and is typically demonstrated through the violation of a suitable Bell inequality. 
However, extending the framework to multipartite systems unveils a substantially richer structure of correlations. 
While bipartite correlations are simply classified as local or nonlocal, multipartite systems admit a hierarchy of nonlocal behaviors depending on the allowed collaborations among subsets of parties. 
Svetlichny~\cite{Svetlichny1987} introduced a notion of multipartite nonlocality that captures correlations irreducible to any hybrid model combining local and bipartite nonlocal resources. This stronger form of \emph{genuine nonlocality}~\cite{bancal2011detecting, bancal2013definitions} is typically witnessed through violations of Svetlichny-type inequalities~\cite{Svetlichny1987,Augusiak2019,singh2024robust}. In a two input $m$-party scenario, the Mermin–Ardehali–Belinskii–Klyshko (MABK)~\cite{Mermin1990, Ardehali1992, Belinski_1993} inequality was proposed, which demonstrates genuine nonlocality only for even $m$. Another notion of multipartite nonlocality, the no-signaling genuine nonlocality, was proposed in Ref.~\cite{bancal2013definitions}, which is strictly weaker than the Svetlichny-type genuine nonlocality~\cite{bancal2013definitions}. In this work, by genuine nonlocality, we refer to the Svetlichny-type, which is the strongest form of multipartite nonlocality.

GMNL has emerged as a key resource for secure multiparty computation~\cite{crepeau2002secure,colbeck2009quantum,sutradhar2021efficient} and decentralized quantum networks~\cite{wehner2018quantum}. 
By distributing entanglement among multiple nodes, such protocols enable a trust structure that prevents any single party from compromising the system’s integrity—an essential feature for secure data exchange~\cite{crepeau2002secure,sutradhar2021efficient}, distributed ledger architectures~\cite{wehner2018quantum}, and privacy-preserving communication~\cite{Broadbent2009,Fitzsimons2017}.

Beyond its cryptographic role, nonlocality also provides a direct route to DI randomness certification. 
In the multipartite scenario, MABK inequalities can certify up to $m$ bits of DI randomness when the number of parties $m$ is odd, and approximately $m-0.4$ bits for even $m$~\cite{Dhara2013max,Wooltorton2025expandingbipartite}. However, for odd $m$, MABK inequalities do not witness GMNL. Building on the construction of the multipartite Bell functional proposed in~\cite{Curchod2019}, maximal $m$ bits of randomness was certified~\cite{Wooltorton2025expandingbipartite}. However, the genuineness of nonlocality is not explored.

Note that the DI randomness is derived solely from observed correlations without assuming detailed knowledge of the internal workings of the measurement. A key tool for this is DI self-testing, which enables one to certify both the underlying quantum state and the measurement operators exclusively from the observed input–output statistics.  Recent work has significantly advanced the self-testing of multipartite entangled states, with notable progress for Greenberger–Horne–Zeilinger (GHZ) states~\cite{Panwar2023,Sarkar2022}.

In realistic scenarios, however, environmental noise is unavoidable and typically lowers the observed Bell inequality violation from its ideal maximum. Understanding how certification protocols perform under such imperfections is therefore essential. Hence, robustness analysis of a self-testing protocol is central to delineating the parameter regimes in which one can still reliably certify quantum states and measurements~\cite{Paul2026}. Early robust self-testing results for the three-qubit $W$-state were obtained via the swap circuit method~\cite{wu2014}, and were later generalized to multipartite states that admit a Schmidt decomposition~\cite{Supic2018}.

Despite these advances, most multipartite self-testing results remain restricted to scenarios involving two measurement settings and two outcomes per party~\cite{pal2014device,yang2014robust,mckague2014self,fadel2017self,vsupic2018self,sekatski2018certifying,baccari2020scalable,adhikary2024self,adhikary2024self1}. In those scenarios, Jordan's lemma~\cite{Jordanlemma} allows measurement operators to be decomposed into block-diagonal form consisting of $2\times2$ blocks, which significantly simplifies the analysis. However, this structural reduction no longer applies when more than two measurement settings per party are considered. Consequently, multipartite self-testing beyond the two-input and dichotomic-output setting becomes substantially more challenging and calls for more general analytical techniques.

In this paper, we introduce a family of Bell functionals in an arbitrary $m$-partite scenario involving an unbounded $n$ number of inputs per party, designed to witness GMNL.  We analytically compute the bi-local value, as well as the upper bounds of the local values of the Bell functionals. For any given $n$ and $m$, the optimal quantum value surpasses the bi-local value, thereby showcasing GMNL.  We derive the optimal quantum values of the Bell functionals in a DI manner through an analytical SOS decomposition technique~\cite{bamps2015} devoid of assuming the dimension of the system. As an application, we demonstrate the generation of maximal multipartite DI randomness ($m$ bits).

The optimization through SOS decomposition enables the development of a fully DI self-testing protocol to characterize the system’s state and measurements without making any assumptions about the underlying dimensions. We present the explicit construction of the swap circuit and demonstrate self-testing of the GHZ state and the associated observables via a local isometry.  Further, to account for experimental imperfections, we derive explicit robustness bounds for the extractability of state and measurement in the presence of noise for the tripartite case and show that the tolerance to such imperfections increases with the number of measurement settings. We extend the robustness analysis for DI randomness generation by considering the imperfections involved in measurement.  Although, for technical reasons, the robustness analysis is explicitly worked out only for the tripartite case, the core framework can straightforwardly be extended to a general $m$-party scenario.

\section{Results}
\subsection{Genuine multipartite nonlocality} Consider a scenario involving $m$ parties, denoted as $Alice^1$ to $Alice^m$. Each party receives an input $i_k\in\{A^k_1,...,A^k_n\}$ and produces the output $a_k\in\{-1,+1\}$. The joint probability distribution $P(a_1, a_2, \ldots, a_m | i_1, i_2, \ldots, i_m)$ is considered \emph{fully local} if it can be expressed as a product of independent probabilities
\begin{eqnarray}\label{eq:local_equation}
P_{L}(a_1, a_2, \ldots, a_m | i_1, i_2, \ldots,i_m) 
 = \sum_{\lambda} \nu(\lambda) \prod_{k=1}^{m} P(a_k | i_k, \lambda)\quad
 \end{eqnarray}
where $\lambda$ is a common classical cause with distribution $\nu(\lambda) \geq 0$ and $\sum_{\lambda} \nu(\lambda) = 1$. 

A distribution is termed \emph{bi-local} if it can be represented as,
\begin{equation}\label{eq:bi_equation}
\begin{split}
 &P_{BL}(a_1, a_2, \ldots, a_m | i_1, i_2, \ldots, i_m)\\
 &= \sum_{\lambda} \nu(\lambda)P_S(\{a_k\}_{k \in S}| \{i_k\}_{k \in S}, \lambda)\,
P_{\bar{S}}(\{a_k\}_{k \in \bar{S}}| \{i_k\}_{k \in \bar{S}}, \lambda)   
\end{split}
\end{equation}
where $\nu(\lambda) \geq 0$ and $\sum_{\lambda} \nu(\lambda) = 1$ and $S \cup \bar{S} = \{1,\ldots,m\}$, $S \cap \bar{S}=\emptyset$, such that each group can share arbitrary correlations internally, but the two groups are correlated only via a classical hidden variable $\lambda$. 

A distribution that defies bi-local representation is classified as \emph{genuine nonlocal} ~\cite{bancal2013definitions}. To demonstrate genuine nonlocality, we introduce a family of  Bell inequalities in the $m$-partite scenario, featuring arbitrary $n$ (odd) inputs per party.

\subsubsection{Arbitrary \texorpdfstring{$n$}{n} input and \texorpdfstring{$m$}{m} party scenario} In a $m$-partite Bell scenario, a common source distributes physical systems to all $m$ parties. Each party receives $n$ inputs ($ n \geq 3 $ and odd), with measurement $i_k \in \{A^k_1, \ldots, A^k_n\},\forall k\in[m] $. Each measurement yields one of two possible outcomes $a_k\in\{+1, -1\},i_k\in[n],\forall k\in[m]$.
Given the scenario, we propose the following multipartite Bell functional,
\begin{widetext}
    \begin{eqnarray}\label{general}
    \Gamma^m_n&=&\sum_{i_m=1}^n\cdot\cdot\sum_{i_3=1}^n \qty(\sum_{{i_1}=1}^n(A^1_{i_1}-A^1_{{i_1}\oplus_n+1})\otimes A^2_{({i_1}+i_3-2)\oplus_n +1 })\otimes_{k=3}^{m-1} A^k_{f_k(i_k,i_{k+1})}\otimes A_{i_m}^m 
\end{eqnarray}
\end{widetext}

where $i_2$ is defined as $i_2\equiv ({i_1}+i_3-2)\oplus_n +1, (\oplus_n x\equiv x \ mod \ n)$ and $f_k$'s are defined as follows. 
\begin{eqnarray}
    && f_k(i_{k+1},i_k)=(i_{k+1}+i_k-2)\oplus_n +1, \forall m>3\label{fkk+1}
\end{eqnarray}

We derive that in fully local model the functional in Eq.~(\ref{general}) is loosely upper bounded by $(\Gamma_n^m)_L \leq 2 (n-1)^{m-1}$. In any bi-local models described by Eq.~(\ref{eq:bi_equation}), we derive $(\Gamma_n^m)_{BL}=2 n^{m-2}(n-1)$. The optimal quantum value is derived using sos decomposition is $(\Gamma_n^m)_Q^{opt} = 2 n^{m-1}\cos\bigl(\tfrac{\pi}{2n}\bigr)$,  which implies a strict relation $(\Gamma_n^m)_Q^{opt} > (\Gamma_n^m)_{BL} > (\Gamma_n^m)_L$. Note that all derivations are analytical. A detailed derivation of local and bi-local models is provided in the Appendix. Here, we provide a sketch of the analytical derivation of $(\Gamma_n^m)_Q^{opt}$.

We introduce the SOS decomposition technique~\cite{bamps2015, Pan2021} to derive  $(\Gamma_n^m)_Q^{opt}$, without assuming the Hilbert-space dimension. The structure of both the observables and the entangled state emerges solely from the optimization conditions. For this, we introduce a positive semi-definite operator
\begin{equation}\label{gamman}
    \gamma_n^m =\frac{1}{2} \sum_{i_m=1}^n\cdot\cdot\sum_{i_3=1}^n \sum_{{i_1}=1}^n w_{i_1i_2..i_m} L_{{i_1}i_2..i_m}^{m^{\dag}} L^m_{{i_1} i_2..i_m},\quad\forall m 
\end{equation}
We take this particular form of $ L^m_{{i_1} i_2..i_m}$ motivated by the structure of the Bell functional as
\begin{eqnarray}\label{sosn}
    L^m_{{i_1} i_2..i_m} &=& \openone_d\otimes  A^2_{({i_1}+i_3-2)\oplus_n +1 }\otimes_{k=3}^{m-1} A^k_{f_k(i_{k+1},i_k)}\otimes A_{i_m}^m \nonumber\\
    &&- \widetilde{A}^1_{{i_1}} \otimes \cdots \otimes \openone_d,\quad \forall m
\end{eqnarray}
with $\widetilde{A}^1_{i_1}= \frac{A^1_{i_1}-A^1_{{i_1}\oplus_n+1}}{w_{i_1i_2..i_m}}$ and  $w_{i_1i_2..i_m} \equiv w_{i_1}= \lVert A^1_{i_1}-A^1_{{i_1}\oplus_n+1} \rVert
=\sqrt{2-\langle \{A^1_{i_1},A^1_{{i_1}\oplus_n+1}\} \rangle}$, where we use the shorthand \(\langle \dots \rangle_{\rho_{A^1\ldots A^m}} = \langle \dots \rangle\). 

Now, substituting $L^m_{{i_1} i_2\ldots i_m}$ from Eq.~(\ref{sosn}) into Eq.~(\ref{gamman}) yields the expectation value of the Bell functional as
\begin{equation}
\langle\gamma^m_n\rangle=\sum_{i_m=1}^n\cdot\cdot\sum_{i_3=1}^n \sum_{{i_1}=1}^n w_{i_1}-
\langle\Gamma^m_n\rangle.
\end{equation}
Since $\gamma^m_n$ is positive semi-definite, $\langle\Gamma^m_n\rangle$ maximizes when $\langle\gamma^m_n\rangle=0$. Therefore,
\begin{eqnarray} (\Gamma_n^m)_Q^{opt}&=&\max\qty(\sum_{i_m=1}^n\cdot\cdot\sum_{i_3=1}^n \sum_{{i_1}=1}^n w_{i_1})=n^{m-2} \max\qty(\sum_{{i_1}=1}^n w_{i_1})
\end{eqnarray}
By optimizing $w_{i_1}$, we obtain $w_{i_1} = w_{i'_1} = 2\cos{\frac{\pi}{2n}},\forall i_1\neq i'_1\in[n]$, which in turn yields $\langle \{A^1_{i_1}, A^1_{{i_1}\oplus_n+1}\} \rangle = -2\cos{\frac{\pi}{n}}$. Consequently, we derive the optimal quantum value as
\begin{eqnarray}
   (\Gamma_n^m)_Q^{opt}=2n^{m-1}\cos {\frac{\pi}{2n}} 
\end{eqnarray}

The detailed derivation is quite lengthy and is therefore deferred to the Appendix.

\subsubsection{Self-testing statements}
When $ (\Gamma_n^m)_Q^{opt}$ is achieved  the following self-testing statements holds.
\begin{enumerate}[(i)]
\item The shared state between the $m$ parties is a pure entangled state with $\langle A^k_{i_k}\rangle=0$, $\langle A^{k_1}_{i_{k_1}}A^{k_2}_{i_{k_2}}\rangle_{k_1\neq k_2}=0$, $\ldots$ , $\langle A^{k_1}_{i_{k_1}}A^{k_2}_{i_{k_2}}\dots A^{k_{m-1}}_{i_{k_{m-1}}}\rangle_{k_1\neq k_2\neq\dots\neq K_{m-1}}=0$ for all $k,k_i\in[m]$.
\item All the parties perform projective measurements on their respective systems, and each party's observable satisfies the following  relations, i.e.,
\begin{eqnarray}
    \langle \{A^k_{i_k}, A^k_{i_k+x}\} \rangle_{\rho_{A^1..A^m}} = 2(-1)^x\cos{\frac{\pi x}{n}}
\end{eqnarray}$ \ \forall k\in[n], i_k\in[n],x\in[n-i_k]$.
\end{enumerate}

\emph{A realization using a local qubit system:} We found that $(\Gamma_n^m)_{Q}^{opt}$ can be achieved for a maximally entangled state $\ket{GHZ}=\frac{1}{\sqrt{2}} \qty(\ket{0}^{\otimes m}+\ket{1}^{\otimes m})$ and the observables 
\begin{eqnarray}
    A^k_{i_k}(\delta^k_{i_k}) = \sin \delta^k_{i_k} \, \sigma_x + \cos \delta^k_{i_k} \, \sigma_y,\forall k\in[m], i_k\in[n]
\end{eqnarray}
where $\delta^1_{i_1} = \frac{(-3n+16+2(n-1)(i_1-1))\pi}{2n}$, $\delta^2_{i_2} = \frac{(6n-15+2(n+1)(i_2-1))\pi}{2n}$, $\delta^3_{i_3} = \frac{(n-1)(i_3-1)\pi}{n}$, $\delta^k_{i_k} =\frac{\pi}{2}- \frac{(n-(-1)^k)(i_k-1)\pi}{n},\forall k\in[4,m], i_k\in[n]$.

\subsection{Multipartite DI randomness}  The Bell nonlocality~\cite{Colbeck2011,Pironio2010} enables the generation of certified randomness in DI way. In an $m$-partite Bell scenario with binary-output measurements, up to $m$ bits of randomness can, in principle be certified. The MABK inequality for odd $m$ allows the generation of $m$ bits of randomness~\cite{Wooltorton2025expandingbipartite}, but such correlations do not exhibit genuine nonlocality\cite{Bancal2011}.

 We demonstrate the certification of $m$ bits DI randomness from our genuinely multipartite nonlocal correlation. The DI randomness in our $m$-partite Bell scenario is quantified by the min-entropy of the observed behavior~\cite{Acin2012,Sasmal2024},
\begin{equation}
\label{random}
\mathcal{R}_{i_1 \dots i_m} = - \log_{2}\left[\max_{a_1,\dots,a_m}P\left(a_1,\dots,a_m|A^1_{i_1},\dots,A^m_{i_m}\right)\right]
\end{equation}
The amount of certified randomness depends explicitly on the specific measurement settings and is thus evaluated across all correlators. For dichotomic observables, the joint probability distribution takes the form
\begin{eqnarray}\label{pbc}
&&P\left(a_1,\dots,a_m|A^1_{i_1},\dots,A^m_{i_m}\right)=\frac{1}{2^m}\Bigg[1+\sum_{k=1}^{m}a_k\langle{A^k_{i_k}}\rangle\nonumber\\&&+\sum_{k_1\neq k_2}^{m}a_{k_1}a_{k_2}\langle{A^{k_1}_{i_{k_1}}A^{k_2}_{i_{k_2}}}\rangle+\dots+a_1a_2\dots a_m\langle{A^1_{i_1}A^2_{{i_2}}\dots A^m_{i_m}}\rangle\Bigg]\quad
\end{eqnarray}

From the SOS conditions, we find that all the marginals, except $\langle{A^1_{i_1}A^2_{{i_2}}\dots A^m_{i_m}}\rangle$, are zero. Hence, we rewrite Eq.\eqref{pbc} as
\begin{eqnarray}\label{pbcrand}
P\left(a_1,\dots,a_m|A^1_{i_1},\dots,A^m_{i_m}\right)=\frac{1}{2^m}\Bigg[1+a_1\dots a_m\langle{A^1_{i_1}\dots A^m_{i_m}}\rangle\Bigg]\quad \ \ \
\end{eqnarray}
Evidently, the randomness is maximized when, for certain choices of indices $\{i_1,\dots,i_m\}$, the correlation function satisfies $\langle A^1_{i_1}A^2_{i_2}\dots A^m_{i_m}\rangle = 0$. Explicitly, the following index choices ensure this requirement.
\begin{eqnarray}
    \Big\langle{A^1_{1+(i_2-i_3+\frac{n+1}{2})\oplus_n}  A^2_{i_2} \prod_{k=3}^{m-1} A^k_{f_k(i_{k+1},i_k)} A_{i_m}^m}\Big\rangle=0
\end{eqnarray}
where $f_k(i_{k+1},i_k)$ is defined in Eq.~(\ref{fkk+1}). Therefore, for these observables, 
\begin{eqnarray}
   P\qty(a^1,a^2,.,a^k,.,a^m|A^1_{1+(i_2-i_3+\frac{n+1}{2})\oplus_n}, A^2_{i_2},., A^k_{f_k(i_{k+1},i_k)},., A_{i_m}^m )=\frac{1}{2^m}\nonumber\\
\end{eqnarray}

where, $k=[3,m-1]$, provide the maximum randomness, $\mathcal{R}^m_{max}=m$. Detailed derivations are provided in Appendix.

\subsection{Self-testing of GHZ state and Observables}\label{self circuit}We now show that achieving the optimal quantum violation of the Bell inequality uniquely certifies both the underlying $m$-partite state and the measurement observables. Our argument is based on a swap-based self-testing protocol, which models the experiment in a fully DI, black-box scenario under the assumption of minimal local dimension. In this framework, a suitable local isometry transfers the properties of the unknown physical system to an ancillary reference system, enabling certification of both state and measurements.

Assuming a physical state $\ket{\psi}_{A^1\dots A^m}$ and the physical measurements are $Z_{A^k}$ and $X_{A^k}$ ($\{Z_{A^k}, X_{A^k}\}=0$) satisfying self-testing conditions, we construct a reference state $\ket{00...0}_{A^{1'}\dots A^{m'}}$ replicating the physical statistics despite unknown dimensions. This is expressed as
\begin{eqnarray}
&\Phi\Bigl(\mathcal{U}_i\ket{\psi}_{A^1\dots A^m} \otimes \ket{00...0}_{A^{1'}\dots A^{m'}}\Bigr)&=\ket{\chi}_{A^1\dots A^m}\otimes \mathcal{T}_i \ket{Ent}_{A^{1'}\dots A^{m'}}\quad \ \
\end{eqnarray}
 where $\mathcal{U}_{i=0}=\openone_2$, $\mathcal{U}_{i=1}=Z_{A^k}$, and $\mathcal{U}_{i=2}=X_{A^k}$ for all $k\in[m]$, $\ket{\chi}_{A^1\dots A^m}$ denotes a junk state, $\ket{Ent}_{A^{1'}\dots A^{m'}}$ denotes the entangled state, and $\mathcal{T}_i$ is a unitary local operator acting on the $i^{th}$ subsystem of the entangled state $\ket{Ent}_{A^{1'}\dots A^{m'}}$.

For a general input $n$, the isometries associated with the swap circuit can be expressed as follows.
{\small\begin{eqnarray}
    X_{A^1} &=& \widetilde{A}^1_1,\ 
    Z_{A^1} = \frac{1}{\lfloor \frac{n}{2} \rfloor}\sum_{i_1=0}^{{\lfloor \frac{n}{2}\rfloor}-1}(-1)^{i_1} \frac{\widetilde{A}^1_{i_1+2}-\widetilde{A}^1_{n-i_1}}{\sqrt{2-\langle\{\widetilde{A}^1_{i_1+2},\widetilde{A}^1_{n-i_1}\}\rangle}}\\
    X_{A^k} &=& A^k_1, \  
    Z_{A^k}= \frac{1}{\lfloor \frac{n}{2} \rfloor}\sum_{i_k=0}^{{\lfloor \frac{n}{2}\rfloor}-1}(-1)^{i_k} \frac{A^k_{i_k+2}-A^k_{n-i_k}}{\sqrt{2-\langle\{A^k_{i_k+2},A^k_{n-i_k}\}\rangle}}, k>1\quad \  \ 
\end{eqnarray}}
where $\{Z_{A^k}, X_{A^k}\}=0, \forall k\in[m]$. (See Appendix for detailed derivation.)

We provide an explicit derivation for $m=3$. We
consider $\ket{\psi}_{A^1A^2A^3}\in \mathcal{H}_{A^1}\otimes \mathcal{H}_{A^2} \otimes \mathcal{H}_{A^3}$ and observables $\widetilde{A}^1_{i_1}$, $A^2_{i_2}$, $A^3_{i_3}$ achieving the optimal value of $(\Gamma^3_n)^{opt}_Q$. It can be shown that there exists a local unitary operation $\Phi$ and ancilla state $\ket{000}_{A^{1'}A^{2'}A^{3'}}$ such that,
\begin{eqnarray}\label{phis}
\Phi(\mathcal{U}_{i}\ket{\psi}_{A^1A^2A^3}\otimes\ket{000}_{A^{1'}A^{2'}A^{3'}})=\ket{\chi}_{A^1A^2A^3}\otimes \mathcal{T}_i\ket{GHZ}_{A^{1'}A^{2'}A^{3'}}\nonumber\\
\end{eqnarray}
 where $\mathcal{U}_{i=0}=\openone_2$, $\mathcal{U}_{i=1}=Z_{A^k}$, $\mathcal{U}_{i=2}=X_{A^k},\forall k\in[3]$ and  $\ket{\chi}_{A^1A^2A^3}=\frac{1}{2}(\openone +X_{A^1}+X_{A^2}+X_{A^3})\ket{\psi}_{A^1A^2A^3}$ is a junk state. We can express $A^k_{i_k}$ in terms of $Z_{A^k}$ and $ X_{A^k}$ for all $k\in[3]$ therefore,
\begin{eqnarray}
 &&\Phi\Bigl((A^1_{i_1}\otimes A^2_{i_2}\otimes A^3_{i_3})\ket{\psi}_{A^1A^2A^3}\otimes\ket{000}_{A^{1'}A^{2'}A^{3'}}\Bigr)\nonumber\\
 &&\hspace{1cm}=\ket{\chi}_{A^1A^2A^3}\otimes(A^{1'}_{i_1}\otimes A^{2'}_{i_2}  \otimes A^{3'}_{i_3}) \ket{GHZ}_{A^{1'}A^{2'}A^{3'}}\label{phim4}
\end{eqnarray}

We note that the detailed proof for arbitrary $n$ is intricate. We provide the detailed derivation for $n = 3$ in the Appendix, which provides sufficient intuitive insights for arbitrary $n$.


\subsection{Robust self-testing using swap circuit in tripartite scenario} In realistic experimental implementations, noise and imperfections prevent one from reaching the optimal quantum violation. It is therefore essential to quantify the robustness of the self-testing protocol against such deviations. Here we analyze the stability of our scheme in the presence of noise leading to a suboptimal Bell violation. Our analysis does not rely on identifying the physical origin of the imperfections and instead models them as arising from imperfectly implemented observables~\cite{bamps2015sum,Acin2020}.

Let $\widetilde{X}_{A^k}$ and $\widetilde{Z}_{A^k}$ denote imperfect versions of the ideal observables $X_{A^k}$ and $Z_{A^k}$, with $k\in[3]$. We quantify the deviation from the ideal implementation by
\begin{eqnarray}
\label{eq:obs_error}
\|(\widetilde{X}_{A^k}-X_{A^k})\ket{\psi}_{A^1A^2A^3}\|\leq \epsilon_{A^k},
\|(\widetilde{Z}_{A^k}-Z_{A^k})\ket{\psi}_{A^1A^2A^3}\|\leq \alpha_{A^k}\nonumber\\
\end{eqnarray}
where $\epsilon_{A^k},\alpha_{A^k}\geq 0$, with $\epsilon_{A^k}=\alpha_{A^k}=0$ in the ideal case. Due to noise, the observables need not be unitary, and consequently the exact self-testing relations are replaced by approximate constraints of the form
\begin{equation}
\|\widetilde{L}^3_{i_1i_2i_3}\ket{\psi}_{A^1A^2A^3}||\leq \xi_{i_1i_2i_3}.
\end{equation}
As a result, for $m=3$ the robust version of Eq.~(\ref{gamman}) can be written as 
\begin{eqnarray}
\label{eq:gamma_noise}
\Tr[\widetilde{\gamma}^3_n\,\rho_{A^1A^2A^3}]&=&\frac{1}{2}\sum_{i_3=1}^{n}\sum_{{i_1}=1}^{n} w^3_{i_1i_2i_3} \bra{\psi}_{A^1A^2A^3}\widetilde{L}^{3\dagger}_{i_1i_2i_3}\widetilde{L}^3_{i_1i_2i_3}\ket{\psi}_{A^1A^2A^3}\nonumber\\
&=&\frac{1}{2}\sum_{i_3=1}^{n}\sum_{{i_1}=1}^{n} w^3_{i_1i_2i_3}  \ \xi_{i_1i_2i_3}^2,
\end{eqnarray}
leading to a suboptimal quantum value of the Bell inequality,
\begin{equation}
\label{Cn_noise}
\widetilde{(\Gamma^3_n)}_Q
=2n^2\cos\!\left(\frac{\pi}{2n}\right)-\xi,
\end{equation}
where $\xi=\frac{1}{2}\sum_{i_3=1}^{n}\sum_{{i_1}=1}^{n} w^3_{i_1i_2i_3}  \ \xi_{i_1i_2i_3}^2\geq 0$ quantifies the total deviation from the ideal violation.
\begin{figure*}
    \centering
    \includegraphics[width=17cm, height=6cm]{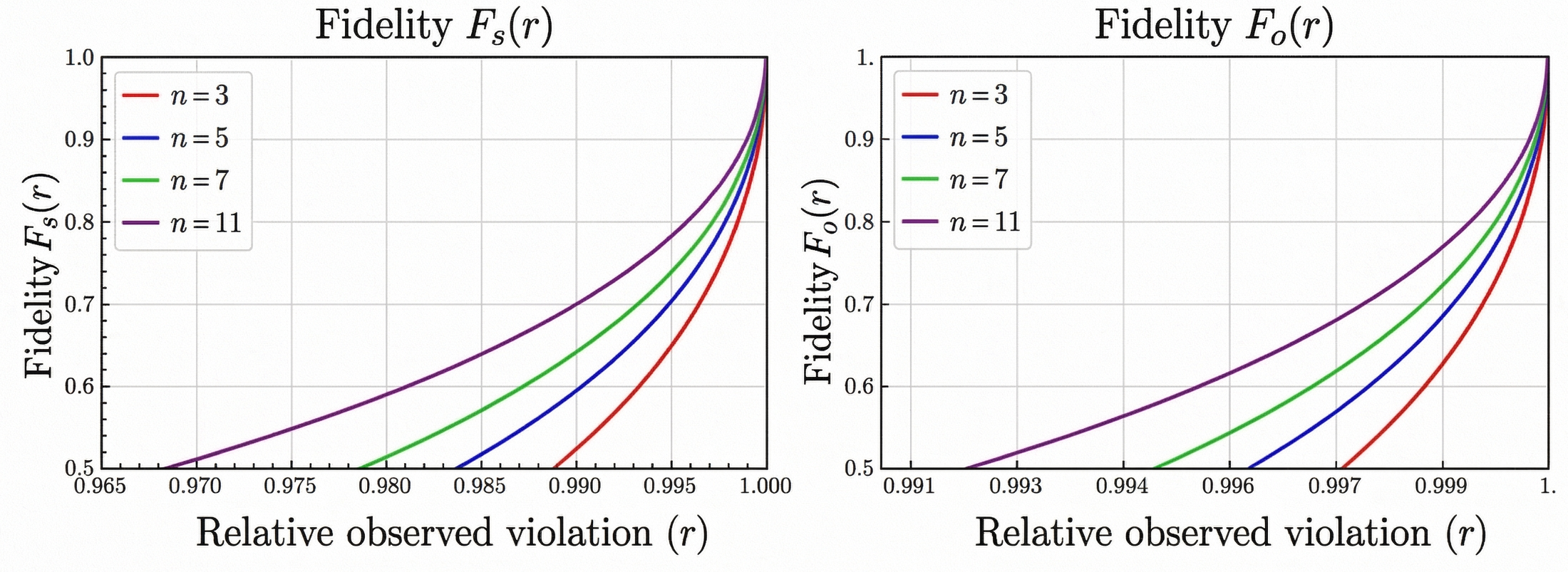}
    \caption{Trade-off between relative observed violation ($r$) and fidelity  $F_s(r)$ and $F_o(r), \forall O\in\{X_{A^3},Z_{A^3}\}$.} 
    \label{osn}
\end{figure*}

To establish robust self-testing of both the state and observables, we compare the outputs of the ideal and noisy swap isometries. Specifically, for $\mathcal{O}_{t=1,2,3}\in\{\openone,\widetilde{X}_{A^k},\widetilde{Z}_{A^k}\}$ we obtain
{\small\begin{eqnarray}
\label{eq:robust_iso}
&&\|\widetilde{\Phi}(\widetilde{\mathcal{O}}_t\ket{\psi}_{A^1A^2A^3}\otimes\ket{000}_{A^{1'}A^{2'}A^{3'}})
-\Phi(\mathcal{O}_t\ket{\psi}_{A^1A^2A^3}\!\otimes\!\ket{000}_{A^{1'}A^{2'}A^{3'}})\|\nonumber\\
&&\leq F_t(\alpha_{A^1},\alpha_{A^2},\alpha_{A^3},\epsilon_{A^1},\epsilon_{A^2},\epsilon_{A^3}),
\end{eqnarray}}
where the functions $F_t=0$ in the ideal self-testing scenario. Explicit expressions and derivations are provided in Appendix.

In practice, imperfections may arise from state preparation, measurements, or both. Since distinguishing these contributions is experimentally challenging, we restrict our attention to errors originating from one party’s measurements. Assuming that only a third party, i.e., $A^3$ implements imperfect observables with identical error $\beta$, we set $\epsilon_{A^3}=\alpha_{A^3}=\beta$. Under this assumption, the deviation of the swap isometry from the ideal one satisfies
\begin{widetext}
\begin{equation}
\begin{aligned}
\|\widetilde{\Phi}(\ket{\psi}_{A^1A^2A^3}\!\otimes\!\ket{000})-\Phi(\ket{\psi}_{A^1A^2A^3}\!\otimes\!\ket{000})\|
&\leq 8\beta +2\beta^2\\
\|\widetilde{\Phi}(\widetilde{O}\ket{\psi}_{A^1A^2A^3}\!\otimes\!\ket{000})
-\Phi(O\ket{\psi}_{A^1A^2A^3}\!\otimes\!\ket{000})\|
&\leq 16\beta+10\beta^2+2\beta^3,\forall O\in\{X_{A^3},Z_{A^3}\}
\end{aligned}
\end{equation}
\end{widetext}

In the Appendix we derive the direct relation between $\beta$ and the observed deviation $\xi$ as
\begin{equation}
\beta=\sqrt{\frac{\xi}{n^2\cos\!\left(\frac{\pi}{2n}\right)}}\label{betab}
\end{equation}
To express the robustness in experimentally accessible terms, we define the relative violation
\begin{equation}
r=\frac{\widetilde{(\Gamma^3_n)}_Q-(\Gamma^3_n)_{BL}}
{(\Gamma^3_n)^{\mathrm{opt}}_Q-(\Gamma^3_n)_{BL}}=1-\frac{\xi}{2 n^2 \cos{\frac{\pi}{2n}}-2n(n-1)},
\end{equation}
where $(\Gamma^3_n)_{BL}=2n(n-1)$ is the bi-local bound. Equivalently,
\begin{equation}
\xi=(1-r)\!\left(2n^2\cos\!\left(\frac{\pi}{2n}\right)-2n(n-1)\right).
\end{equation}

Using this relation, one can bound the trace distance between the extracted and ideal states in terms of the observed relative value $r$ as
\begin{equation}
\label{eq:state_trace}
\|\widetilde{\Phi}(\ket{\psi}_{A^1A^2A^3}\!\otimes\!\ket{000})-\Phi(\ket{\psi}_{A^1A^2A^3}\!\otimes\!\ket{000})\|
\leq f_s(r),
\end{equation}
with an explicit expression for $f_s(r)$ given in Appendix. Finally, applying the Fuchs-van de Graaf inequalities~\cite{wilde2013quantum}, we obtain a lower bound on the fidelity,
\begin{equation}
F_q(r)\geq \left(1-\frac{1}{2}f_q(r)\right)^2,
\forall q\in\{s,o\}
\end{equation}
where $q=s$ and $q=o$ correspond to the state and observables, respectively. Explicit expressions for $f_o(r)$ and their derivations are provided in Appendix.

Figure~\ref{osn} depicts the relationship between the lower bound on the fidelity and the relative observed violation. Employing the swap circuit, both the state and the measurement observables can be robustly extracted whenever the fidelity exceeds $50\%$. For $n=11$, this condition is met at $r=0.968$ $(0\leq \beta \leq 0.0358)$ for the state and at $r=0.992$ $(0\leq \beta \leq 0.0358)$ for the observables. At the point where the state is extracted, the maximum allowed tolerance $\xi$ is limited to $0.04$ for $n=3$, but increases to $0.62$ for $n=11$. This shows that increasing the number of measurement settings $n$ allows for successful extraction even in the presence of smaller observed violations, indicating that relatively weak Bell inequality violations can still be adequate to guarantee the required fidelity threshold. Detailed derivations are provided in Appendix.

We now quantify the robustness of the generated randomness in the presence of noise. The certified randomness is derived as
\begin{equation}
\widetilde{\mathcal{R}}^m_{\max}
=\log_2\left(\frac{8\sqrt{1+\beta^2}}{\sqrt{1+\beta^2}+\beta}\right),
\end{equation}
where the noise parameter $\beta$ captures the deviation of the implemented observables from the ideal ones via
\begin{equation}
\|(\widetilde{A}^3_{i_3}-A^3_{i_3})\ket{\psi}_{A^1A^2A^3}\|\leq \beta,
\quad \forall i_3\in[n].
\end{equation}
Where $\beta$ is defined in Eq.~(\ref{betab}), the complete derivations are presented in the Appendix.


\section{Discussion} In summary, we construct a multipartite Bell inequality with an arbitrary number of measurement settings that detects genuine nonlocality and exhibits a separation between classical, bi-local, and quantum correlations that increases monotonically with $n$. Using an analytic SOS decomposition, we derive the optimal quantum violation without imposing constraints on the Hilbert space dimension, thereby bypassing the need for Jordan's lemma.

A central contribution of this work is the certification of maximal $m$ bits of randomness. A couple of studies have addressed this problem. Based on MABK inequalities, $m$ bits of randomness  can be certified for odd $m$ party~\cite{Dhara2013max,Wooltorton2025expandingbipartite}, but in that case, the correlations do not exhibit GMNL. In ~\cite{Wooltorton2025expandingbipartite}, $m$ bits of randomness were certified, but there remains a question about the genuineness of the nonlocal correlation. Our work closes this gap as we demonstrated the generation $m$ bits of DI randomness at the optimal quantum violation of our inequality, thereby showcasing the maximal multipartite  randomness when the correlations are genuinely nonlocal.

Further, to connect with experimental implementations, we analyze the tripartite scenario and employ a swap circuit based isometry to realize the local maps required for DI self-testing of both the shared state and the measurement observables. In realistic experimental settings, noise can lead to a sub-optimal violation of Bell inequalities. This noise may affect the state, the observables, or both. To strengthen the reliability of our swap-circuit approach, we analyzed the robustness of our protocol in a tripartite scenario in which only one party’s observables are imperfect. This is due to the fact that the bi-local and quantum values become large enough with the number of inputs. As a result, the certification procedure gains resilience against experimental imperfections. We have further demonstrated the robustness of the generated randomness. Although robust self-testing is  explicitly demonstrated for the tripartite case, the framework naturally extends to an arbitrary $m$ number of parties.

Despite these advances, important challenges persist. The current inequality for genuine non-locality applies only to an odd number of measurement settings, motivating future work to generalize the framework to even numbers of settings. These open questions underscore the need for further study to fully harness multipartite quantum certification. As quantum technologies move toward practical deployment, stringent certification tools like the one introduced here will be increasingly crucial. The results presented provide a foundation for future progress in multipartite quantum systems, strengthening the connection between quantum foundations and emerging technologies and supporting advances in secure quantum communication and other quantum applications.
\section{Data availability}
In this study, no data was created or analyzed.
\section{Acknowledgments} R.P. acknowledges the financial support from the Council of Scientific and Industrial Research (CSIR, 09/1001(12429)/2021-EMR-I), Government of India. R.A. acknowledges financial support from the National Science and Technology Council, Taiwan (Grants No. 112-2628-M-006-007-MY4, 114-2811-M-006-069-MY2). A.K.P. acknowledges the support of the research grant I-HUB/PDF/2022-23/06, Government of India.
\section{Author contributions}
R.P., R.A., and A.K.P. conceptualized and designed the study. R.P. and R.A. conducted the technical components of the proofs. A.K.P. supervised the overall progress of the project. All authors contributed substantially to the analysis and interpretation of the results and to the preparation and revision of the manuscript.
\section{Competing interests}
The authors declare no conflict of interest.
\appendix
\onecolumngrid
\section{Derivation of the optimal quantum values for arbitrary \texorpdfstring{$m$}{m} party with \texorpdfstring{$n$}{n} input}
In a multi-partite Bell scenario, a common source distributes physical systems to all $m$ parties. Each party receives $n$ inputs ($ n \geq 3 $ and odd), with measuring $i_k \in \{A^k_1, \ldots, A^k_n\},\forall k\in[m] $. Each measurement yields one of two possible outcomes: $a_k\in\{+1, -1\}$.
Given the scenario, we propose the following multipartite Bell functional,
\begin{eqnarray}\label{generala}
    \Gamma^m_n&=&\sum_{i_m=1}^n\cdot\cdot\sum_{i_3=1}^n \qty(\sum_{{i_1}=1}^n(A^1_{i_1}-A^1_{{i_1}\oplus_n+1})\otimes A^2_{({i_1}+i_3-2)\oplus_n +1 })\otimes_{k=3}^{m-1} A^k_{f_k(i_k,i_{k+1})}\otimes A_{i_m}^m, 
\end{eqnarray}
where, we introduce $i_2$ in the later section via the definition $i_2\equiv ({i_1}+i_3-2)\oplus_n +1$ and $f_k$s are defined as follows
\begin{eqnarray}
    && f_k(i_{k+1},i_k)=(i_{k+1}+i_k-2) \oplus_n +1, \text{for} \  m>3
\end{eqnarray}
We use an analytical method, the sum-of-squares approach \cite{bamps2015, Pan2021} to determine the optimal quantum value of $\Gamma_n^m$. Define a positive semi-definite operator $\gamma_n^m$, such that $ \gamma_n^m = \beta_n^m \openone_d - \Gamma_n^m$, where $\beta_n^m > 0$. Considering a suitable set  operators $L_{i_1 i_2 \dots i_m}^m$, we can write,
\begin{equation}\label{gammana}
    \gamma_n^m =\frac{1}{2} \sum_{i_m=1}^n\cdot\cdot\sum_{i_3=1}^n \sum_{{i_1}=1}^n w_{i_1i_2\dots i_m} L_{i_1i_2\dots i_m}^{m^{\dag}} L^m_{i_1 i_2 \dots i_m}
\end{equation}
where the $w_{i_1 i_2 \dots i_m}$ are suitable normalizers. We define the operators $L^m_{i_1 i_2 \dots i_m}$ as follows, keeping in mind that $L^m_{i_1 i_2 \dots i_m}$ is not unique. However, in this work we adopt this particular form motivated by the structure of the Bell
functional.
\begin{eqnarray}\label{SOSn}
    L^m_{i_1 i_2 \dots i_m} &=& \openone_d\otimes  A^2_{(i_1+i_3-2)\oplus_n +1 }\otimes_{k=3}^{m-1} A^k_{f_k(i_{k+1},i_k)}\otimes A_{i_m}^m - \widetilde{A}^1_{i_1} \otimes \cdots \otimes \openone_d 
\end{eqnarray}
with $\widetilde{A}^1_{i_1}= \frac{A^1_{i_1}-A^1_{{i_1}\oplus_n+1}}{w_{i_1i_2\dots i_m}}$ and $w_{i_1i_2\dots i_m} = \lVert A^1_{i_1}-A^1_{{i_1}\oplus_n+1} \rVert_{\rho_{A^1..A^m}}$. 

Putting Eq.~(\ref{SOSn}) in Eq.~(\ref{gammana}) and further re-arranging,
\begin{equation}
   \Tr[\gamma_n^m \ \rho_{A^1..A^m}] = \sum_{i_m=1}^n\cdot\cdot\sum_{i_3=1}^n \sum_{{i_1}=1}^n  w_{i_1i_2\dots i_m}^m - \Tr[\Gamma_n^m \ \rho_{A^1..A^m}] 
\end{equation}

Clearly, the optimal quantum value of $\Gamma_n^m$ can be obtained when $\Tr[\gamma_n^m \ \rho_{A^1..A^m}]=0$. And this further implies 
\begin{eqnarray}\label{lili}
   \sum_{i_m=1}^n\cdot\cdot\sum_{i_3=1}^n \sum_{{i_1}=1}^n \Tr[  L^{m\dagger}_{i_1 i_2 \dots i_m} L^m_{i_1 i_2 \dots i_m}\rho_{A^1..A^m}]=0
\end{eqnarray}
As $L^{m\dagger}_{i_1 i_2 \dots i_m} L^m_{i_1 i_2 \dots i_m}$ is positive semi-definite operators, Eq.~(\ref{lili}) leads to $\Tr[L^m_{i_1 i_2 \dots i_m} \rho_{A^1..A^m}]=0,\forall i_k\in[n],k\in[m]$, which further implies 
\begin{eqnarray}\label{SOSc}
    \widetilde{A}^1_{{i_1}}\otimes  A^2_{({i_1}+i_3-2)\oplus_n +1 }\otimes_{k=3}^{m-1} A^k_{f_k(i_{k+1},i_k)}\otimes A_{i_m}^m\ket{\psi}_{A^1..A^m}=\ket{\psi}_{A^1..A^m}
\end{eqnarray}
Therefore, the optimal quantum value is given by
\begin{eqnarray}
(\Gamma_n^m)_Q^{\text{opt}} = \Tr[\Gamma_n^m \,\rho_{A^1\ldots A^m}] 
= \max\qty(\sum_{i_m=1}^n \cdots \sum_{i_3=1}^n \sum_{i_1=1}^n w_{i_1 i_2 \dots i_m})
\end{eqnarray}
where the coefficients \(w_{i_1 i_2 \dots i_m}\) take the form
\(
w_{i_1 i_2 \dots i_m} = \sqrt{2 - \langle \{A^1_{i_1}, A^1_{i_1 \oplus_n + 1}\} \rangle},
\)
and we use the shorthand \(\langle \dots \rangle_{\rho_{A^1\ldots A^m}} = \langle \dots \rangle\). In a nutshell, we can write $w_{i_1i_2\dots i_m}\equiv w_{i_1}$, where $i_1\equiv i_1i_2\dots i_m, \ \forall {i_k}\in[n],k\in[m]$. Applying the concavity inequality $\sum_{i_1} w_{i_1} \leq \sqrt{n\sum_{i_1} w_{i_1}^2}$, we obtain
\vspace{-0.18cm}
\begin{equation}
 (\Gamma_n^m)_Q =\max\qty(\sum_{i_m=1}^n \cdots \sum_{i_3=1}^n \sum_{i_1=1}^n w_{i_1})\leq n^{m-2}\sqrt{n\left( \sum_{{i_1}=1}^{n} w_{i_1}^2\right) }= n^{m-2}\sqrt{n \left(2n-\left(\sum_{{i_1}=1}^{n}\langle \{A_{{i_1}}^1,A_{i_1+1}^1\} \rangle\right)\right)}
\end{equation}
where we used $\langle \{A^1_{i_1},A^1_{{i_1}\oplus_n+1}\} \rangle_{\rho_{A^1..A^m}}=\langle \{A^1_{i_1},A^1_{i_1+1}\} \rangle_{\rho_{A^1..A^m}},\forall i_1\in[n], A^1_{n+1}=A^1_1$.
Considering $\Delta^m_n = \sum_{i_1=1}^{n} A_{i_1}^1$, we obtain
\[
(\Delta^m_n)^2 = n\,\openone_d + \sum_{i_1=1}^{n} \{A^1_{i_1},A^1_{i_1+1}\} + \sum_{i=3}^{n-1}\{A_1,A_i\} + \sum_{i_1=2}^{n-2}\sum_{i=i_1+2}^{n} \{A^1_{i_1},A^1_i\}.
\]
Further we define
\[
\delta^m_{1,n} = \sum_{i_1=1}^{n} \{A^1_{i_1},A^1_{i_1+1}\}, \quad
\delta^m_{2,n} = \sum_{i=3}^{n-1}\{A_1,A_i\} + \sum_{i_1=2}^{n-2}\sum_{i=i_1+2}^{n} \{A^1_{i_1},A^1_i\}.
\]
Then it follows that
\[
-\delta^m_{1,n} = n\,\openone_d + \delta^m_{2,n} - (\Delta^m_n)^2.
\] Hence, we have,
\vspace{-0.2cm}
\begin{equation}
\begin{split}
 (\Gamma^m_n)_Q &\leq  n^{m-2}\sqrt{n \left(2n+\langle n \cdot \openone_d+\delta^m_{2,n} - (\Delta^m_{n})^2 \rangle\right)} = n^{m-2}\sqrt{n \left(3n + \langle \delta^m_{2,n} \rangle - \langle (\Delta^m_{n})^2 \rangle\right)}
\end{split}
\end{equation}

Now, we will obtain the optimal value of $(\Gamma^m_n)_Q$ when $(\Delta^m_{n})^2=0$, i.e., $\Delta^m_{n}=0$. Further optimizing $\langle \delta^m_{2,n} \rangle$, for ${i_1} \in [n]$ and $x \in [n-{i_1}]$ we get,
\vspace{-0.25cm}
\begin{eqnarray}
    &&\langle \{A^1_{i_1}, A^1_{i_1+x}\} \rangle= 2(-1)^x\cos{\frac{\pi x}{n}}
\end{eqnarray}
For the remaining parties, we find the following outcomes (a detailed discussion for $m=3$ and $4$ with $n=3$ and $4$ settings is provided in Sec.~\ref{sec34f35}).
\begin{eqnarray}
    &&\langle \{A^k_{i_k}, A^k_{i_k+x}\} \rangle= 2(-1)^x\cos{\frac{\pi x}{n}},\forall k\in[m],i_k\in[n]
\end{eqnarray}
The optimal value we get $(\Gamma_n^m)_Q^{opt}=2n^{m-1}\cos {\frac{\pi}{2n}}$. The derivation is sufficiently intricate that we have added the detailed steps in the following sections.
\section{Local and Bi-local value}\label{lbi}
Here we proved that,
in a fully local model, the inequality in Eq.~(\ref{generala}) is loosely upper bounded by $(\Gamma_n^m)_L \leq 2 (n-1)^{m-1}$, whereas in bi-local scenarios, the maximum value it can attain is $2 n^{m-2}(n-1)$. By contrast, the optimal quantum value is $(\Gamma_n^m)_Q^{opt} = 2 n^{m-1}\cos\bigl(\tfrac{\pi}{2n}\bigr)$, which implies a strict difference $(\Gamma_n^m)_Q^{opt} > (\Gamma_n^m)_{BL} > (\Gamma_n^m)_L$.  \\

\subsection{The derivation of local value of \texorpdfstring{$\Gamma_n^m$}{Gamma\_n\^{}m}} To derive the local value of $\Gamma_n^m$, we first focus on the simplest bipartite Bell scenario, with two spatially separated parties, each choosing from a finite set of dichotomic measurements. In this framework, the Bell expression under study can be expressed as
\begin{eqnarray}
\label{biparty}
\Gamma^{2}_n
&=& (A^1_1 - A^1_2) \otimes A^2_1
   + (A^1_2 - A^1_3)\otimes A^2_2
   + \cdots + (A^1_{n-1} - A^1_n)\otimes A^2_{n-1}
   + (A^1_n - A^1_1)\otimes A^2_n ,
\end{eqnarray}
We next determine the maximal value of $\Gamma^{2}_n$ attainable by local hidden variable (LHV) model. Because the set of local correlations forms a convex polytope, its extreme points are deterministic strategies, where all measurement outcomes are predetermined. Thus, it is enough to restrict attention to deterministic assignments $A^1_{i_1}, A^2_{i_2} \in \{-1,+1\},\forall i_1,i_2\in[n]$.

For any such deterministic assignment, the absolute value of $\Gamma^{2}_n$ can be bounded as
\begin{eqnarray}
|\Gamma^{2}_n|
&\leq&
|A^1_1 - A^1_2|
+ |A^1_2 - A^1_3|
+ \cdots
+ |A^1_n - A^1_1| .
\end{eqnarray}
Each term $|A^1_{i_1} - A^1_{i_1+1}|,\forall i_1\in[n], A^1_{n+1}=A^1_1$ can have a maximum value $2$, since $A^1_{i_1}= \pm 1$. However, due to the cyclic structure of the expression and the fact that $n$ is odd, it is impossible for all $n$ terms to simultaneously reach the maximum value of $2$. Indeed, for any deterministic assignment, at most $n-1$ of these terms can be equal to $2$, while the remaining term must vanish. As a result, the local bound of $\Gamma_n^2$ is given by
\begin{equation}\label{bigmn}
|\Gamma^{2}_n|_{L} \leq 2(n-1).
\end{equation}

This bound is tight and can be saturated by an explicit deterministic strategy. One such assignment is obtained by choosing
\begin{equation}
A^1_{i_1} =
\begin{cases}
+1, & \text{if $i_1$ is odd} \\
-1, & \text{if $i_1$ is even}
\end{cases}
,\qquad
A^2_{i_2} =
\begin{cases}
+1, & \text{if $i_2$ is odd} \\
-1, & \text{if $i_2$ is even}
\end{cases}
\end{equation}
In this configuration, precisely $n-1$ of the differences $|A^1_{i_1} - A^1_{i_1+1}|$ equal $2$, so $|\Gamma^{2}_n|_{L} = 2(n-1)$, which is maximal.

We now extend the bipartite expression $\Gamma_n^m$ in Eq.~(\ref{bigmn})  to the tripartite scenario, following a recursive construction. The idea is to build higher-party Bell expressions by combining relabeled copies of lower-party Bell expressions. Hence, the tripartite Bell expression is defined as
\begin{eqnarray}
\label{triparty}
\Gamma^3_n
&=& \Gamma^{2,1}_n \otimes A^3_1
  + \Gamma^{2,2}_n \otimes A^3_2
  + \cdots
  + \Gamma^{2,n}_n \otimes A^3_n ,
\end{eqnarray}
where $A^3_{i_3}$, represents the $i_3$-th dichotomic observable of the third party for all $i_3\in[n]$. Note that, the operators $\Gamma^{2,i_3}_n$ are relabeled versions of the bipartite expression $\Gamma^2_n$, obtained by cyclically shifting the measurement settings of the second party, while the observables of the first party remain unchanged..

Explicitly, these relabeled expressions take the form
\begin{eqnarray}
\Gamma^{2,1}_n
&=& (A^1_1 - A^1_2)\otimes A^2_1
   + (A^1_2 - A^1_3)\otimes A^2_2
   + \cdots
   + (A^1_n - A^1_1)\otimes A^2_n , \nonumber \\
\Gamma^{2,2}_n
&=& (A^1_1 - A^1_2)\otimes A^2_2
   + (A^1_2 - A^1_3)\otimes A^2_3
   + \cdots
   + (A^1_n - A^1_1)\otimes A^2_1 , \nonumber \\
&\vdots& \nonumber \\
\Gamma^{2,n}_n
&=& (A^1_1 - A^1_2)\otimes A^2_n
   + (A^1_2 - A^1_3)\otimes A^2_1
   + \cdots
   + (A^1_n - A^1_1)\otimes A^2_{n-1} .
\end{eqnarray}

We now analyze the maximum value of $\Gamma^3_n$ achievable by fully local hidden variable models. As before, it is sufficient to restrict attention to deterministic strategies. Since the local correlation set is convex and its extremal points are deterministic strategies. For any fixed deterministic assignment of outcomes, each individual expression $\Gamma^{2,i_3}_n$ is bounded by
\begin{equation}
|\Gamma^{2,i_3}_n| \leq 2(n-1).
\end{equation}
However, due to the cyclic relabeling structure, there does not exist a single deterministic assignment for which all $\Gamma^{2,i_3}_n$ simultaneously attain this maximal value. Indeed, for any fixed assignment of the observables $A^1$ and $A^2$, at least one of the expressions $\Gamma^{2,i_3}_n$ necessarily evaluates to zero. Intuitively, the pattern of signs that maximizes one cyclic shift forces destructive cancellations in at least one other shift.

Consequently, at most $n-1$ of the terms in Eq.~(\ref{triparty}) can contribute non-trivially. Using the fact that $|A^3_{i_3}|=1$ for all $i_3$, we obtain the following upper bound on the local value:
\begin{equation}
|\Gamma^3_n|_{L}
\leq (n-1)\times 2(n-1)
= 2(n-1)^2 .
\end{equation}
We emphasize that this bound is generally not tight; it merely provides an upper bound on the fully local value of the tripartite inequality.

The recursive construction naturally generalizes to an arbitrary number of parties. For $m$ parties, we define the Bell expression
\begin{eqnarray}
\label{general1}
\Gamma^m_n
&=& \Gamma^{m-1,1}_n \otimes A^m_1
  + \Gamma^{m-1,2}_n \otimes A^m_2
  + \cdots
  + \Gamma^{m-1,n}_n \otimes A^m_n ,
\end{eqnarray}
where each $\Gamma^{m-1,i_m}_n$ is obtained from $\Gamma^{m-1}_n$ by an appropriate cyclic relabeling of measurement settings, and $A^m_{i_m}$ denotes the $i_m$-th dichotomic observable of the $m$-th party.

Applying the same reasoning recursively, one finds that no deterministic assignment can simultaneously maximize all the $(m-1)$-party expressions $\Gamma^{m-1,i_m}_n$. At each recursive step, at least one relabeled term must vanish, leading to an additional factor of $(n-1)$ in the maximal local contribution.

As a result, the fully local hidden variable value of $\Gamma^m_n$ is upper-bounded by
\begin{equation}
|\Gamma^m_n|_{L}
\leq 2 (n-1)^{m-1}.
\end{equation}
This establishes a general upper bound on the local correlations achievable for the recursively defined multipartite Bell expressions $\Gamma^m_n$.
\subsection{The derivation of Bi-local value of \texorpdfstring{$\Gamma_n^m$}{Gamma	extsubscript{n}	extasciicircum m}} We are now in a position to evaluate the maximum bi-local value of the multipartite Bell expression $\Gamma^m_n$. Recall that a bi-local model corresponds to correlations that are local with respect to a fixed bi-partition of the $m$ parties. We show, by mathematical induction on the number of parties $m$, that the maximal bi-local value of $\Gamma^m_n$ is given by
\begin{equation}
|\Gamma^m_n|_{BL} \le 2 n^{m-2} (n-1).
\end{equation}

\noindent\textbf{Base case: $m=3$.}
We first establish the claim for the tripartite scenario, namely that
\begin{equation}
|\Gamma^3_n|_{BL}\leq 2n(n-1)
\end{equation}
for any bi-local model.

Consider the bipartition $A^1 / A^2 A^3$, where $A^2A^3$ are allowed to share arbitrary correlations while remaining local with respect to $A^1$. In this setting, $A^2A^3$ can be viewed as a single effective party implementing a coordinated strategy. The Bell operator admits the decomposition
\begin{equation}\label{eq:trigen}
\Gamma^3_n = \sum_{i_3=1}^n \Gamma^{2,i_3}_n \otimes A^3_{i_3},
\end{equation}
where the measurement choice $i_3$ of $A^3$ selects the corresponding bipartite operator $\Gamma^{2,i_3}_n$ acting on $A^1A^2$. Since $A^2$ and $A^3$ share arbitrary correlations, the value of $i_3$ can be treated as classical side information available to $A^2$. Consequently, $A^2$ can condition its deterministic strategy on $i_3$ and optimize each term $\Gamma^{2,i_3}_n$ independently.

Since each $|\Gamma^{2,i_3}_n|\leq 2(n-1)$, all $n$ contributions in Eq.~\eqref{eq:trigen} can be simultaneously maximized under the bipartition $A^1 / A^2A^3$. This yields the bound
\begin{equation}
|\Gamma^3_n|_{BL} = 2n(n-1).
\end{equation}
The bound is achievable via a deterministic strategy. For example, let $(A^1_{\text{odd}},A^1_{\text{even}})=(+1,-1)$ and $A^3_{i_3}=+1$. Exploiting shared correlations with $A^3$, party $A^2$ conditions its outputs on the index $i_3$: for odd $i_3$, choose $(A^2_{\text{odd}},A^2_{\text{even}})=(+1,-1)$, while for even $i_3$, choose $(A^2_{\text{odd}},A^2_{\text{even}})=(-1,+1)$. This strategy ensures that each $\Gamma^{2,i_3}_n$ attains its maximal value $2(n-1)$.

By symmetry of the Bell expression $\Gamma^3_n$ under permutations of the parties, the same maximal value is obtained for any other bipartition of the three parties. This completes the base case. Similarly  for $m=4$ the Bell operator admits the following decomposition 
\begin{equation}\label{eq:tetragen}
\Gamma^4_n = \sum_{i_4=1}^n \Gamma^{3,i_4}_n \otimes A^4_{i_4},
\end{equation}
 where $\Gamma^{3,i_4}_n$ attains the maximum value $2n(n-1)$ and Eq.~(\ref{eq:tetragen}) maximizes when $A^3_{i_3}=+1$, which gives
 \begin{equation}
|\Gamma^4_n|_{BL} = 2n^2(n-1).
\end{equation}

\noindent\textbf{Induction hypothesis.}
Suppose that for an $(m-1)$-partite Bell expression $\Gamma^{m-1}_n$, the maximum bi-local value obeys
\begin{equation}
|\Gamma^{m-1}_n|_{BL} = 2 n^{m-3} (n-1)
\end{equation}
for any bipartition of the $(m-1)$ parties.

\noindent\textbf{Induction step.}
We extend the argument to $m$ parties by induction. Consider an arbitrary bipartition $(A^1\cdots A^k)/(A^{k+1}\cdots A^{m})$ and the recursive decomposition
\begin{equation}
\Gamma^m_n=\sum_{i_m=1}^n \Gamma^{m-1,i_m}_n \otimes A^m_{i_m}.
\end{equation}
Since $A^m$ is grouped with $A^{k+1}\cdots A^{m-1}$, the joint system on this side can treat the measurement choice $i_m$ as classical side information selecting the corresponding $(m\!-\!1)$-party operator $\Gamma^{m-1,i_m}_n$. They can therefore condition their deterministic strategy on $i_m$ and optimize each term independently. By the induction hypothesis, each term is bounded by $|\Gamma^{m-1,i_m}_n|_{BL}\leq 2\,n^{m-3}(n-1)$. Summing over the $n$ choices of $i_m$ yields
\begin{equation}
|\Gamma^m_n|_{BL} \leq 2n^{m-2}(n-1).
\end{equation}


Once again, the permutation symmetry of $\Gamma^m_n$ ensures that the same bound holds for any bipartition of the $m$ parties. This completes the induction step.
We therefore conclude that, for all $m \geq 3$, the maximal value of $\Gamma^m_n$ attainable by bi-local models is
\begin{equation}
|\Gamma^m_n|_{BL} = 2 n^{m-2} (n-1),
\end{equation}
which establishes the desired result.

\section{Derivation of the optimal quantum values for \texorpdfstring{$m=3,4$}{m=3,4} and \texorpdfstring{$n=3,5$}{n=3,5}}\label{sec34f35}
We will present detailed calculations for the tripartite and quadripartite cases, considering three and five measurement settings, respectively.

\subsection{Detailed calculation for \texorpdfstring{$m=3$ and $n=3$}{m=3 and n=3}}
Consider $Alice^1$, $Alice^2$, and $Alice^3$ have three possible measurement choices $x_1 \in \{A^1_1, A^1_2, A^1_3\}$, $x_2 \in \{A^2_1, A^2_2, A^2_3\}$ and $x_3 \in \{A^3_1, A^3_2, A^3_3\}$ respectively. Now,  the Bell functional $\Gamma^3_3$ has the following form,

\begin{equation}\label{ineq1}
\begin{split}
   \Gamma^3_3 = ((A^1_1-A^1_2)\otimes A^2_1+(A^1_2-A^1_3)\otimes A^2_2+(A^1_3-A^1_1)\otimes A^2_3)\otimes A^3_1&\\+((A^1_1-A^1_2)\otimes A^2_2+(A^1_2-A^1_3)\otimes A^2_3+(A^1_3-A^1_1)\otimes A^2_1)\otimes A^3_2&\\+((A^1_1-A^1_2)\otimes A^2_3+(A^1_2-A^1_3)\otimes A^2_1+(A^1_3-A^1_1)\otimes A^2_2)\otimes A^3_3    
\end{split}
\end{equation}

Now, we will derive the optimal quantum value of $\Gamma^3_3$ using the sum-of-squares approach. Note that throughout the derivation, there is no restriction on the dimension of the underlying Hilbert space, which makes it device-independent.  We define the positive semi-definite operator $\gamma^3_3$, such that $ \gamma^3_3 = \beta^3_3 \openone_d - \Gamma^3_3$, where $\beta^3_3>0$. Considering a suitable set of operators $L^3_{i_1i_2i_3}$, we can write,

\begin{equation}
    \gamma^3_3 = \frac{1}{2}\sum_{i_3=1}^{3}\sum_{{i_1}=1}^{3} w_{i_1i_2i_3 } L^{3\dag}_{i_1i_2i_3} L^3_{i_1i_2i_3 }
\end{equation}

where $w_{i_1i_2i_3}$'s are the suitable normalizers and the positive operator $L^3_{i_1i_2i_3 }$ defined as
\begin{equation}
    L^3_{i_1i_2i_3 } = \openone_d \otimes A^2_{(i_1+i_3-2)\oplus_3 +1}\otimes A^3_{i_3}  - \widetilde{A}^1_{i_1} \otimes \openone_d \otimes \openone_d 
\end{equation}

with $\widetilde{A}^1_{i_1} = \left(\frac{A^1_{i_1}-A^1_{i_1\oplus_3+1}}{w_{i_1{i_2}i_3}}\right)$ and $w_{i_1i_2i_3} = \lVert A^1_{i_1}-A^1_{i_1\oplus_3+1} \rVert_{\rho_{A^1 A^2 A^3}}$, where $\lVert . \rVert$ is the Frobenious norm, given by $\lVert \mathcal{O} \rVert=\sqrt{\Tr[\mathcal{O}^{\dagger} \mathcal{O} \ \rho]}$.

So, $w_{i_1i_2i_3}$ becomes $\lVert A^1_{i_1}-A^1_{i_1\oplus_3+1} \rVert_{\rho_{A^1 A^2 A^3}} = \sqrt{2-\langle \{A^1_{i_1},A^1_{i_1\oplus_3+1}\} \rangle}$, where $\langle \cdots \rangle_{\rho_{A^1A^2A^3}}=\langle \cdots \rangle$. Note that as $w_{i_1i_2i_3}=w_{i_1},\forall i_1,i_2,i_3\in[3]$,  which implies $w_1=\lVert A^1_1-A^1_2 \rVert_{\rho_{A^1 A^2 A^3}},w_2=\lVert A^1_2-A^1_3 \rVert_{\rho_{A^1 A^2 A^3}}, w_3=\lVert A^1_3-A^1_1 \rVert_{\rho_{A^1 A^2 A^3}}$. Hence, $\Tr[\gamma^3_3 \ \rho_{A^1 A^2 A^3}]  = 3(w_1+w_2+w_3) - \Tr[\Gamma^3_3 \ \rho_{A^1 A^2 A^3}]$. Hence, the optimal value of $\Gamma^3_3$ is obtained when $\Tr[\gamma^3_3 \ \rho_{A^1 A^2 A^3}] = 0$, which further implies 
\begin{eqnarray}
    \widetilde{A}^1_{i_1}\otimes  A^2_{(i_1+i_3-2)\oplus_3+1 }\otimes A^3_{i_3}\ket{\psi}_{A^1A^2A^3}=\ket{\psi}_{A^1A^2A^3},\forall i_1,i_3\in[3]\label{slf3}
\end{eqnarray}
Hence we will have, $\langle \Gamma^3_3 \rangle^{opt}_Q = \max [3(w_1+w_2+w_3)]$. Using the concavity inequality, we can have,
\begin{equation}
\begin{split}
  \langle \Gamma^3_3 \rangle_Q & \leq 3\sqrt{3(w_1^2 + w_2^2 + w_3^2)}= 3\sqrt{3[6-(\langle \{A^1_1,A^1_2\} \rangle+ \langle \{A^1_2,A^1_3\} \rangle + \langle \{A^1_3,A^1_1\} \rangle)]}
\end{split}   
\end{equation}

Now let us consider $\Delta^3_3 = A^1_1 + A^1_2 + A^1_3$, then, $(\Delta^3_3)^2 = 3\openone_d+\{A^1_1,A^1_2\}+\{A^1_2,A^1_3\}+\{A^1_3,A^1_1\}$. Eventually, we have $\langle \Gamma^3_3 \rangle_Q \leq 3\sqrt{3[9-\langle (\Delta^3_3)^2 \rangle]}$. $\langle \Gamma^3_3 \rangle^{opt}_Q $ will have a maximum value of $9\sqrt{3}$ if and only if $\langle (\Delta^3_3)^2 \rangle = 0\Rightarrow A^1_1 + A^1_2 + A^1_3=0$; i.e., 
\begin{eqnarray}\label{a13}
    \langle \{A^1_{i_1},A^1_{i_1+x}\} \rangle = 2(-1)^x \cos{\frac{\pi x}{3}},\quad \forall i_1\in[3], x\in[3-i_1]
\end{eqnarray}
Furthermore, using the relations above, we can also obtain $\langle \{\widetilde{A}^1_{i_1},\widetilde{A}^1_{i_1+1}\} \rangle = 2(-1)^x \cos{\frac{\pi x}{3}},\forall i_1\in[3], x\in[3-i_1],\widetilde{A}^1_4=\widetilde{A}^1_1$. Analogously, to determine the relation between the observables of the remaining parties, we put $i_3=1$ in Eq.~(\ref{slf3}) and get
\begin{eqnarray}\label{slf341}
    &&\widetilde{A}^1_{i_1}\otimes A^2_{(i_1-1)\oplus_3 +1} \otimes  A^3_{1}\ket{\psi}_{A^1A^2A^3}=\ket{\psi}_{A^1A^2A^3},\quad i_1\in[3]\nonumber\\
    &&\widetilde{A}^1_{i_1}\otimes A^2_{i_1} \otimes  A^3_{1}\ket{\psi}_{A^1A^2A^3}=\ket{\psi}_{A^1A^2A^3},\quad \text{As $i_1-1<3$}
\end{eqnarray}
Now putting $i_1=i_2$ and $i_1=i_2+x$ in Eq.~(\ref{slf341}) we get
\begin{eqnarray}
    &&\begin{cases}
        \openone_d\otimes A^2_{i_2} \otimes \openone_d\ket{\psi}_{A^1A^2A^3} = \widetilde{A}^1_{i_2}\otimes \openone_d \otimes A^3_1\ket{\psi}_{A^1A^2A^3} \\
        \openone_d\otimes A^2_{i_2+x} \otimes \openone_d\ket{\psi}_{A^1A^2A^3} = \widetilde{A}^1_{i_2+x}\otimes \openone_d \otimes A^3_1\ket{\psi}_{A^1A^2A^3}
    \end{cases} \hspace{-0.8cm}\Rightarrow \langle \{A^2_{i_2},A^2_{i_2+x}\}\rangle=2(-1)^x\cos{\frac{\pi x}{3}},\forall i_2\in[3],x\in[3-i_2]\qquad
    \end{eqnarray}
Similarly putting $i_1=1$ in Eq.~(\ref{slf3}) we get
\begin{eqnarray}\label{slf33}
    \widetilde{A}^1_{1}\otimes A^2_{i_3} \otimes  A^3_{i_3}\ket{\psi}_{A^1A^2A^3}=\ket{\psi}_{A^1A^2A^3}
\end{eqnarray}
Again putting $i_3=i_3+x$ in Eq.~(\ref{slf33}) and using previous approach we get $\langle \{A^3_{i_3},A^3_{i_3+x}\}\rangle=2(-1)^x\cos{\frac{\pi x}{3}},\forall x\in[3-i_3]$.Hence, we may express the relationships between the observables for any given party as
\begin{eqnarray}
    \langle \{A^k_{i_k},A^k_{i_k+x}\}\rangle=2(-1)^x\cos{\frac{\pi x}{3}},\forall x\in[3-i_k],k\in[3]
\end{eqnarray}

\subsection{Detailed calculation for \texorpdfstring{$m=3$ and $n=5$}{m=3 and n=5}}\label{sec35}
Next, we examine the inequality in the case of five inputs. In this setting, all three parties $Alice^k,\ \forall k \in [3]$ each have five possible measurement settings, denoted by $i_k \in \{A^k_1, \ldots, A^k_5\}$, respectively. In this scenario, our inequality takes the form,

\begin{equation}\label{ineq}
\begin{split}
   \Gamma^3_5 = &( (A^1_1-A^1_2)\otimes A^2_1+ (A^1_2-A^1_3)\otimes A^2_2+ \ldots + (A^1_5-A^1_1)\otimes A^2_5)\otimes A^3_1 \\+& ((A^1_1-A^1_2)\otimes A^2_2 + (A^1_2-A^1_3)\otimes A^2_3+ \ldots + (A^1_5-A^1_1)\otimes A^2_1)\otimes A^3_2\\+&( (A^1_1-A^1_2)\otimes A^2_3+ (A^1_2-A^1_3)\otimes A^2_4+ \ldots + (A^1_5-A^1_1)\otimes A^2_2)\otimes A^3_3 \\+& ((A^1_1-A^1_2)\otimes A^2_4 + (A^1_2-A^1_3)\otimes A^2_5+ \ldots + (A^1_5-A^1_1)\otimes A^2_3)\otimes A^3_4\\+& ((A^1_1-A^1_2)\otimes A^2_5 + (A^1_2-A^1_3)\otimes A^2_1+ \ldots + (A^1_5-A^1_1)\otimes A^2_4)\otimes  A^3_5 
\end{split}
\end{equation}

It can be shown that a violation of the inequality by local correlation is upper bounded by $32$, and any bi-local correlation can provide a violation of at most $40$. Now, we will find the optimal quantum violation using the same approach as before for the $n=3$ case.  We define the positive semi-definite operator $\gamma^3_5$, such that $ \gamma^3_5 = \beta^3_5 \openone_d - \Gamma^3_5$, where $\beta^3_5$ is a positive quantity. Considering a suitable set of operators $L^3_{i_1i_2i_3}$, we can write,

\begin{equation}
    \gamma^3_5 = \frac{1}{2}\sum_{i_3=1}^{5}\sum_{{i_1}=1}^{5} w_{i_1i_2i_3 } L{^{3\dag}_{i_1i_2i_3}} L^3_{i_1i_2i_3}
\end{equation}

where $w_{i_1i_2i_3}$'s are the suitable normalizers and the positive operator $L^3_{i_1i_2i_3 }$ defined as \begin{equation}
    L^3_{i_1{i_2}i_3 } = \openone_d \otimes A^2_{(i_1+i_3-2)\oplus_5 +1}\otimes A^3_{i_3}  - \widetilde{A}^1_{i_1} \otimes \openone_d \otimes \openone_d 
\end{equation}
with $\widetilde{A}^1_{i_1} = \left(\frac{A^1_{i_1}-A^1_{i_1\oplus_5+1}}{w_{i_1i_2i_3}}\right)$ and $w_{i_1i_2i_3} = \lVert A^1_{i_1}-A^1_{i_1\oplus_5+1} \rVert_{\rho_{A^1 A^2 A^3}}$, where $\lVert . \rVert$ is the Frobenius norm, given by $\lVert \mathcal{O} \rVert=\sqrt{\Tr[\mathcal{O}^{\dagger} \mathcal{O} \ \rho]}$.

So, $w_{i_1i_2i_3}$ becomes $\lVert A^1_{i_1}-A^1_{i_1\oplus_5+1} \rVert_{\rho_{A^1 A^2 A^3}} = \sqrt{2-\langle \{A^1_{i_1},A^1_{i_1\oplus_5+1}\} \rangle}$. In a nutshell we can write, $w_{i_1i_2i_3}=w_{i_1}$,$\forall i_1,i_2,i_3\in[5]$. Hence, $\Tr[\gamma^3_5 \ \rho_{A^1 A^2 A^3}]  = 5\sum_{i_1=1}^5 w_{i_1} - \Tr[\Gamma^3_5 \ \rho_{A^1 A^2 A^3}]$. Hence the optimal value of $\Gamma^3_5$ is obtained when $\Tr[\gamma^3_5 \ \rho_{A^1 A^2 A^3}] = 0$, which further implies 
\begin{eqnarray}\label{slf5}
    \widetilde{A}^1_{i_1}\otimes  A^2_{(i_1+i_3-2)\oplus_5 +1 }\otimes A^3_{i_3}\ket{\psi}_{A^1A^2A^3}=\ket{\psi}_{A^1A^2A^3},\forall i_1,i_3\in[5]
\end{eqnarray}
Hence we will have, $\langle \Gamma^3_5 \rangle^{opt}_Q = \max [5(w_1+w_2+w_3+w_4+w_5)]$. Using the concavity inequality, we can have,
\begin{equation}
\begin{split}
  \langle \Gamma^3_5 \rangle_Q  \leq 5\sqrt{5(w_1^2 +\ldots+ w_5^2)} = 5\sqrt{5\left[10-\sum_{i_1=1}^{5}\langle \{A^1_{i_1},A^1_{i_1+1}\} \rangle\right]}
\end{split}   
\end{equation}

Now let us consider,
\begin{equation}
       \Delta^3_5 =A^1_1+A^1_2+A^1_3+A^1_4+A^1_5
\end{equation}

Then we have,
\begin{equation}
(\Delta^3_5)^2 =5\openone_d+\sum_{i_1=1}^{5}\{A^1_{i_1},A^1_{i_1+1}\}+\sum_{i_1=3}^{4}\{A_1,A_{i_1}\}+\sum_{i_1=2}^{3}\sum_{i=i_1+2}^{5} \{A^1_{i_1},A^1_i\}
\end{equation}

Considering $\delta^3_{1,5}=\sum_{i_1=1}^{5}\{A^1_{i_1},A^1_{i_1+1}\}$ and $\delta^3_{2,5}=\sum_{i_1=3}^{4}\{A_1,A_{i_1}\}+\sum_{i_1=2}^{3}\sum_{i=i_1+2}^{5} \{A^1_{i_1},A^1_i\}$, we have $(\Delta^3_5)^2 =5\openone_d+\delta^3_{1,5}+\delta^3_{2,5}$.

Hence, we have,
\begin{equation}
  \langle \Gamma^3_5 \rangle_Q\leq 5 
  \sqrt{5\left[10+\langle(5\openone_d+\delta^3_{2,5}-(\Delta^3_5)^2)\rangle\right]} 
\end{equation}

$\langle \Gamma^3_5 \rangle_Q$ has maximum value when $(\Delta^3_5)^2 =0$. This implies $\Delta^3_5 =0$. Now, $\delta^3_{2,5}=\{A^1_1,A^1_3+A^1_4\} +\{A^1_2,A^1_4+A^1_5\} +\{A^1_3,A^1_5\}$. Considering $A^1_1=\frac{A^1_3+A^1_4}{\nu_{34}}$ and $A^1_2=\frac{A^1_4+A^1_5}{\nu_{45}}$, where $\nu_{34} = \lVert (A^1_3+A^1_4) \rVert_{\rho_{A^1A^2A^3}} = \sqrt{2+\langle \{A^1_3,A^1_4\} \rangle}$ and $\nu_{45} = \lVert (A^1_4+A^1_5) \rVert_{\rho_{A^1A^2A^3}} = \sqrt{2+\langle \{A^1_4,A^1_5\} \rangle}$ we have,

\begin{equation}
\begin{split}\label{d1253}
  \langle \delta^3_{2,5} \rangle &= 2(\nu_{34}+\nu_{45})+\langle \{A^1_3,A^1_5\} \rangle\leq 2\sqrt{2(\nu_{34}^2+\nu_{45}^2)}+\langle \{A^1_3,A^1_5\} \rangle = 2\sqrt{2(4+\langle \{A^1_3,A^1_4\} \rangle+\langle \{A^1_4,A^1_5\} \rangle)}+\langle \{A^1_3,A^1_5\} \rangle\\
& = 2\sqrt{2(4+\langle \{A^1_4,A^1_3+A^1_5\} \rangle)}+\langle \{A^1_3,A^1_5\} \rangle
\end{split}  
\end{equation}

As the maximization condition gives $\Delta^3_5 =0$, we consider $A^1_4 = \frac{A^1_3+A^1_5}{v_{35}}$, where $w_{35}  = \sqrt{2+\langle \{A^1_3,A^1_5\} \rangle}$. Hence, we get
\begin{equation}\label{d253}
\begin{split}
  \langle \delta^3_{2,5} \rangle &= 2\sqrt{2(4+2\sqrt{2+\langle \{A^1_3,A^1_5\} \rangle})}+\langle \{A^1_3,A^1_5\} \rangle
\end{split}  
\end{equation}
Now, Eq.~(\ref{d253}) does not exhibit any maxima within the interval $\langle \{A^1_3,A^1_5\} \rangle\in\{-2,2\}$. Therefore, we set $A^1_4 = -\frac{A^1_3+A^1_5}{v_{35}}$, substitute this expression into Eq.~(\ref{d1253}), and obtain
\begin{equation}\label{d2153}
\begin{split}
  \langle \delta^3_{2,5} \rangle &= 2\sqrt{2(4-2\sqrt{2+\langle \{A^1_3,A^1_5\} \rangle})}+\langle \{A^1_3,A^1_5\} \rangle
\end{split}  
\end{equation}
Now Eq.~(\ref{d2153}) exhibits two singularities within the interval $\langle \{A^1_3,A^1_5\} \rangle\in\{-2,2\}$, and the function attains its maximum at $\langle \{A^1_3,A^1_5\} \rangle=0.618034\approx 2\cos{\frac{2\pi}{5}}$, which yields
\begin{eqnarray}
     \langle \Gamma^3_5 \rangle^{opt}_Q=47.5528=50\cos{\frac{\pi}{10}}
\end{eqnarray}
Further we can compute $\nu_{35}=\sqrt{2+\langle \{A^1_3,A^1_5\} \rangle}=2\cos{\frac{\pi}{5}}$, which leads to $A^1_4=-\frac{A^1_3+A^1_5}{2\cos{\frac{\pi}{5}}}$, from this we can derive that $\langle \{A^1_3,A^1_4\} \rangle=\langle \{A^1_5,A^1_4\} \rangle=-2\cos{\frac{\pi}{5}}$. Similarly we get $\nu_{34}=\nu_{45}=2\cos{\frac{2\pi}{5}}$, which implies $A^1_1=\frac{A^1_3+A^1_4}{2\cos{\frac{2\pi}{5}}}$ and $A^1_2=\frac{A^1_4+A^1_5}{2\cos{\frac{2\pi}{5}}}$, from which we get $\langle \{A^1_1,A^1_3\} \rangle=\langle \{A^1_2,A^1_4\} \rangle=2\cos{\frac{2\pi}{5}}, \langle \{A^1_1,A^1_4\} \rangle=\langle \{A^1_2,A^1_5\} \rangle=-2\cos{\frac{3\pi}{5}}, \langle \{A^1_1,A^1_2\} \rangle=-2\cos{\frac{\pi}{5}}, \langle \{A^1_1,A^1_5\} \rangle=2\cos{\frac{4\pi}{5}}$ and $ \langle \{A^1_2,A^1_3\} \rangle=-2\cos{\frac{\pi}{5}}$. In short, we can write it in the following way
\begin{eqnarray}\label{a15}
    &&\langle \{A^1_{i_1}, A^1_{i_1+x}\} \rangle= 2(-1)^x\cos{\frac{\pi x}{5}},\forall x\in[5-i_1]
\end{eqnarray}
Furthermore, from the above relations we also find that $\langle \{\widetilde{A}^1_{i_1},\widetilde{A}^1_{i_1+x}\} \rangle = 2(-1)^x\cos{\frac{\pi x}{5}},\forall x\in[5-i_1]$, with $\widetilde{A}^1_6 = \widetilde{A}^1_1$.
Therefore, using Eqs.~(\ref{a13}) and (\ref{a15}), we can express for an arbitrary number of $n$ settings 
\begin{eqnarray}
    &&\langle \{A^1_{i_1}, A^1_{i_1+x}\} \rangle= \langle \{\widetilde{A}^1_{i_1},\widetilde{A}^1_{i_1+x}\} \rangle=2(-1)^x\cos{\frac{\pi x}{n}},\forall x\in[n-i_1]
\end{eqnarray}
 To determine the relations between the observables for the remaining parties, in the tripartite scenario with arbitrary settings $n$, Eq.~(\ref{SOSc}) can be recast as
\begin{eqnarray}
    \widetilde{A}^1_{i_1}\otimes  A^2_{(i_1+i_3-2)\oplus_n+1 }\otimes A^3_{i_3}\ket{\psi}_{A^1A^2A^3}=\ket{\psi}_{A^1A^2A^3},\forall i_1,i_3\in[n]\label{slf3n}
\end{eqnarray}
Now, putting $i_3=1$ in Eq.~(\ref{slf3n}) we get
\begin{eqnarray}
    \widetilde{A}^1_{i_1}\otimes  A^2_{(i_1-1)\oplus_n+1 }\otimes A^3_{1}\ket{\psi}_{A^1A^2A^3}=\ket{\psi}_{A^1A^2A^3},\forall i_1\in[n]\label{slf3n1}
\end{eqnarray}
Since $(i_1 - 1) < n$, we can therefore express $(i_1 - 1)\oplus_n + 1 = i_1$. Hence, from Eq.~(\ref{slf3n1}) we obtain the following.
\begin{eqnarray}
    \widetilde{A}^1_{i_1}\otimes  A^2_{i_1}\otimes A^3_{1}\ket{\psi}_{A^1A^2A^3}=\ket{\psi}_{A^1A^2A^3},\forall i_1\in[n]\label{slf3n2}
\end{eqnarray}
For $i_1=i_2$ and $i_1 = i_2 + x$, Eq.~(\ref{slf3n2}) can be expressed as
\begin{eqnarray}
    \widetilde{A}^1_{i_2+x}\otimes  A^2_{i_2+x}\otimes A^3_{1}\ket{\psi}_{A^1A^2A^3}=\ket{\psi}_{A^1A^2A^3},\forall i_2\in[n]\label{slf3n3}
\end{eqnarray}
Hence from Eq.~(\ref{slf3n2}) and (\ref{slf3n3}) we get
\begin{eqnarray}
    &&\begin{cases}
        \openone_d\otimes A^2_{i_2} \otimes\openone_d\ket{\psi}_{A^1A^2A^3} = \widetilde{A}^1_{i_2}\otimes\openone_d \otimes A^3_1\ket{\psi}_{A^1A^2A^3}\\
        \openone_d\otimes A^2_{i_2+x} \otimes\openone_d\ket{\psi}_{A^1A^2A^3} = \widetilde{A}^1_{i_2+x}\otimes\openone_d \otimes A^3_1\ket{\psi}_{A^1A^2A^3}
    \end{cases} \hspace{-0.8cm}\Rightarrow \langle \{A^2_{i_2},A^2_{i_2+x}\}\rangle=2(-1)^x\cos{\frac{\pi x}{n}},\forall i_2\in[n], x\in[n-i_2]\qquad
    \end{eqnarray}
Similarly, for the third party that puts $i_1=1$ in Eq.~(\ref{slf3n}), we get
\begin{eqnarray}\label{slfn14}
    &&\widetilde{A}^1_1\otimes  A^2_{(i_3-1)\oplus_n+1 }\otimes A^3_{i_3}\ket{\psi}_{A^1A^2A^3}=\ket{\psi}_{A^1A^2A^3},\forall i_3\in[n]\nonumber\\
    &&\widetilde{A}^1_1\otimes  A^2_{i_3}\otimes A^3_{i_3}\ket{\psi}_{A^1A^2A^3}=\ket{\psi}_{A^1A^2A^3},(\quad\text{As $i_3-1<n$})
\end{eqnarray}
Now, putting $i_3=i_3+x$ we get
\begin{eqnarray}
    &&\begin{cases}
        \openone_d\otimes \openone_d \otimes A^3_{i_3}\ket{\psi}_{A^1A^2A^3} = \widetilde{A}^1_{1}\otimes A^2_{i_3} \otimes \openone_d\ket{\psi}_{A^1A^2A^3} \\
        \openone_d\otimes \openone_d \otimes A^3_{i_3+x}\ket{\psi}_{A^1A^2A^3} = \widetilde{A}^1_{1}\otimes A^2_{i_3+x} \otimes \openone_d\ket{\psi}_{A^1A^2A^3} 
    \end{cases} \hspace{-0.8cm}\Rightarrow \langle \{A^3_{i_3},A^3_{i_3+x}\}\rangle=2(-1)^x\cos{\frac{\pi x}{n}},\forall i_3\in[n], x\in[n-i_3]\qquad
    \end{eqnarray}
Hence, for $m=3$ in arbitrary $n$ settings, we can write 
\begin{eqnarray}\label{oc3n}
    \langle \{A^k_{i_k},A^k_{i_k+x}\}\rangle=2(-1)^x\cos{\frac{\pi x}{n}},\forall x\in[n-i_k], k\in[3]\qquad
\end{eqnarray}

\subsection{Detailed calculation for \texorpdfstring{$m=4$ and $n=3$}{m=4 and n=3}}
Consider $Alice^k$ having three possible measurement choices $i_k \in \{A^k_1, A^k_2, A^k_3\}, \forall k\in[4]$, respectively. Now, $\Gamma^4_3$ has the following form,

\begin{eqnarray}
    \Gamma^4_3= \Gamma^{3,1}_{3}\otimes A^4_1+\Gamma^{3,2}_{3}\otimes A^4_2+\Gamma^{3,3}_{3}\otimes A^4_3
\end{eqnarray}
where 
\begin{eqnarray}\label{gamma3s}
   \Gamma^{3,1}_3 &=& ((A^1_1-A^1_2)\otimes A^2_1+(A^1_2-A^1_3)\otimes A^2_2+(A^1_3-A^1_1)\otimes A^2_3)\otimes A^3_1\nonumber\\
   &+&((A^1_1-A^1_2)\otimes A^2_2+(A^1_2-A^1_3)\otimes A^2_3+(A^1_3-A^1_1)\otimes A^2_1)\otimes A^3_2\nonumber\\
   &+&((A^1_1-A^1_2)\otimes A^2_3+(A^1_2-A^1_3)\otimes A^2_1+(A^1_3-A^1_1)\otimes A^2_2)\otimes A^3_3 \\
    \Gamma^{3,2}_3 &=& ((A^1_1-A^1_2)\otimes A^2_1+(A^1_2-A^1_3)\otimes A^2_2+(A^1_3-A^1_1)\otimes A^2_3)\otimes A^3_2\nonumber\\
   &+&((A^1_1-A^1_2)\otimes A^2_2+(A^1_2-A^1_3)\otimes A^2_3+(A^1_3-A^1_1)\otimes A^2_1)\otimes A^3_3\nonumber\\
   &+&((A^1_1-A^1_2)\otimes A^2_3+(A^1_2-A^1_3)\otimes A^2_1+(A^1_3-A^1_1)\otimes A^2_2)\otimes A^3_1 \\
    {\Gamma^{3,3}_3} &=& ((A^1_1-A^1_2)\otimes A^2_1+(A^1_2-A^1_3)\otimes A^2_2+(A^1_3-A^1_1)\otimes A^2_3)\otimes A^3_3\nonumber\\
   &+&((A^1_1-A^1_2)\otimes A^2_2+(A^1_2-A^1_3)\otimes A^2_3+(A^1_3-A^1_1)\otimes A^2_1)\otimes A^3_1\nonumber\\
   &+&((A^1_1-A^1_2)\otimes A^2_3+(A^1_2-A^1_3)\otimes A^2_1+(A^1_3-A^1_1)\otimes A^2_2)\otimes A^3_2 
\end{eqnarray}

Now, we will derive the optimal quantum value of $\Gamma^4_3$ using the sum-of-squares approach. Note that throughout the derivation, there is no restriction on the dimension of the underlying Hilbert space, which makes it device-independent.  We define the positive semi-definite operator $\gamma^4_3$, such that $ \gamma^4_3 = \beta^4_3 \openone_d - \Gamma^4_3$, where $\beta^4_3$ is a positive quantity. Taking into account a suitable set of operators $L_{i_1i_2i_3i_4}$, we can write

\begin{equation}
    \gamma^4_3 = \frac{1}{2}\sum_{i_4=1}^{3}\sum_{i_3=1}^{3}\sum_{{i_1}=1}^{3} w^3_{i_1i_2i_3 i_4} L{^{4^\dag}_{i_1i_2i_3 i_4}} L^4_{i_1i_2i_3 i_4}
\end{equation}

where $w_{i_1 i_2 i_3 i_4}$'s are the suitable normalizers and the positive operator $L^4_{i_1i_2i_3 I_4}$ defined as
\begin{equation}
    L^4_{i_1 i_2 i_3 i_4} =\openone_d\otimes A^2_{(i_1+i_3-2)\oplus_3 +1} \otimes  A^3_{(i_3+i_4-2)\oplus_3+1}\otimes A^4_{i_4} - \widetilde{A}^1_{i_1} \otimes \openone_d \otimes \openone_d \otimes \openone_d
\end{equation}

with $\widetilde{A}^1_{i_1} = \left(\frac{A^1_{i_1}-A^1_{i_1\oplus_3+1}}{w_{i_1 i_2i_3i_4}}\right)$ and $w_{i_1i_2i_3i_4} = \lVert A^1_{i_1}-A^1_{i_1\oplus_3+1} \rVert_{\rho_{A^1 A^2 A^3 A^4}}$, where $\lVert . \rVert$ is the Frobenius norm, given by $\lVert \mathcal{O} \rVert=\sqrt{\Tr[\mathcal{O}^{\dagger} \mathcal{O} \ \rho]}$.

So, $w_{i_1i_2i_3i_4}$ becomes $\lVert A^1_{i_1}-A^1_{i_1\oplus_3+1} \rVert_{\rho_{A^1 A^2 A^3 A^4}} = \sqrt{2-\langle \{A^1_{i_1},A^1_{i_1\oplus_3+1}\} \rangle}$, where $\langle\cdots\rangle_{\rho_{A^1A^2A^3A^4}}=\langle\cdots\rangle$. In a nutshell, we can write $w_{i_1 i_2 i_3 i_4}\equiv w_{i_1},\forall i_1,i_2,i_3,i_4\in[3]$. Hence, $\Tr[\gamma^4_3 \ \rho_{A^1 A^2 A^3 A^4}]  = 9(w_1+w_2+w_3) - \Tr[\Gamma^4_3 \ \rho_{A^1 A^2 A^3 A^4}]$. Hence, the optimal value of $\Gamma^4_3$ is obtained when $\Tr[\gamma^4_3 \ \rho_{A^1 A^2 A^3A^4}] = 0$. From where we can further write 
\begin{eqnarray}\label{slf43}
    \widetilde{A}^1_{i_1}\otimes A^2_{(i_1+i_3-2)\oplus_3 +1} \otimes  A^3_{(i_3+i_4-2)\oplus_3+1}\otimes A^4_{i_4}\ket{\psi}_{A^1A^2A^3A^4}=\ket{\psi}_{A^1A^2A^3A^4}
\end{eqnarray}
Hence, we will have $\langle \Gamma^4_3 \rangle^{opt}_Q = \max [9(w_1+w_2+w_3)]$. Using the concavity inequality, we can have
\begin{equation*}
\begin{split}
  \langle \Gamma^4_3 \rangle_Q& \leq 9\sqrt{3(w_1^2 + w_2^2 + w_3^2)}=9\sqrt{3[6-(\langle \{A^1_1,A^1_2\} \rangle + \langle \{A^1_2,A^1_3\} \rangle + \langle \{A^1_3,A^1_1\} \rangle)]}
\end{split}   
\end{equation*}

Now, consider $\Delta^4_3 = A^1_1 + A^1_2 + A^1_3$, then, $(\Delta^4_3)^2 = 3\openone_d+\{A^1_1,A^1_2\}+\{A^1_2,A^1_3\}+\{A^1_3,A^1_1\}$. Eventually, we have $\langle \Gamma^4_{3} \rangle _Q\leq 3\sqrt{3[9-\langle (\Delta^4_3)^2 \rangle]}$. $\langle \Gamma^4_{3} \rangle_Q^{opt}$ will have the maximum value $27\sqrt{3}$ if and only if $\langle (\Delta^4_3)^2 \rangle = 0\Rightarrow A^1_1 + A^1_2 + A^1_3=0$, i.e., $\langle \{A^1_{i_1},A^1_{i_1+x}\} \rangle= 2(-1)^x \cos{\frac{\pi x}{3}},\forall i_1\in[3],x\in[3-i_1]$, which also implies $\langle \{\widetilde{A}^1_{i_1},\widetilde{A}^1_{i_1+x}\} \rangle= 2(-1)^x \cos{\frac{\pi x}{3}}$.
Analogously, to determine the relation between the observables of the remaining parties, we put $i_4=1$ and $i_3=1$ in Eq.~(\ref{slf43}) and get
\begin{eqnarray}\label{slf431}
    &&\widetilde{A}^1_{i_1}\otimes A^2_{(i_1-1)\oplus_3 +1} \otimes  A^3_{1}\otimes A^4_{1}\ket{\psi}_{A^1A^2A^3A^4}=\ket{\psi}_{A^1A^2A^3A^4},\quad i_1\in[3]\nonumber\\
    &&\widetilde{A}^1_{i_1}\otimes A^2_{i_1} \otimes  A^3_{1}\otimes A^4_{1}\ket{\psi}_{A^1A^2A^3A^4}=\ket{\psi}_{A^1A^2A^3A^4},\quad \text{As $i_1-1<3$}
\end{eqnarray}
Now, putting $i_1=i_2$ and $i_1=i_2+x$ in Eq.~(\ref{slf431}), we get $i_1=i_2$ and $i_1=i_2+x$ in Eq.~(\ref{slf431}), we get the following.
\begin{eqnarray}
    &&\begin{cases}
        \openone_d\otimes A^2_{i_2} \otimes \openone_d\otimes \openone_d\ket{\psi}_{A^1A^2A^3A^4} = \widetilde{A}^1_{i_2}\otimes \openone_d \otimes A^3_1\otimes A^4_1\ket{\psi}_{A^1A^2A^3A^4} \\
        \openone_d\otimes A^2_{i_2+x} \otimes \openone_d\otimes \openone_d\ket{\psi}_{A^1A^2A^3A^4} = \widetilde{A}^1_{i_2+x}\otimes \openone_d \otimes A^3_1\otimes A^4_1\ket{\psi}_{A^1A^2A^3A^4}
    \end{cases} \hspace{-0.8cm}\Rightarrow \langle \{A^2_{i_2},A^2_{i_2+x}\}=2(-1)^x\cos{\frac{\pi x}{3}},\forall x\in[3-i_2]\qquad
    \end{eqnarray}
Similarly putting $i_1=1$ and $i_4=1$ in Eq.~(\ref{slf43}) we get
\begin{eqnarray}\label{slf433}
    \widetilde{A}^1_{1}\otimes A^2_{i_3} \otimes  A^3_{i_3}\otimes A^4_{1}\ket{\psi}_{A^1A^2A^3A^4}=\ket{\psi}_{A^1A^2A^3A^4}
\end{eqnarray}
Again, putting $i_3=i_3+x$ in Eq.~(\ref{slf433}) and using the previous approach, we get $\langle \{A^3_{i_3},A^3_{i_3+x}\}\rangle=2(-1)^x\cos{\frac{\pi x}{3}},\forall x\in[3-i_3]$. Similarly, putting $i_1=1$ and $i_3=1$ in Eq.~(\ref{slf43}) we get $\langle \{A^4_{i_4},A^4_{i_4+x}\}\rangle=2(-1)^x\cos{\frac{\pi x}{3}},\forall x\in[3-i_4]$. Hence we can write in general as 
\begin{eqnarray}
    \langle \{A^k_{i_k},A^k_{i_k+x}\}\rangle=2(-1)^x\cos{\frac{\pi x}{3}},\forall i_k\in [3], x\in[3-i_k],k\in[4]
\end{eqnarray}

\subsection{Detailed calculation for \texorpdfstring{$m=4$ and $n=5$}{m=4 and n=5}}
Consider four parties $Alice^k, \forall k\in[4]$ to have three possible measurement choices $i_k \in \{A^k_1, A^k_2, A^k_3\}$ respectively. Now, $\Gamma^4_5$ has the following form,

\begin{eqnarray}
\Gamma^4_5&=&\sum_{i_4=1}^5\sum_{i_3=1}^5 \qty(\sum_{{i_1}=1}^5(A^1_{i_1}-A^1_{{i_1}\oplus_5+1})\otimes A^2_{({i_1}+i_3-2)\oplus_5 +1 })\otimes A^3_{(i_3+i_4-2)\oplus_5 +1} \otimes A_{i_4}^4  
\end{eqnarray}

Now, we will derive the optimal quantum value of $\Gamma^4_5$ using the sum-of-squares approach. Note that throughout the derivation, there is no restriction on the dimension of the underlying Hilbert space, which makes it device-independent.  We define the positive semi-definite operator $\gamma^4_5$, such that $ \gamma^4_5 = \beta^4_5 \openone_d - \Gamma^4_5$, where $\beta^4_5$ is a positive quantity. Taking into account a suitable set of operators $L_{i_1i_2i_3i_4}$, we can write

\begin{equation}
    \gamma^5_3 = \frac{1}{2}\sum_{i_4=1}^{5}\sum_{i_3=1}^{5}\sum_{{i_1}=1}^{5} w^3_{i_1i_2i_3 i_4} L{^{4^\dag}_{i_1i_2i_3 i_4}} L^4_{i_1i_2i_3 i_4}
\end{equation}

where $w_{i_1 i_2 i_3 i_4}$'s are the suitable normalizers and the positive operator $L^4_{i_1i_2i_3 i_4}$ defined as
\begin{equation}
    L^4_{i_1 i_2 i_3 i_4} =\openone_d\otimes A^2_{(i_1+i_3-2)\oplus_5 +1} \otimes  A^3_{(i_3+i_4-2)\oplus_5+1}\otimes A^4_{i_4} - \widetilde{A}^1_{i_1} \otimes \openone_d \otimes \openone_d \otimes \openone_d
\end{equation}

with $\widetilde{A}^1_{i_1} = \left(\frac{A^1_{i_1}-A^1_{i_1\oplus_3+1}}{w_{i_1 i_2i_3i_4}}\right)$ and $w_{i_1i_2i_3i_4} = \lVert A^1_{i_1}-A^1_{i_1\oplus_3+1} \rVert_{\rho_{A^1 A^2 A^3 A^4}}$, where $\lVert . \rVert$ is the Frobenius norm, given by $\lVert \mathcal{O} \rVert=\sqrt{\Tr[\mathcal{O}^{\dagger} \mathcal{O} \ \rho]}$.

So, $w_{i_1i_2i_3i_4}$ becomes $\lVert A^1_{i_1}-A^1_{i_1\oplus_5+1} \rVert_{\rho_{A^1 A^2 A^3 A^4}} = \sqrt{2-\langle \{A^1_{i_1},A^1_{i_1\oplus_5+1}\} \rangle}=\sqrt{2-\langle \{A^1_{i_1},A^1_{i_1+1}\} \rangle},\forall i_1<5$, where $\langle\cdots\rangle_{\rho_{A^1A^2A^3A^4}}=\langle\cdots\rangle$. In a nutshell, we can write $w_{i_1 i_2 i_3 i_4}\equiv w_{i_1},\forall i_1,i_2,i_3,i_4\in[5]$. Hence, $\Tr[\Gamma^4_5 \ \rho_{A^1 A^2 A^3 A^4}]  = 25\sum_{i_1=1}^5 w_{i_1} - \Tr[\gamma^4_5 \ \rho_{A^1 A^2 A^3 A^4}]$. Hence, the optimal value of $\Gamma^4_5$ is obtained when $\Tr[\gamma^4_5 \ \rho_{A^1 A^2 A^3A^4}] = 0$. From where we can further write 
\begin{eqnarray}\label{slf45}
    \widetilde{A}^1_{i_1}\otimes A^2_{(i_1+i_3-2)\oplus_5 +1} \otimes  A^3_{(i_3+i_4-2)\oplus_5+1}\otimes A^4_{i_4}\ket{\psi}_{A^1A^2A^3A^4}=\ket{\psi}_{A^1A^2A^3A^4}
\end{eqnarray}
Hence, we will have $\langle \Gamma^4_5 \rangle^{opt}_Q = \max [25(w_1+w_2+w_3+w_4+w_5)]$. Using the concavity inequality, we can have
\begin{equation}
\begin{split}
  \langle \Gamma^4_5 \rangle_Q \leq 25\sqrt{5(w_1^2 +\ldots+ w_5^2)} = 25\sqrt{5\left[10-\sum_{i_1=1}^{5}\langle \{A^1_{i_1},A^1_{i_1+1}\} \rangle\right]}
\end{split}   
\end{equation}

Now, let us consider
\begin{equation}
       \Delta^4_5 =A^1_1+A^1_2+A^1_3+A^1_4+A^1_5
\end{equation}

Then we have
\begin{equation}
(\Delta^4_5)^2 =5\openone_d+\sum_{i_1=1}^{5}\{A^1_{i_1},A^1_{i_1+1}\}+\sum_{i_1=3}^{4}\{A_1,A_{i_1}\}+\sum_{i_1=2}^{3}\sum_{i=i_1+2}^{5} \{A^1_{i_1},A^1_i\}
\end{equation}

Taking into account $\delta^4_{1,5}=\sum_{i_1=1}^{5}\{A^1_{i_1},A^1_{i_1+1}\}$ and $\delta^4_{2,5}=\sum_{i_1=3}^{4}\{A_1,A_{i_1}\}+\sum_{i_1=2}^{3}\sum_{i=i_1+2}^{5} \{A^1_{i_1},A^1_i\}$, we have $(\Delta^4_5)^2 =5\openone_d+\delta^4_{1,5}+\delta^4_{2,5}$.

Hence, we have
\begin{equation}
  \langle \Gamma^4_5 \rangle_Q\leq 5 
  \sqrt{5\left[10+\langle(5\openone_d+\delta^4_{2,5}-(\Delta^4_5)^2)\rangle\right]} 
\end{equation}

$\langle \Gamma^4_5 \rangle_Q$ has maximum value when $(\Delta^4_5)^2 =0$. This implies $\Delta^4_5 =0$. Now, $\delta^4_{2,5}=\{A^1_1,A^1_3+A^1_4\} +\{A^1_2,A^1_4+A^1_5\} +\{A^1_3,A^1_5\}$. Taking into account $A^1_1=\frac{A^1_3+A^1_4}{\nu_{34}}$ and $A^1_2=\frac{A^1_4+A^1_5}{\nu_{45}}$, where $\nu_{34} = \lVert (A^1_3+A^1_4) \rVert_{\rho_{A^1A^2A^3}} = \sqrt{2+\langle \{A^1_3,A^1_4\} \rangle}$ and $\nu_{45} = \lVert (A^1_4+A^1_5) \rVert_{\rho_{A^1A^2A^3}} = \sqrt{2+\langle \{A^1_4,A^1_5\} \rangle}$ we have

\begin{equation}
\begin{split}
  \langle \delta^4_{2,5} \rangle &= 2(\nu_{34}+\nu_{45})+\langle \{A^1_3,A^1_5\} \rangle\leq 2\sqrt{2(\nu_{34}^2+\nu_{45}^2)}+\langle \{A^1_3,A^1_5\} \rangle = 2\sqrt{2(4+\langle \{A^1_3,A^1_4\} \rangle+\langle \{A^1_4,A^1_5\} \rangle)}+\langle \{A^1_3,A^1_5\} \rangle\\
& = 2\sqrt{2(4+\langle \{A^1_4,A^1_3+A^1_5\} \rangle)}+\langle \{A^1_3,A^1_5\} \rangle
\end{split}  
\end{equation}
From now on, the optimization is similar to sec.~\ref{sec35}, which gives the optimal value of $ \langle \delta^4_{2,5} \rangle$ at $\langle \{A^1_3,A^1_5\} \rangle=0.618034\approx 2\cos{\frac{2\pi}{5}}$, which yields
\begin{eqnarray}
     \langle \Gamma^4_5 \rangle^{opt}_Q=250\cos{\frac{\pi}{10}}
\end{eqnarray}
In summary, the relationship between the observables can be expressed as follows
\begin{eqnarray}\label{a45}
    &&\langle \{A^1_{i_1}, A^1_{i_1+x}\} \rangle= 2(-1)^x\cos{\frac{\pi x}{5}},\forall x\in[5-i_1]
\end{eqnarray}
Furthermore, from the above relations we also find that $\langle \{\widetilde{A}^1_{i_1},\widetilde{A}^1_{i_1+x}\} \rangle = 2(-1)^x\cos{\frac{\pi x}{5}},\forall x\in[5-i_1]$, with $\widetilde{A}^1_6 = \widetilde{A}^1_1$.
Therefore, using Eqs.~(\ref{a13}) and (\ref{a45}), we can express for an arbitrary number of $n$ settings 
\begin{eqnarray}
    &&\langle \{A^1_{i_1}, A^1_{i_1+x}\} \rangle= \langle \{\widetilde{A}^1_{i_1},\widetilde{A}^1_{i_1+x}\} \rangle=2(-1)^x\cos{\frac{\pi x}{n}},\forall x\in[n-i_1]
\end{eqnarray}
 To analyze how the observables of the remaining parties are related, we can rewrite the self-testing condition for $n$ settings from Eq.~(\ref{slf5}) as
\begin{eqnarray}
    \widetilde{A}^1_{i_1}\otimes A^2_{(i_1+i_3-2)\oplus_n +1} \otimes  A^3_{(i_3+i_4-2)\oplus_n+1}\otimes A^4_{i_4}\ket{\psi}_{A^1A^2A^3A^4}=\ket{\psi}_{A^1A^2A^3A^4}\label{slf4n}
\end{eqnarray}
Analogously, to determine the relation between the observables of the remaining parties, we put $i_4=1$ and $i_3=1$ in Eq.~(\ref{slf4n}) and get
\begin{eqnarray}\label{slf4n1}
    &&\widetilde{A}^1_{i_1}\otimes A^2_{(i_1-1)\oplus_n +1} \otimes  A^3_{1}\otimes A^4_{1}\ket{\psi}_{A^1A^2A^3A^4}=\ket{\psi}_{A^1A^2A^3A^4},\quad i_1\in[n]\nonumber\\
    &&\widetilde{A}^1_{i_1}\otimes A^2_{i_1} \otimes  A^3_{1}\otimes A^4_{1}\ket{\psi}_{A^1A^2A^3A^4}=\ket{\psi}_{A^1A^2A^3A^4},\quad \text{As $i_1-1<n$}
\end{eqnarray}
Now, putting $i_1=i_2$ and $i_1=i_2+x$ in Eq.~(\ref{slf4n1}) we get
\begin{eqnarray}
    &&\begin{cases}
        \openone_d\otimes A^2_{i_2} \otimes \openone_d\otimes \openone_d\ket{\psi}_{A^1A^2A^3A^4} = \widetilde{A}^1_{i_2}\otimes \openone_d \otimes A^3_1\otimes A^4_1\ket{\psi}_{A^1A^2A^3A^4} \\
        \openone_d\otimes A^2_{i_2+x} \otimes \openone_d\otimes \openone_d\ket{\psi}_{A^1A^2A^3A^4} = \widetilde{A}^1_{i_2+x}\otimes \openone_d \otimes A^3_1\otimes A^4_1\ket{\psi}_{A^1A^2A^3A^4}
    \end{cases} \hspace{-0.8cm}\Rightarrow \langle \{A^2_{i_2},A^2_{i_2+x}\}=2(-1)^x\cos{\frac{\pi x}{n}},\forall x\in[n-i_2]\qquad
    \end{eqnarray}
Similarly putting $i_1=1$ and $i_4=1$ in Eq.~(\ref{slf4n}) we get
\begin{eqnarray}\label{slf4nn}
    \widetilde{A}^1_{1}\otimes A^2_{i_3} \otimes  A^3_{i_3}\otimes A^4_{1}\ket{\psi}_{A^1A^2A^3A^4}=\ket{\psi}_{A^1A^2A^3A^4}
\end{eqnarray}
Again, putting $i_3=i_3+x$ in Eq.~(\ref{slf4nn}) and using the previous approach, we get $\langle \{A^3_{i_3},A^3_{i_3+x}\}\rangle=2(-1)^x\cos{\frac{\pi x}{n}},\forall x\in[n-i_3]$. Similarly, putting $i_1=1$ and $i_3=1$ in Eq.~(\ref{slf4n}) we get $\langle \{A^4_{i_4},A^4_{i_4+x}\}\rangle=2(-1)^x\cos{\frac{\pi x}{n}},\forall x\in[n-i_4]$. Hence we can write in general as 
\begin{eqnarray}\label{oc4n1}
    \langle \{A^k_{i_k},A^k_{i_k+x}\}\rangle=2(-1)^x\cos{\frac{\pi x}{n}},\forall x\in[n-i_k],k\in[4]
\end{eqnarray}
From Eq.~(\ref{oc3n}) and (\ref{oc4n1}) for arbitrary party $k$, we can write 
\begin{eqnarray}\label{ockn}
    \langle \{A^k_{i_k},A^k_{i_k+x}\}\rangle=2(-1)^x\cos{\frac{\pi x}{n}},\forall i_k\in[n], x\in[n-i_k],k\in[m]
\end{eqnarray}
This is Eq.~(10) in the main text.
\section{Derivation of maximum Randomness}
Consider an initial tripartite  scenario with an arbitrary number of odd settings \(n\) where the observed behavior \(\mathcal{P} \equiv \{P(a^1,a^2,a^3 \mid A^1_{i_1},A^2_{i_2},A^3_{i_3})\}\). Then, the certified randomness \(\mathcal{R}_{i_1 i_2 i_3}\) associated with the measurement choices \((A^1_{i_1}, A^2_{i_2}, A^3_{i_3})\) is given by:
\begin{equation}\label{randoma}
\mathcal{R}_{i_1{i_2}i_3} = - \log_{2}\left[\max_{a^1,a^2,a^3}P\left(a^1,a^2,a^3|A^1_{i_1},A^2_{i_2},A^3_{i_3}\right)\right]
\end{equation}
It is important to note that $\mathcal{R}_{i_1{i_2}i_3}$ is explicitly dependent on the measurement choices $A^1_{i_1}$, $A^2_{i_2}$, and $A^3_{i_3}$. The maximum certified randomness is defined as $\mathcal{R}_{max}$, which is the maximum value of $\mathcal{R}_{i_1{i_2}i_3}$ in all measurement choices. Note that for a given measurement setting $A^1_{i_1}$, $A^2_{i_2}$, and $A^3_{i_3}$, $P\left(a^1,a^2,a^3|A^1_{i_1},A^2_{i_2},A^3_{i_3}]\right)$ defined as
\vspace{-0.3cm}
{\small\begin{eqnarray}\label{pabc}
    P\left(a^1,a^2,a^3|A^1_{i_1},A^2_{i_2},A^3_{i_3}\right)=\frac{1}{8}\Bigg[1+a^1\langle{A^1_{i_1}}\rangle+a^2\langle{A^2_{i_2}}\rangle+a^3\langle{A^3_{i_3}}\rangle+a^1a^2\langle{A^1_{i_1}A^2_{i_2}}\rangle+a^2a^3\langle{A^2_{{i_2}}A^3_{i_3}}\rangle+a^1a^3\langle{A^1_{i_1}A^3_{i_3}}\rangle+a^1a^2a^3\langle{A^1_{i_1}A^2_{{i_2}}A^3_{i_3}}\rangle\Bigg]\quad\ \ \
\end{eqnarray}}
Using the SOS conditions, we proceed to compute the marginals. Therefore, we examine the SOS condition in the tripartite setting.
\begin{eqnarray}
    \widetilde{A}^1_{i_1}\otimes  A^2_{(i_1+i_3-2)\oplus_n +1 }\otimes A^3_{i_3}\ket{\psi}_{A^1A^2A^3}=\ket{\psi}_{A^1A^2A^3}
\end{eqnarray}
From this it is straightforward that each of $\widetilde{A}^1_{i_1}\otimes  A^2_{(i_1+i_3-2)\oplus_n +1 }\otimes A^3_{i_3}$ is commuting for all $i_1,i_3\in[n]$ w.r.t the state $\ket{\psi}_{A^1A^2A^3}$, and they stabilize the state $\ket{\psi}_{A^1A^2A^3}$. Hence, $\widetilde{A}^1_{i_1}\otimes  A^2_{(i_1+i_3-2)\oplus_n +1 }\otimes A^3_{i_3}$  are stabilizers, therefore, we can write 
\begin{eqnarray}\label{Sm}
    S_{i_1i_2i_3}=\widetilde{A}^1_{i_1}\otimes  A^2_{i_2}\otimes A^3_{i_3}
\end{eqnarray}

where $i_2=(i_1+i_3-2)\oplus_n +1$ implies $S_{i_1i_2i_3}\ket{\psi}_{A^1A^2A^3}=\ket{\psi}_{A^1A^2A^3}$. We know that all the observables are positive and Hermitian, and from that we can write $S^{\dagger}_{i_1i_2i_3}=S_{i_1i_2i_3}$.  Again, we know from the self-testing condition that the operators obey $\langle\{A^k_{i_k}, A^k_{i_k+x}\}\rangle_{\rho_{A^1A^2A^3}} = 2c_x,\forall k \in[3]$, where $c_x = (-1)^x \cos(\pi x/n)$ and $\langle..\rangle=\langle..\rangle_{\rho_{A^1A^2A^3}}$. This relation implies two key identities:
\begin{itemize}
    \item \textbf{Property 1:}  $(A^k_{i_k})^2 = \openone_d$.
    \item \textbf{Property 2:}  $A^k_{i_k} A^k_x A^k_{i_k} \ket{\psi}_{A^1A^2A^3} = 2c_{x-i_k} A^k_{i_k} \ket{\psi}_{A^1A^2A^3}- A^k_x\ket{\psi}_{A^1A^2A^3}$.
\end{itemize}
Furthermore, for any operator $O$, the stabilizer properties imply $\langle O \rangle = \langle S_{i_1i_2i_3} O \ S_{i_1i_2i_3} \rangle$.

\subsection{Single-Party Marginals}
We demonstrate that $\langle A^k_{i_k} \rangle = 0$. This is achieved by employing stabilizer characteristics and Property 2.
\begin{eqnarray}\label{sm1}
    \langle A^2_x \rangle=\langle  S_{i_1i_2 i_3}A^2_x S_{i_1i_2 i_3}\rangle=\langle A^2_{i_2}A^2_xA^2_{i_2}\rangle=2c_{x-i_2} \langle A^2_{i_2}\rangle-\langle A^2_x\rangle\implies \langle A^2_x \rangle=c_{x-i_2} \langle A^2_{i_2}\rangle\qquad
\end{eqnarray}
Putting $i_2=x$ in Eq.~(\ref{Sm}), we obtain a new stabilizer operator $S'_{i_1i_2i_3}= \widetilde{A}^1_{i_1'}\otimes  A^2_x\otimes A^3_{i_3'}$. It is worth mentioning that if we change the value of $i_2$, then the value of $i_1,i_3$ also changes to $i_1',i_3'$. Similarly, using the previous approach, we obtain the following.
\begin{eqnarray}\label{sm2}
    \langle A^2_{i_2}\rangle=c_{i_2-x}\langle A^2_x\rangle
\end{eqnarray}
As $c_{x-j}=c_{j-x}$, we can write from Eq.~(\ref{sm1}) and (\ref{sm2}) $\langle A^2_x\rangle=\langle A^2_{i_2}\rangle=0,\forall x\neq i_2\in[n]$. Similarly, we can write it for $\langle \widetilde{A}^1_{i_1}\rangle=\langle A^3_{i_3}\rangle=0,\forall i_1,i_3\in[n]$. The next section will explore the complexities of $\langle A^1_{i_1}\rangle$. Now, Eq.~(\ref{Sm}) can be rewritten as
\begin{eqnarray}\label{sm3}
    \langle A^1_{i_1}\rangle=\langle  S_{i_1i_2 i_3}A^1_{i_1}\  S_{i_1i_2 i_3}\rangle=\langle \widetilde{A}^1_{i_1} A^1_{i_1} \widetilde{A}^1_{i_1}\rangle
\end{eqnarray}
Again, 
\begin{eqnarray}\label{sm4}
    \langle \{\widetilde{A}^1_{i_1}, A^1_{i_1}\}\rangle = \Bigg\langle \Bigg\{\frac{A^1_i+A^1_{i\ \oplus_n+1}}{w_{i_1}},A^1_{i_1}\Bigg\}\Bigg\rangle=\frac{2-2\cos{\frac{\pi}{n}}}{w_{i_1}}\implies   A^1_{i_1} \widetilde{A}^1_{i_1} \ket{\psi}_{A^1A^2A^3}= \frac{2-2\cos{\frac{\pi}{n}}}{w_{i_1}}\ket{\psi}_{A^1A^2A^3}-\widetilde{A}^1_{i_1} A^1_{i_1}\ket{\psi}_{A^1A^2A^3}\qquad
\end{eqnarray}
Now, putting Eq.~(\ref{sm4}) in Eq.~(\ref{sm3}) we get
\begin{eqnarray}
     \langle A^1_{i_1}\rangle=\langle \widetilde{A}^1_{i_1} A^1_{i_1} \widetilde{A}^1_{i_1}\rangle=\frac{2-2\cos{\frac{\pi}{n}}}{w_{i_1}} \langle \widetilde{A}^1_{i_1}\rangle -\langle A^1_{i_1} \rangle \implies \langle A^1_{i_1}\rangle=0\quad (\text{As} \ \langle \widetilde{A}^1_{i_1}\rangle=0)
\end{eqnarray}

\subsection{Double-Party Marginals}
In the preceding section, we established that all single marginals are equal to zero. We now show that the double marginal must also be zero. In this regard, our aim is to prove $f(b,c) = \langle \openone_d\otimes A^2_b \otimes A^3_c \rangle = 0,\forall b,c\in[n]$. We conjugate the operator $O = \openone_d \otimes A^2_b \otimes A^3_c$ with $S_{i_1i_2i_3}=\widetilde{A}^1_{i_1}\otimes  A^2_{i_2}\otimes A^3_{i_3}$.
\begin{equation}
    f(b,c) = \langle S_{i_1i_2i_3}  (\openone_d\otimes A^2_b \otimes A^3_c)  S_{i_1i_2i_3}\rangle=\langle \openone_d\otimes A^2_{i_2} A^2_b A^2_{i_2} \otimes A^3_{i_3} A^3_c A^3_{i_3} \rangle
\end{equation}
Applying Property 2 to both terms and expanding gives a functional equation:
\begin{equation}
    2c_{i_2-b}c_{i_3-c} f(i_2, i_3) - c_{i_2-b} f(i_2,c) - c_{i_3-c} f(b, i_3) = 0
\end{equation}
By setting $i_3 = c$, the relation becomes simpler. 
\begin{eqnarray}\label{fbc}
   f(b,c) = c_{i_2-b} f(i_2,c) 
\end{eqnarray}
 Now we have the value of $i_2=(i_1+i_3-2)\oplus_n +1$ and we already set the value of $i_3=c$, so we choose the value of $i_1$ in a way that gives $i_2=1$ and $2$. This gives 
\begin{eqnarray}
    f(b,c)=c_{1-b} f(1,c)=c_{2-b} f(2,c)\label{bc3}
\end{eqnarray}

Now, putting $b=1$ and $2$ in Eq.~(\ref{bc3}) we get
\begin{eqnarray}
     f(1,c)=c_{1} f(2,c);\quad f(2,c)=c_{-1}f(1,c); \implies f(1,c)=f(2,c)=0
\end{eqnarray}

Therefore, we obtain  $f(b,c)=\langle \openone_d\otimes A^2_b \otimes A^3_c \rangle = 0$.

Analogously, it can be demonstrated that all double marginal probabilities equate to zero. This approach is remarkably sophisticated, allowing one to deduce results for any arbitrary marginal distributions across any given party. And by doing that, we get $\langle A^k_{i_k}\rangle=0$, $\langle A^{k_1}_{i_{k_1}}A^{k_2}_{i_{k_2}}\rangle_{k_1\neq k_2}=0$ for all $k,k_i\in[3]$.

Substituting the parameters associated with both double and single marginal effects into Eq.~(\ref{pabc}), it is observed that the resultant expression can be derived as follows:
\begin{eqnarray}\label{pabc1}
    P\left(a^1,a^2,a^3|A^1_{i_1},A^2_{i_2},A^3_{i_3}\right)=\frac{1}{8}\Bigg[1+a^1a^2a^3\langle{A^1_{i_1}A^2_{{i_2}}A^3_{i_3}}\rangle\Bigg]
\end{eqnarray}
In the context of an arbitrary m-partite scenario, all marginals except marginals consisting of $m$ terms become zero and it is possible to articulate the probability distribution function as

\begin{eqnarray}
    P\qty(a^1,a^2,..,a^m|A^1_{i_1}, A^2_{i_2},.., A^m_{i_m})=\frac{1}{2^m}\qty(1+a^1a^2..a^m \langle A^1_{i_1}A^2_{i_2}..A^m_{i_m}\rangle)
\end{eqnarray}
The subsequent section will examine the conditions under which the observables yield maximal randomness. 
\subsection{Randomness evaluation for any odd \texorpdfstring{$n$}{n} input and  \texorpdfstring{$m$}{m} party}\label{rand} From the SOS condition given in Eq.~(\ref{SOSc}), we get
. 
\begin{eqnarray}\label{s1}
    && \widetilde{A}^1_{i_1}\otimes  A^2_{(i_1+i_3-2)\oplus_n +1 }\otimes_{k=3}^{m-1} A^k_{f_k(i_{k+1},i_k)}\otimes A_{i_m}^m \ket{\psi}_{A^1..A^m} =\ket{\psi}_{A^1..A^m}, \quad \forall i_{k\in [3,m-1]}\in [n]\\
    &&\frac{A^1_{i_1}-A^1_{i_1\oplus_n+1}}{w_{i_1}} \otimes  A^2_{(i_1+i_3-2)\oplus_n +1 }\otimes_{k=3}^{m-1} A^k_{f_k(i_{k+1},i_k)}\otimes A_{i_m}^m  \ket{\psi}_{A^1..A^m} =\ket{\psi}_{A^1..A^m},\quad \text{where} \ w_{i_1} = \lVert A^1_{i_1}-A^1_{i_1\oplus_n+1} \rVert_{\rho_{A^1..A^m}}\\
    &&\frac{A^1_{i_1}-A^1_{i_1\oplus_n+1}}{\sqrt{2-\langle\{A^1_{i_1},A^1_{i_1\oplus_n+1}\}\rangle}} \otimes  A^2_{(i_1+i_3-2)\oplus_n +1 }\otimes_{k=3}^{m-1} A^k_{f_k(i_{k+1},i_k)} \otimes A_{i_m}^m \ket{\psi}_{A^1..A^m} =\ket{\psi}_{A^1..A^m}\\
    &&\frac{A^1_{i_1}-A^1_{i_1\oplus_n+1}}{2\cos\frac{\pi}{2n}} \otimes  A^2_{(i_1+i_3-2)\oplus_n +1 }\otimes_{k=3}^{m-1} A^k_{f_k(i_{k+1},i_k)}\otimes A_{i_m}^m  \ket{\psi}_{A^1..A^m} =\ket{\psi}_{A^1..A^m},\quad \text{where} \ \langle\{A^1_{i_1},A^1_{i_1\oplus_n+1}\}\rangle=-2\cos\frac{\pi}{n},\forall i_1\leq n\in \text{odd}\nonumber\\
    &&\openone_d \otimes  A^2_{(i_1+i_3-2)\oplus_n +1 }\otimes_{k=3}^{m-1} A^k_{f_k(i_{k+1},i_k)}\otimes A_{i_m}^m \ket{\psi}_{A^1..A^m} =\frac{A^1_{i_1}-A^1_{i_1\oplus_n+1}}{2\cos\frac{\pi}{2n}}\otimes_{k=2}^m\openone_d \ket{\psi}_{A^1..A^m} \label{s2}\\
    &&\bra{\psi}_{A^1..A^m}\openone_d \otimes  A^2_{(i_1+i_3-2)\oplus_n +1 }\otimes_{k=3}^{m-1} A^k_{f_k(i_{k+1},i_k)}\otimes A_{i_m}^m  =\bra{\psi}_{A^1..A^m}\frac{A^1_{i_1}-A^1_{i_1\oplus_n+1}}{2\cos\frac{\pi}{2n}}\otimes_{k=2}^m\openone_d\label{s3}
\end{eqnarray}
Using Eqs.~(\ref{s2}) and (\ref{s3}), we calculate $\bra{\psi}_{A^1..A^m} A^1_{i_1'}\otimes  A^2_{(i_1+i_3-2)\oplus_n +1 }\otimes_{k=3}^{m-1} A^k_{f_k(i_{k+1},i_k)}\otimes A_{i_m}^m   \ket{\psi}_{A^1..A^m}$ as follows 
\begin{eqnarray}\label{s4}
    \bra{\psi}_{A^1..A^m} A^1_{i_1'}\otimes  A^2_{(i_1+i_3-2)\oplus_n +1 } \otimes_{k=3}^{m-1} A^k_{f_k(i_{k+1},i_k)}\otimes A_{i_m}^m  \ket{\psi}_{A^1..A^m} &=&
    \begin{cases}
        \bra{\psi}_{A^1..A^m} \left(\frac{A^1_{i_1}-A^1_{i_1\oplus_n+1}}{2\cos\frac{\pi}{2n}}A^1_{i_1'} \otimes_{k=2}^m\openone_d\right)\ket{\psi}_{A^1..A^m} \nonumber\\
        \bra{\psi}_{A^1..A^m}\left( A^1_{i_1'}\frac{A^1_{i_1}-A^1_{i_1\oplus_n+1}}{2\cos\frac{\pi}{2n}} \otimes_{k=2}^m\openone_d\right)\ket{\psi}_{A^1..A^m} \nonumber
    \end{cases}\\
    &&=\frac{1}{2}\bra{\psi}_{A^1..A^m} \left(\frac{\{A^1_{i_1'},A^1_{i_1}\}-\{A^1_{i_1'}, A^1_{i_1\oplus_n+1}\}}{2\cos{\frac{\pi}{2 n}}} \otimes_{k=2}^m\openone_d\right)\ket{\psi}_{A^1..A^m} \nonumber\\
    &&=\frac{\langle\{A^1_{i_1'},A^1_{i_1}\}\rangle-\langle\{A^1_{i_1'}, A^1_{i_1\oplus_n+1}\}\rangle}{4\cos{\frac{\pi}{2 n}}}
\end{eqnarray}
Now for the three party scenario, i.e., $m=3$, Eq.~(\ref{s4}) becomes 
\begin{eqnarray}\label{s48}
    &&\bra{\psi}_{A^1A^2A^3} A^1_{i_1'} \otimes A^2_{(i_1+i_3-2)\oplus_n +1 }\otimes A^3_{i_3}  \ket{\psi}_{A^1A^2A^3}= \frac{\langle\{A^1_{i_1'},A^1_{i_1}\}\rangle-\langle\{A^1_{i_1'}, A^1_{i_1\oplus_n+1}\}\rangle}{4\cos{\frac{\pi}{2 n}}}\\
    &&\bra{\psi}_{A^1A^2A^3} A^1_{i_1'}\otimes A^2_{i_2'}\otimes A^3_{i_3} \ket{\psi}_{A^1A^2A^3}= \frac{\langle\{A^1_{i_1'},A^1_{i_1}\}\rangle-\langle\{A^1_{i_1'}, A^1_{i_1\oplus_n+1}\}\rangle}{4\cos{\frac{\pi}{2 n}}}
\end{eqnarray}

Where, $i_2=(i_1+i_3-2)\oplus_n +1 , \forall i_1,i_3\in[n]$. Now, for maximum randomness, the choice of measurement settings for the 1st, 2nd, and 3rd parties is $A^1_{1+(i_2-i_3+\frac{n+1}{2})\oplus_n}, A^2_{i_2}$ and $A^3_{i_3}$, respectively. Hence, putting $i_1'=1+(i_2-i_3+\frac{n+1}{2})\oplus_n$ in Eq.~(\ref{s48}), we get
\begin{eqnarray}\label{sy5}
    \bra{\psi}_{A^1A^2A^3}A^1_{1+(i_2-i_3+\frac{n+1}{2})\oplus_n}\otimes A^2_{i_2} \otimes A^3_{i_3} \ket{\psi}_{A^1A^2A^3}
    &&=\frac{\langle\{A^1_{1+(i_2-i_3+\frac{n+1}{2})\oplus_n},A^1_{i_1}\}\rangle-\langle\{A^1_{1+(i_2-i_3+\frac{n+1}{2})\oplus_n}, A^1_{i_1\oplus_n +1}\}\rangle}{4\cos{\frac{\pi}{2n}}}\quad
\end{eqnarray}
Now, for arbitrary $n$, the derivation of the RHS of Eq.~(\ref{sy5}) is challenging. Therefore, consider the example with $n=3$, $i_3=1$, and $i_2=1$. From the relation $i_2=(i_1+i_3-2)\oplus_n +1$ and the constraint $i_1\leq 3$, it follows that $i_1=1$. Putting this in Eq.~(\ref{sy5}), we get
\begin{eqnarray}
    \bra{\psi}_{A^1A^2A^3}A^1_{3} \otimes A^2_1 \otimes A^3_1\ket{\psi}_{A^1A^2A^3}
    &&=\frac{\langle\{A^1_{3},A^1_1\}\rangle-\langle\{A^1_{3}, A^1_{2}\}\rangle}{4\cos{\frac{\pi}{2n}}}=0\quad \left(\text{As $\langle \{A^1_{i_1}, A^1_{i_1+x}\} \rangle = 2(-1)^x\cos{\frac{\pi x}{3}}$}\right)\quad\quad
\end{eqnarray}
For any odd $n$ in a tripartite scenario, we can similarly prove
\begin{eqnarray}\label{maxraibjck}
    \bra{\psi}_{A^1A^2A^3}\left( A^1_{1+(i_2-i_3+\frac{n+1}{2})\oplus_n}\otimes A^2_{i_2} \otimes A^3_{i_3}\right)\ket{\psi}_{A^1A^2A^3}
    &=&0
\end{eqnarray}
Thus, for this particular selection of observables, we obtain  
\(P\left(a^1,a^2,a^3 \mid A^1_{1+(i_2-i_3+\frac{n+1}{2})\oplus_n}, A^2_{i_2}, A^3_{i_3}\right) = \frac{1}{8}\),  
which yields maximal randomness, \(\mathcal{R}^3_{\max} = 3\).

Our analysis then shifts to this context, analogous to the tri-party case, delineating the settings for any $m$-parties across any $n$-settings.
\begin{eqnarray}
    \bra{\psi}_{A^1..A^m} A^1_{1+(i_2-i_3+\frac{n+1}{2})\oplus_n}\otimes A^2_{i_2} \otimes_{k=3}^{m-1} A^k_{f_k(i_{k+1},i_{k})}\otimes A_{i_m}^m   \ket{\psi}_{A^1..A^m}=0,\quad\forall m\geq3
\end{eqnarray}
where
\begin{eqnarray}
    &&f_k(i_{k+1},i_k)=(i_{k}+i_{k+1}-2)\oplus_n +1, \quad  m>3
\end{eqnarray}
For these observables, $P\qty(a^1,a^2,..,a^k,..,a^m|A^1_{1+(i_2-i_3+\frac{n+1}{2})\oplus_n}, A^2_{i_2},.., A^k_{f_k(i_{k+1},i_k)},.., A_{i_m}^m )=\frac{1}{2^m},\forall k=[3,m-1]$, provide the maximum randomness, $\mathcal{R}^m_{max}=m$. That is, Eq.~(16) in the main text.

\section{Self-testing using swap circuit in tripartite scenario}
\subsection{For \texorpdfstring{$m=3$ and  $n=3$}{m=3 and n=3}}\label{sSelf3}
\begin{figure}
    \centering
    \includegraphics[width=12cm, height=5.5cm]{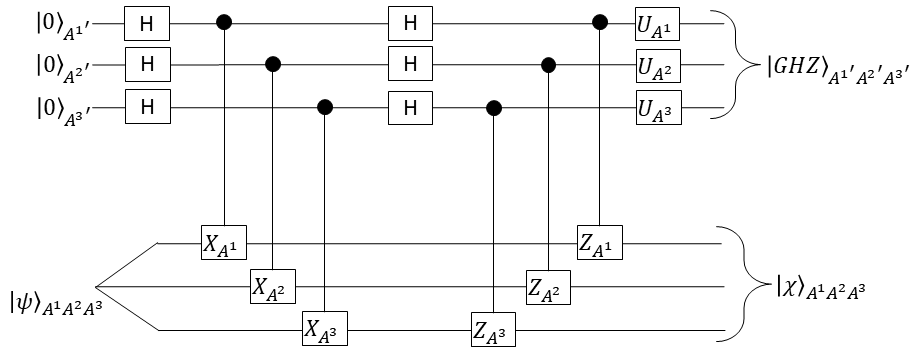}
    \caption{Self-testing circuit for n=3}\label{n=3c}
\end{figure}
For circuit implementation, note that
\begin{eqnarray}
&&\widetilde{A}^1_1=X_{A^1},\ \ \
    \widetilde{A}^1_2=\sin{\frac{\pi}{3}}Z_{A^1}-\cos{\frac{\pi}{3}}X_{A^1},\ \ \
    \widetilde{A}^1_3=-\sin{\frac{\pi}{3}}Z_{A^1}-\cos{\frac{\pi}{3}}X_{A^1}\label{C3}\\
    &&A^2_1=X_{A^2}, \ \ \
    A^2_2=-\sin{\frac{\pi}{6}}X_{A^2}+\cos{\frac{\pi}{6}}Z_{A^2},\ \ \
    A^2_3=-\sin{\frac{\pi}{6}}X_{A^2}-\cos{\frac{\pi}{6}}Z_{A^2}\label{B3}\\
    &&A^3_1=X_{A^3},\ \ \
    A^3_2=\sin{\frac{\pi}{3}}Z_{A^3}-\cos{\frac{\pi}{3}}X_{A^3},\ \ \
    A^3_3=-\sin{\frac{\pi}{3}}Z_{A^3}-\cos{\frac{\pi}{3}}X_{A^3}\label{A3}
\end{eqnarray}
By solving the previous equations, we get $X_{A^1}=\widetilde{A}^1_1$, $Z_{A^1}=\frac{\widetilde{A}^1_2-\widetilde{A}^1_3}{\sqrt{2-\langle\{\widetilde{A}^1_2,\widetilde{A}^1_3\}\rangle}}$, $X_{A^2}=A^2_1$, $Z_{A^2}=\frac{A^2_2-A^2_3}{\sqrt{2-\langle\{A^2_2,A^2_3\}\rangle}}$,$X_{A^3}=A^3_1$, and $Z_{A^3}=\frac{A^3_2-A^3_3}{\sqrt{2-\langle\{A^3_2,A^3_3\}\rangle}}$.
Hence, putting Eq.~(\ref{C3}), (\ref{B3}), and (\ref{A3}) in Eq.~(\ref{slf3}), we simplified it and obtained the following conditions, which we will use later.
\begin{eqnarray}
    &&X_{A^1}\otimes X_{A^2}\otimes X_{A^3}\ket{\psi}_{A^1A^2A^3}=
    X_{A^1}\otimes Z_{A^2} \otimes Z_{A^3} \ket{\psi}_{A^1A^2A^3}=
    Z_{A^1} \otimes Z_{A^2} \otimes X_{A^3} \ket{\psi}_{A^1A^2A^3}=\ket{\psi}_{A^1A^2A^3}\quad \quad \label{C34}\\
   && -
   Z_{A^1}  \otimes X_{A^2} \otimes Z_{A^3}\ket{\psi}_{A^1A^2A^3}= \ket{\psi}_{A^1A^2A^3},\ \openone_d\otimes X_{A^2} \otimes X_{A^3}\ket{\psi}_{A^1A^2A^3}=\openone_d\otimes Z_{A^2} \otimes Z_{A^3}\ket{\psi}_{A^1A^2A^3}\label{C35}
\end{eqnarray}
If $\ket{\psi}_{A^1A^2A^3}\in\mathcal{H}_{A^1}\otimes\mathcal{H}_{A^2}\otimes \mathcal{H}_{A^3}$ is the appropriate state and $A^1_{i_1}\in \mathbf{L}(\mathcal{H}_{A^1})$, $A^2_{i_2}\in\mathbf{L}(\mathcal{H}_{A^2})$ and $A^3_{i_3}\in\mathbf{L}(\mathcal{H}_{A^3})$ are the corresponding observables to obtain the optimal value of the inequality, then from Fig.~\ref{n=3c} it can be proved that there exists a local unitary operation, say $\phi$ and Ancillary state $\ket{000}_{A^{1'} A^{2'} A^{3'}}$ such that
\begin{eqnarray}\label{phipsi}
    &&\phi\qty(\ket{\psi}_{A^1A^2A^3}\otimes\ket{000}_{A^{1'} A^{2'} A^{3'}} 
 )\nonumber\\
 &&= \frac{1}{8}\Big[(1+X_{A^1})(1+X_{A^2})(1+X_{A^3})\ket{\psi}_{A^1A^2A^3}\otimes\ket{000}_{A^{1'} A^{2'} A^{3'}}+(1+X_{A^1})(1+X_{A^2})Z_{A^3}(1-X_{A^3})\ket{\psi}_{A^1A^2A^3}\otimes\ket{001}_{A^{1'} A^{2'} A^{3'}}\nonumber\\
 &&+(1+X_{A^1})Z_{A^2}(1-X_{A^2})(1+X_{A^3})\ket{\psi}_{A^1A^2A^3}\otimes\ket{010}_{A^{1'} A^{2'} A^{3'}}+(1+X_{A^1})Z_{A^2}(1-X_{A^2})Z_{A^3}(1-X_{A^3})\ket{\psi}_{A^1A^2A^3}\otimes\ket{011}_{A^{1'} A^{2'} A^{3'}}\nonumber\\
 &&+Z_{A^1}(1-X_{A^1})(1+X_{A^2})(1+X_{A^3})\ket{\psi}_{A^1A^2A^3}\otimes\ket{100}_{A^{1'} A^{2'} A^{3'}}+Z_{A^1}(1-X_{A^1})(1+X_{A^2})Z_{A^3}(1-X_{A^3})\ket{\psi}_{A^1A^2A^3}\otimes\ket{101}_{A^{1'} A^{2'} A^{3'}}\nonumber\\
 &&+Z_{A^1}(1-X_{A^1})Z_{A^2}(1-X_{A^2})(1+X_{A^3})\ket{\psi}_{A^1A^2A^3}\otimes\ket{110}_{A^{1'} A^{2'} A^{3'}}\nonumber\\
 &&+Z_{A^1}(1-X_{A^1})Z_{A^2}(1-X_{A^2})Z_{A^3}(1-X_{A^3})\ket{\psi}_{A^1A^2A^3}\otimes\ket{111}_{A^{1'} A^{2'} A^{3'}}\Big]
\end{eqnarray}
The calculation of the first term in Eq.~(\ref{phipsi})
\begin{eqnarray}
    && (1+X_{A^1})(1+X_{A^2})(1+X_{A^3})\ket{\psi}_{A^1A^2A^3}\nonumber\\
    &&= (1+X_{A^1}+X_{A^2}+X_{A^1} X_{A^2}+X_{A^3}+X_{A^1} X_{A^3} + X_{A^2} X_{A^3} + X_{A^1} X_{A^2} X_{A^3})\ket{\psi}_{A^1A^2A^3}\nonumber\\
    &&= 2(1+X_{A^1}+X_{A^2}+X_{A^3})\ket{\psi}_{A^1A^2A^3}\quad \text{[using Eq.~(\ref{C34})]}
\end{eqnarray}
In the following manner, from the second to the eighth term, we get
\begin{eqnarray}
 &&(1+X_{A^1})(1+X_{A^2})Z_{A^3}(1-X_{A^3})\ket{\psi}_{A^1A^2A^3}=0\\
    &&(1+X_{A^1})Z_{A^2}(1-X_{A^2})(1+X_{A^3})\ket{\psi}_{A^1A^2A^3}=0\\
    && (1+X_{A^1})Z_{A^2}(1-X_{A^2})Z_{A^3}(1-X_{A^3})\ket{\psi}_{A^1A^2A^3}=2(1+X_{A^3}+X_{A^2}+X_{A^1})\ket{\psi}_{A^1A^2A^3}\\
    &&Z_{A^1}(1-X_{A^1})(1+X_{A^2})(1+X_{A^3})\ket{\psi}_{A^1A^2A^3}=0\\
    && Z_{A^1}(1-X_{A^1})(1+X_{A^2})Z_{A^3}(1-X_{A^3})\ket{\psi}_{A^1A^2A^3}=-2(1+X_{A^3}+X_{A^2}+X_{A^1})\ket{\psi}_{A^1A^2A^3}\\
    &&Z_{A^1}(1-X_{A^1})Z_{A^2}(1-X_{A^2})(1+X_{A^3})\ket{\psi}_{A^1A^2A^3}=2(1+X_{A^3}+X_{A^2}+X_{A^1})\ket{\psi}_{A^1A^2A^3}\\
    && Z_{A^1}(1-X_{A^1})Z_{A^2}(1-X_{A^2})Z_{A^3}(1-X_{A^3})\ket{\psi}_{A^1A^2A^3}=0
\end{eqnarray}

Now, putting all the first to eighth terms in Eq.~(\ref{phipsi}) we get
\begin{eqnarray}
    \phi(\ket{\psi}_{A^1A^2A^3}\otimes\ket{000}_{A^{1'} A^{2'} A^{3'}} 
 )&=&\frac{(1+X_{A^1}+X_{A^2}+X_{A^3})}{4}(\ket{000}+\ket{011}-\ket{101}+\ket{110})\ket{\psi}_{A^1A^2A^3}
\end{eqnarray}
Now $Alice^k$, use local unitary $U_{A^{k'}}, \forall k\in[3]$ on their part in such a way that it extracts the GHZ states in the output. Hence, 
\begin{eqnarray}
&&\Phi(\ket{\psi}_{A^1A^2A^3}\otimes\ket{000}_{A^{1'} A^{2'} A^{3'}})=\frac{(1+X_{A^1}+X_{A^2}+X_{A^3})}{4}\ket{\psi}_{A^1A^2A^3}(U_{A^{1'}}\otimes U_{A^{2'}} \otimes U_{A^{3'}})(\ket{000}+\ket{011}-\ket{101}+\ket{110})_{A^{1'} A^{2'} A^{3'}}\nonumber\\
 &&\hspace{3.9cm}=\frac{(1+X_{A^1}+X_{A^2}+X_{A^3})}{2}\ket{\psi}_{A^1A^2A^3}\frac{(\ket{000}_{A^{1'} A^{2'} A^{3'}}+\ket{111}_{A^{1'} A^{2'} A^{3'}})}{\sqrt{2}}\nonumber\\
&&\hspace{3.9cm}=\ket{\chi}_{A^1A^2A^3}\otimes \ket{GHZ}_{A^{1'} A^{2'} A^{3'}}
\end{eqnarray}
Where $\Phi(\ket{\psi}_{A^1A^2A^3}\otimes\ket{000}_{A^{1'} A^{2'} A^{3'}})= (U_{A^{1'}}\otimes U_{A^{2'}} \otimes U_{A^{3'}})\phi(\ket{\psi}_{A^1A^2A^3}\otimes\ket{000}_{A^{1'} A^{2'} A^{3'}})$ and $\ket{\chi}_{A^1A^2A^3}=\frac{(1+X_{A^1}+X_{A^2}+X_{A^3})}{2}\ket{\psi}_{A^1A^2A^3}$ which is the junk state. This implies the self-testing of a GHZ state using the optimal quantum value of the corresponding inequality. And the local unitaries are $U_{A^{1'}}=\frac{1}{\sqrt{2}}\begin{pmatrix}
    1&-i\\-i&1
\end{pmatrix}$, $U_{A^{2'}}=\frac{1}{\sqrt{2}}\begin{pmatrix}
    1&i\\-1&i
\end{pmatrix}$ and $U_{A^{3'}}=\frac{1}{\sqrt{2}}\begin{pmatrix}
    1&-i\\-i&1
\end{pmatrix}$. Now, for self-testing the measurements, we write $A^1_i, A^2_{i_2}, A^3_k$ just-like Eq.~(\ref{C3}),(\ref{B3}),(\ref{A3}). Hence, it is enough to demonstrate how the local isometry works for $X_{A^1}, X_{A^2}, X_{A^3}, Z_{A^1}, Z_{A^2}, Z_{A^3}$. We thus show the following using the self-testing circuit:
{\small\begin{eqnarray}\label{X_{A^3}}
    &&\phi(X_{A^3}\ket{\psi}_{A^1A^2A^3}\otimes\ket{000}_{A^{1'} A^{2'} A^{3'}} 
 )\nonumber\\
 &&= \frac{1}{8}\Big[(1+X_{A^1})(1+X_{A^2})(1+X_{A^3})X_{A^3}\ket{\psi}_{A^1A^2A^3}\otimes\ket{000}_{A^{1'} A^{2'} A^{3'}}+(1+X_{A^1})(1+X_{A^2})Z_{A^3}(1-X_{A^3})X_{A^3}\ket{\psi}_{A^1A^2A^3}\otimes\ket{001}_{A^{1'} A^{2'} A^{3'}}\nonumber\\
 &&+(1+X_{A^1})Z_{A^2}(1-X_{A^2})(1+X_{A^3})X_{A^3}\ket{\psi}_{A^1A^2A^3}\otimes\ket{010}_{A^{1'} A^{2'} A^{3'}}+(1+X_{A^1})Z_{A^2}(1-X_{A^2})Z_{A^3}(1-X_{A^3})X_{A^3}\ket{\psi}_{A^1A^2A^3}\otimes\ket{011}_{A^{1'} A^{2'} A^{3'}}\nonumber\\
 &&+Z_{A^1}(1-X_{A^1})(1+X_{A^2})(1+X_{A^3})X_{A^3}\ket{\psi}_{A^1A^2A^3}\otimes\ket{100}_{A^{1'} A^{2'} A^{3'}}+Z_{A^1}(1-X_{A^1})(1+X_{A^2})Z_{A^3}(1-X_{A^3})X_{A^3}\ket{\psi}_{A^1A^2A^3}\otimes\ket{101}_{A^{1'} A^{2'} A^{3'}}\nonumber\\
 &&+Z_{A^1}(1-X_{A^1})Z_{A^2}(1-X_{A^2})(1+X_{A^3})X_{A^3}\ket{\psi}_{A^1A^2A^3}\otimes\ket{110}_{A^{1'} A^{2'} A^{3'}}+Z_{A^1}(1-X_{A^1})Z_{A^2}(1-X_{A^2})Z_{A^3}(1-X_{A^3})X_{A^3}\ket{\psi}_{A^1A^2A^3}\otimes\ket{111}_{A^{1'} A^{2'} A^{3'}}\Big]\nonumber\\
\end{eqnarray}}
 Calculation of the first term in Eq.~(\ref{phipsi})
\begin{eqnarray}
    && (1+X_{A^1})(1+X_{A^2})(1+X_{A^3})X_{A^3}\ket{\psi}_{A^1A^2A^3}\nonumber\\
    &&= (1+X_{A^1}+X_{A^2}+X_{A^1} X_{A^2}+X_{A^3}+X_{A^1} X_{A^3} + X_{A^2} X_{A^3} + X_{A^1} X_{A^2} X_{A^3})\ket{\psi}_{A^1A^2A^3}\nonumber\\
    &&= 2(1+X_{A^1}+X_{A^2}+X_{A^3})\ket{\psi}_{A^1A^2A^3}\quad \text{[Using Eq.~(\ref{C34})]}
\end{eqnarray}
In the following manner, using Eq.~(\ref{C34}) to Eq.~(\ref{C35}) from the second to seventh term, we get
\begin{eqnarray}
     &&(1+X_{A^1})(1+X_{A^2})Z_{A^3}(1-X_{A^3})X_{A^3}\ket{\psi}_{A^1A^2A^3}=0\nonumber\\
     && (1+X_{A^1})Z_{A^2}(1-X_{A^2})(1+X_{A^3})X_{A^3}\ket{\psi}_{A^1A^2A^3}=0\nonumber\\
     && (1+X_{A^1})Z_{A^2}(1-X_{A^2})Z_{A^3}(1-X_{A^3})X_{A^3}\ket{\psi}_{A^1A^2A^3}=-2(1+X_{A^3}+X_{A^2}+X_{A^1})\ket{\psi}_{A^1A^2A^3}\nonumber\\
    && Z_{A^1}(1-X_{A^1})(1+X_{A^2})(1+X_{A^3})X_{A^3}\ket{\psi}_{A^1A^2A^3}=0\nonumber\\
     && Z_{A^1}(1-X_{A^1})(1+X_{A^2})Z_{A^3}(1-X_{A^3})X_{A^3}\ket{\psi}_{A^1A^2A^3}=2(1+X_{A^3}+X_{A^2}+X_{A^1})\ket{\psi}_{A^1A^2A^3}\nonumber\\
    &&Z_{A^1}(1-X_{A^1}) Z_{A^2}(1-X_{A^2})(1+X_{A^3})X_{A^3}\ket{\psi}_{A^1A^2A^3}=2(1+X_{A^3}+X_{A^2}+X_{A^1})\ket{\psi}_{A^1A^2A^3}\nonumber\\
      &&Z_{A^1}(1-X_{A^1})Z_{A^2}(1-X_{A^2})Z_{A^3}(1-X_{A^3})X_{A^3} \ket{\psi}_{A^1A^2A^3}=0
\end{eqnarray}

Now, putting all the first to eighth terms in Eq.~(\ref{phipsi}) we get
\begin{eqnarray}
    \phi(X_{A^3}\ket{\psi}_{A^1A^2A^3}\otimes\ket{000}_{A^{1'} A^{2'} A^{3'}} 
 )&=&\frac{(1+X_{A^1}+X_{A^2}+X_{A^3})}{4}(\ket{000}-\ket{011}+\ket{101}+\ket{110})\ket{\psi}_{A^1A^2A^3}
\end{eqnarray}
Now $A^1$, $A^2$ and $A^3$ uses local unitary $U_{A^{1'}}, U_{A^{2'}}, U_{A^{3'}}$ on their part, and get the output as
\begin{eqnarray}\label{Xa}
    &&(U_{A^{1'}}\otimes U_{A^{2'}} \otimes U_{A^{3'}})\phi(X_{A^3}\ket{\psi}_{A^1A^2A^3}\otimes\ket{000}_{A^{1'} A^{2'} A^{3'}} 
 )=\frac{(1+X_{A^1}+X_{A^2}+X_{A^3})}{2}\frac{(-i\ket{000}_{A^{1'} A^{2'} A^{3'}}+i\ket{011}_{A^{1'} A^{2'} A^{3'}})}{\sqrt{2}}\ket{\psi}_{A^1A^2A^3}\nonumber\\
&&\Phi(X_{A^3}\ket{\psi}_{A^1A^2A^3}\otimes\ket{000}_{A^{1'} A^{2'} A^{3'}})=\frac{(1+X_{A^1}+X_{A^2}+X_{A^1})}{3}\ket{\psi}_{A^1A^2A^3}\otimes(\openone_2\otimes \openone_2\otimes -\sigma_y)\frac{(\ket{000}_{A^{1'} A^{2'} A^{3'}}+\ket{111}_{A^{1'} A^{2'} A^{3'}})}{\sqrt{2}}\nonumber\\
 &&\hspace{4.5cm}= \ket{\chi}_{A^1A^2A^3}\otimes(\openone_2\otimes \openone_2\otimes -\sigma_y) \ket{GHZ}_{A^{1'} A^{2'} A^{3'}}
\end{eqnarray}
Where $\ket{\chi}_{A^1A^2A^3}=\frac{(1+X_{A^3}+X_{A^2}+X_{A^1})}{2}\ket{\psi}_{A^1A^2A^3}$ is the junk state. In a similar way we can write

\begin{eqnarray}
(U_{A^{1'}}\otimes U_{A^{2'}} \otimes U_{A^{3'}})\phi(X_{A^1}\ket{\psi}_{A^1A^2A^3}\otimes\ket{000}_{A^{1'} A^{2'} A^{3'}} 
 )
 &=& \Phi(X_{A^1}\ket{\psi}_{A^1A^2A^3}\otimes\ket{000}_{A^{1'} A^{2'} A^{3'}})\nonumber\\
 &=&\ket{\chi}_{A^1A^2A^3}\otimes(-\sigma_y\otimes \openone_2\otimes \openone_2) \ket{GHZ}_{A^{1'} A^{2'} A^{3'}}\label{Xc}\\
    (U_{A^{1'}}\otimes U_{A^{2'}} \otimes U_{A^{3'}})\phi(X_{A^2}\ket{\psi}_{A^1A^2A^3}\otimes\ket{000}_{A^{1'} A^{2'} A^{3'}} 
 )
 &=&\Phi(X_{A^2}\ket{\psi}_{A^1A^2A^3}\otimes\ket{000}_{A^{1'} A^{2'} A^{3'}})\nonumber\\
 &=&\ket{\chi}_{A^1A^2A^3}\otimes(\openone_2\otimes -\sigma_x \otimes \openone_2) \ket{GHZ}_{A^{1'} A^{2'} A^{3'}}\label{Xb}\\
 (U_{A^{1'}}\otimes U_{A^{2'}} \otimes U_{A^{3'}})\phi(Z_{A^1}\ket{\psi}_{A^1A^2A^3}\otimes\ket{000}_{A^{1'} A^{2'} A^{3'}} 
 )
 &=& \Phi(Z_{A^1}\ket{\psi}_{A^1A^2A^3}\otimes\ket{000}_{A^{1'} A^{2'} A^{3'}})\nonumber\\
 &=&\ket{\chi}_{A^1A^2A^3}\otimes(\sigma_x\otimes \openone_2\otimes \openone_2) \ket{GHZ}_{A^{1'} A^{2'} A^{3'}}\label{Zc}\\
 (U_{A^{1'}}\otimes U_{A^{2'}} \otimes U_{A^{3'}})\phi(Z_{A^2}\ket{\psi}_{A^1A^2A^3}\otimes\ket{000}_{A^{1'} A^{2'} A^{3'}} 
 )
 &=&\Phi(Z_{A^2}\ket{\psi}_{A^1A^2A^3}\otimes\ket{000}_{A^{1'} A^{2'} A^{3'}})\nonumber\\
 &=&\ket{\chi}_{A^1A^2A^3}\otimes(\openone_2\otimes \sigma_y \otimes \openone_2) \ket{GHZ}_{A^{1'} A^{2'} A^{3'}}\label{Zb}\\
 (U_{A^{1'}}\otimes U_{A^{2'}} \otimes U_{A^{3'}})\phi(Z_{A^3}\ket{\psi}_{A^1A^2A^3}\otimes\ket{000}_{A^{1'} A^{2'} A^{3'}} 
 )
 &=& \Phi(Z_{A^3}\ket{\psi}_{A^1A^2A^3}\otimes\ket{000}_{A^{1'} A^{2'} A^{3'}})\nonumber\\
 &=&\ket{\chi}_{A^1A^2A^3}\otimes(\openone_2\otimes \openone_2\otimes \sigma_x) \ket{GHZ}_{A^{1'} A^{2'} A^{3'}}\label{Za}
\end{eqnarray}
Using Eq.~(\ref{Xa}),(\ref{Xc}), (\ref{Xb}) and Eq.~ (\ref{Zc}), (\ref{Zb}),(\ref{Za}) it is straightforward that
\begin{eqnarray}\label{selfm}
\Phi(A^1_{i_1}\ket{\psi}_{A^1A^2A^3}\otimes\ket{000}_{A^{1'} A^{2'} A^{3'}})
 &=& \ket{\chi}_{A^1A^2A^3}\otimes(A^{1'}_{i_1}\otimes\openone_2\otimes \openone_2) \ket{GHZ}_{A^{1'} A^{2'} A^{3'}}\\
\Phi(A^2_{i_2}\ket{\psi}_{A^1A^2A^3}\otimes\ket{000}_{A^{1'} A^{2'} A^{3'}})
 &=& \ket{\chi}_{A^1A^2A^3}\otimes(\openone_2\otimes A^{2'}_{i_2} \otimes \openone_2) \ket{GHZ}_{A^{1'} A^{2'} A^{3'}}\\
\Phi(A^3_{i_3}\ket{\psi}_{A^1A^2A^3}\otimes\ket{000}_{A^{1'} A^{2'} A^{3'}})
 &=& \ket{\chi}_{A^1A^2A^3}\otimes( \openone_2\otimes \openone_2\otimes A^{3'}_{i_3}\otimes) \ket{GHZ}_{A^{1'} A^{2'} A^{3'}}
\end{eqnarray}
Similarly, with a few more steps, we can show that
\begin{eqnarray}\label{selfm1}
    \Phi(A^1_{i_1}\otimes A^2_{i_2}\otimes A^3_{i_3}\ket{\psi}_{A^1A^2A^3}\otimes\ket{000}_{A^{1'} A^{2'} A^{3'}})
 &=& \ket{\chi}_{A^1A^2A^3}\otimes(A^{1'}_{i_1}\otimes A^{2'}_{i_2} \otimes A^{3'}_{i_3}) \ket{GHZ}_{A^{1'} A^{2'} A^{3'}}
\end{eqnarray}
That corresponds to Eq.~(21) in the main text. Hence, we have self-tested the measurements.

\section{Detailed calculations for robust self-testing of state and observables}
We have assumed that the deviated observables $\widetilde{Z}_{A^k}, \widetilde{X}_{A^k}$  are closed to the real observables $Z_{A^k}, X_{A^k}$ ($||X_{A^k}\ket{\psi}_{A^1A^2A^3}||=||Z_{A^k}\ket{\psi}_{A^1A^2A^3}||=1$) with $\epsilon_{A^k}$ and $\alpha_{A^k}$ margin respectively, mathematically it can be written as
\begin{eqnarray}\label{rob}
    ||(\widetilde{X}_{A^k}-X_{A^k})\ket{\psi}_{A^1A^2A^3}||\leq \epsilon_{A^k}\Rightarrow  \widetilde{X}_{A^k}\approx \epsilon_{A^k} \openone_d+ X_{A^k} \ ; \  ||(\widetilde{Z}_{A^k}-Z_{A^k})\ket{\psi}_{A^1A^2A^3}||\leq \alpha_{A^k} \Rightarrow  \widetilde{Z}_{A^k}\approx \alpha_{A^k} \openone_d+ Z_{A^k}, \forall k\in[3]\qquad
\end{eqnarray}
Using the Eq.~(\ref{rob}) following relations are derived
\begin{eqnarray}\label{robxx}
    ||(\widetilde{X}_{A^i} \widetilde{X}_{A^j}-X_{A^i} X_{A^j})\ket{\psi}_{A^1A^2A^3}||&\approx& ||(\widetilde{X}_{A^i} (\epsilon_{A^j} \openone_d+ X_{A^j})-X_{A^i} X_{A^j})\ket{\psi}_{A^1A^2A^3}||\nonumber\\
    &\leq& \epsilon_{A^j} ||\widetilde{X}_{A^i}\ket{\psi}_{A^1A^2A^3}||+||(\widetilde{X}_{A^i}-X_{A^i})X_{A^j}\ket{\psi}_{A^1A^2A^3}||\nonumber\\
    &\leq& \epsilon_{A^j} ||\widetilde{X}_{A^i}\ket{\psi}_{A^1A^2A^3}||+\epsilon_{A^i}
\end{eqnarray}
\begin{eqnarray}\label{robxxx}
    ||(\widetilde{X}_{A^i} \widetilde{X}_{A^j}\widetilde{Z}_{A^k}-X_{A^i} X_{A^j} X_{A^k})\ket{\psi}_{A^1A^2A^3}||&\approx& ||(\widetilde{X}_{A^i} \widetilde{X}_{A^j}(\epsilon_{A^k} \openone_d+ X_{A^k})-X_{A^i} X_{A^j} X_{A^k})\ket{\psi}_{A^1A^2A^3}||\nonumber\\
    &\leq& \epsilon_{A^k} ||\widetilde{X}_{A^i}\widetilde{X}_{A^j}\ket{\psi}_{A^1A^2A^3}||+||(\widetilde{X}_{A^i}\widetilde{X}_{A^j}-X_{A^i}X_{A^j})X_{A^k}\ket{\psi}_{A^1A^2A^3}||\nonumber\\
    &\leq& \epsilon_{A^k} ||\widetilde{X}_{A^i}\widetilde{X}_{A^j}\ket{\psi}_{A^1A^2A^3}||+\epsilon_{A^j} ||\widetilde{X}_{A^i}\ket{\psi}_{A^1A^2A^3}||+\epsilon_{A^i} \ \ \ (\text{From Eq.~(\ref{robxx})})
\end{eqnarray}
\begin{eqnarray}\label{robzx}
    ||(\widetilde{Z}_{A^i} \widetilde{X}_{A^j}-Z_{A^i} X_{A^j})\ket{\psi}_{A^1A^2A^3}||&\approx& ||(\widetilde{Z}_{A^i} (\epsilon_{A^j} \openone_d+ X_{A^j})-Z_{A^i} X_{A^j})\ket{\psi}_{A^1A^2A^3}||\nonumber\\
    &\leq& \epsilon_{A^j} ||\widetilde{Z}_{A^i}\ket{\psi}_{A^1A^2A^3}||+||(\widetilde{Z}_{A^i}-Z_{A^i})X_{A^j}\ket{\psi}_{A^1A^2A^3}||\nonumber\\
    &\leq& \epsilon_{A^j} ||\widetilde{Z}_{A^i}\ket{\psi}_{A^1A^2A^3}||+\alpha_{A^i}
\end{eqnarray}
\begin{eqnarray}\label{robzxx}
    ||(\widetilde{Z}_{A^i} \widetilde{X}_{A^j}\widetilde{Z}_{A^k}-Z_{A^i} X_{A^j} X_{A^k})\ket{\psi}_{A^1A^2A^3}||&\approx& ||(\widetilde{Z}_{A^i} \widetilde{X}_{A^j}(\epsilon_{A^k} \openone_d+ X_{A^k})-Z_{A^i} X_{A^j} X_{A^k})\ket{\psi}_{A^1A^2A^3}||\nonumber\\
    &\leq& \epsilon_{A^k} ||\widetilde{Z}_{A^i}\widetilde{X}_{A^j}\ket{\psi}_{A^1A^2A^3}||+||(\widetilde{Z}_{A^i}\widetilde{X}_{A^j}-Z_{A^i}X_{A^j})X_{A^k}\ket{\psi}_{A^1A^2A^3}||\nonumber\\
    &\leq& \epsilon_{A^k} ||\widetilde{Z}_{A^i}\widetilde{X}_{A^j}\ket{\psi}_{A^1A^2A^3}||+\epsilon_{A^j} ||\widetilde{Z}_{A^i}\ket{\psi}_{A^1A^2A^3}||+\alpha_{A^i} \ \ \ (\text{From Eq.~(\ref{robzx})})
\end{eqnarray}
\begin{eqnarray}\label{robzxxx}
    ||(\widetilde{Z}_{A^i} \widetilde{X}_{A^j}\widetilde{Z}_{A^k}\widetilde{Z}_{A^l}-Z_{A^i} X_{A^j} X_{A^k} X_{A^l})\ket{\psi}_{A^1A^2A^3}||&\approx& ||(\widetilde{Z}_{A^i} \widetilde{X}_{A^j} \widetilde{Z}_{A^k}(\epsilon_{A^l} \openone_d+ X_{A^l})-Z_{A^i} X_{A^j} X_{A^k} X_{A^l})\ket{\psi}_{A^1A^2A^3}||\nonumber\\
    &\leq& \epsilon_{A^l} ||\widetilde{Z}_{A^i}\widetilde{X}_{A^j}\widetilde{Z}_{A^k} \ket{\psi}_{A^1A^2A^3}||+||(\widetilde{Z}_{A^i}\widetilde{X}_{A^j}\widetilde{Z}_{A^k} -Z_{A^i}X_{A^j} X_{A^k})X_{A^l}\ket{\psi}_{A^1A^2A^3}||\nonumber\\
    &\leq& \epsilon_{A^l} ||\widetilde{Z}_{A^i}\widetilde{X}_{A^j} \widetilde{Z}_{A^k}\ket{\psi}_{A^1A^2A^3}||+\epsilon_{A^k} ||\widetilde{Z}_{A^i}\widetilde{X}_{A^j}\ket{\psi}_{A^1A^2A^3}||\nonumber\\
    &&+\epsilon_{A^j} ||\widetilde{Z}_{A^i}\ket{\psi}_{A^1A^2A^3}||+\alpha_{A^i} \ \ \ (\text{From Eq.~(\ref{robzxx})})
\end{eqnarray}
\begin{eqnarray}\label{robZZ}
    ||(\widetilde{Z}_{A^i} \widetilde{Z}_{A^j}-Z_{A^i} Z_{A^j})\ket{\psi}_{A^1A^2A^3}||&\approx& ||(\widetilde{Z}_{A^i} (\alpha_{A^j} \openone_d+ Z_{A^j})-Z_{A^i} Z_{A^j})\ket{\psi}_{A^1A^2A^3}||\nonumber\\
    &\leq& \alpha_{A^j} ||\widetilde{Z}_{A^i}\ket{\psi}_{A^1A^2A^3}||+||(\widetilde{Z}_{A^i}-Z_{A^i})Z_{A^j}\ket{\psi}_{A^1A^2A^3}||\nonumber\\
    &\leq& \alpha_{A^j} ||\widetilde{Z}_{A^i}\ket{\psi}_{A^1A^2A^3}||+\alpha_{A^i}
\end{eqnarray}
\begin{eqnarray}\label{robZZX}
    ||(\widetilde{Z}_{A^i} \widetilde{Z}_{A^j}\widetilde{Z}_{A^k}-Z_{A^i} Z_{A^j} X_{A^k})\ket{\psi}_{A^1A^2A^3}||&\approx& ||(\widetilde{Z}_{A^i} \widetilde{Z}_{A^j}(\epsilon_{A^k} \openone_d+ X_{A^k})-Z_{A^i} Z_{A^j} X_{A^k})\ket{\psi}_{A^1A^2A^3}||\nonumber\\
    &\leq& \epsilon_{A^k} ||\widetilde{Z}_{A^i} \widetilde{Z}_{A^j}\ket{\psi}_{A^1A^2A^3}||+||(\widetilde{Z}_{A^i} \widetilde{Z}_{A^j}-Z_{A^i} Z_{A^j}) X_{A^k}\ket{\psi}_{A^1A^2A^3}||\nonumber\\
    &\leq& \epsilon_{A^k} ||\widetilde{Z}_{A^i} \widetilde{Z}_{A^j}\ket{\psi}_{A^1A^2A^3}||+\alpha_{A^j} ||\widetilde{Z}_{A^i}\ket{\psi}_{A^1A^2A^3}||+\alpha_{A^i} \ \ \ (\text{From Eq.~(\ref{robZZ})})
\end{eqnarray}
\begin{eqnarray}\label{robZZXX}
    ||(\widetilde{Z}_{A^i} \widetilde{Z}_{A^j}\widetilde{Z}_{A^k}\widetilde{Z}_{A^l}-Z_{A^i} Z_{A^j} X_{A^k} X_{A^l})\ket{\psi}_{A^1A^2A^3}||&\approx& ||(\widetilde{Z}_{A^i} \widetilde{Z}_{A^j}\widetilde{Z}_{A^k}(\epsilon_{A^l} \openone_d+ X_{A^l})-Z_{A^i} Z_{A^j} X_{A^k} X_{A^l})\ket{\psi}_{A^1A^2A^3}||\nonumber\\
    &\leq& \epsilon_{A^l} ||\widetilde{Z}_{A^i} \widetilde{Z}_{A^j}\widetilde{Z}_{A^k}\ket{\psi}_{A^1A^2A^3}||+||(\widetilde{Z}_{A^i} \widetilde{Z}_{A^j}\widetilde{Z}_{A^k}-Z_{A^i} Z_{A^j} X_{A^k}) X_{A^l}\ket{\psi}_{A^1A^2A^3}||\nonumber\\
    &\leq& \epsilon_{A^l} ||\widetilde{Z}_{A^i} \widetilde{Z}_{A^j} \widetilde{Z}_{A^k}\ket{\psi}_{A^1A^2A^3}||+\epsilon_{A^k} ||\widetilde{Z}_{A^i} \widetilde{Z}_{A^j}\ket{\psi}_{A^1A^2A^3}||\nonumber\\
    &&+\alpha_{A^j} ||\widetilde{Z}_{A^i}\ket{\psi}_{A^1A^2A^3}||+\alpha_{A^i} \ \ \ (\text{From Eq.~(\ref{robZZX})})
\end{eqnarray}
\begin{eqnarray}\label{robZZXXX}
    ||(\widetilde{Z}_{A^i} \widetilde{Z}_{A^j}\widetilde{Z}_{A^k}\widetilde{Z}_{A^l}\widetilde{Z}_{A^m}-Z_{A^i} Z_{A^j} X_{A^k} X_{A^l} X_{A^m})\ket{\psi}_{A^1A^2A^3}||&\approx& ||(\widetilde{Z}_{A^i} \widetilde{Z}_{A^j}\widetilde{Z}_{A^k}\widetilde{Z}_{A^l}(\epsilon_{A^m} \openone_d+ X_{A^m})-Z_{A^i} Z_{A^j} X_{A^k} X_{A^l} X_{A^m})\ket{\psi}_{A^1A^2A^3}||\nonumber\\
    &\leq& \epsilon_{A^m} ||\widetilde{Z}_{A^i} \widetilde{Z}_{A^j}\widetilde{Z}_{A^k}\widetilde{Z}_{A^l}\ket{\psi}_{A^1A^2A^3}||\nonumber\\
    &&+||(\widetilde{Z}_{A^i} \widetilde{Z}_{A^j}\widetilde{Z}_{A^k}\widetilde{Z}_{A^l}-Z_{A^i} Z_{A^j} X_{A^k} X_{A^l}) X_{A^m}\ket{\psi}_{A^1A^2A^3}||\nonumber\\
    &\leq& \epsilon_{A^m} ||\widetilde{Z}_{A^i} \widetilde{Z}_{A^j} \widetilde{Z}_{A^k}\widetilde{Z}_{A^l}\ket{\psi}_{A^1A^2A^3}||+\epsilon_{A^l} ||\widetilde{Z}_{A^i} \widetilde{Z}_{A^j} \widetilde{Z}_{A^k}\ket{\psi}_{A^1A^2A^3}||\nonumber\\
    &&+\epsilon_{A^k} ||\widetilde{Z}_{A^i} \widetilde{Z}_{A^j}\ket{\psi}_{A^1A^2A^3}||+\alpha_{A^j} ||\widetilde{Z}_{A^i}\ket{\psi}_{A^1A^2A^3}||\nonumber\\
    &&+\alpha_{A^i}(\text{From Eq.~(\ref{robZZXX}})
\end{eqnarray}
\begin{eqnarray}\label{robZZZXXX}
    &&\hspace{-1cm}||(\widetilde{Z}_{A^i} \widetilde{Z}_{A^j}\widetilde{Z}_{A^k}\widetilde{Z}_{A^l}\widetilde{Z}_{A^m}\widetilde{Z}_{A^n}-Z_{A^i} Z_{A^j} Z_{A^k} X_{A^l} X_{A^n})\ket{\psi}_{A^1A^2A^3}||\nonumber\\&&\approx ||(\widetilde{Z}_{A^i} \widetilde{Z}_{A^j}\widetilde{Z}_{A^k}\widetilde{Z}_{A^l}\widetilde{Z}_{A^m}(\epsilon_{A^n} \openone_d+ X_{A^n})-Z_{A^i} Z_{A^j} Z_{A^k} X_{A^l} X_{A^m} X_{A^n})\ket{\psi}_{A^1A^2A^3}||\nonumber\\
    &&\leq \epsilon_{A^n} ||\widetilde{Z}_{A^i} \widetilde{Z}_{A^j}\widetilde{Z}_{A^k}\widetilde{Z}_{A^l} \widetilde{Z}_{A^m}\ket{\psi}_{A^1A^2A^3}||+||(\widetilde{Z}_{A^i} \widetilde{Z}_{A^j}\widetilde{Z}_{A^k}\widetilde{Z}_{A^l}\widetilde{Z}_{A^m}-Z_{A^i} Z_{A^j} X_{A^k} X_{A^l} X_{A^m}) X_{A^n}\ket{\psi}_{A^1A^2A^3}||\nonumber\\
    &&\leq \epsilon_{A^n} ||\widetilde{Z}_{A^i} \widetilde{Z}_{A^j}\widetilde{Z}_{A^k}\widetilde{Z}_{A^l} \widetilde{Z}_{A^m}\ket{\psi}_{A^1A^2A^3}||+\epsilon_{A^m} ||\widetilde{Z}_{A^i} \widetilde{Z}_{A^j}\widetilde{Z}_{A^k}\widetilde{Z}_{A^l} \ket{\psi}_{A^1A^2A^3}||+\epsilon_{A^l}||\widetilde{Z}_{A^i} \widetilde{Z}_{A^j}\widetilde{Z}_{A^k}\ket{\psi}_{A^1A^2A^3}||\nonumber\\
    &&\ +\alpha_{A^k}||\widetilde{Z}_{A^i} \widetilde{Z}_{A^j}\ket{\psi}_{A^1A^2A^3}||+\alpha_{A^j}||\widetilde{Z}_{A^i}\ket{\psi}_{A^1A^2A^3}||+\alpha_{A^i}\ \  (\text{Using previous patterns})\quad \ \ 
\end{eqnarray}
\subsection{Robust self-testing of state}
\noindent
For perfect and imperfect implementation, the output of the isometry is
\begin{eqnarray}\label{robs}
    &&\hspace{-1cm}\Phi(\ket{\psi}_{A^1A^2A^3}\otimes\ket{000}_{A^{1'}A^{2'}A^{3'}})\nonumber\\
    &&=\frac{1}{8}\sum_{a^1,a^2,a^3\in\{0,1\}} \hspace{-0.5cm}(Z_{A^1})^{a^1} (Z_{A^2})^{a^2} (Z_{A^3})^{a^3} (1+(-1)^{a^1} X_{A^1}) (1+(-1)^{a^2} X_{A^2}) (1+(-1)^{a^3} X_{A^3}) \ket{\psi}_{A^1A^2A^3} \ket{a^1a^2a^3}_{A^{1'}A^{2'}A^{3'}}\\
    &&\hspace{-1cm}\widetilde{\Phi}(\ket{\psi}_{A^1A^2A^3}\otimes\ket{000}_{A^{1'}A^{2'}A^{3'}})\nonumber\\
    &&=\frac{1}{8}\sum_{a^1,a^2,a^3\in\{0,1\}}\hspace{-0.5cm} (\widetilde{Z}_{A^1})^{a^1} (\widetilde{Z}_{A^2})^{a^2} (\widetilde{Z}_{A^3})^{a^3} (1+(-1)^{a^1} \widetilde{X}_{A^1}) (1+(-1)^{a^2} \widetilde{X}_{A^2}) (1+(-1)^{a^3} \widetilde{X}_{A^3}) \ket{\psi}_{A^1A^2A^3} \ket{a^1a^2a^3}_{A^{1'}A^{2'}A^{3'}}
\end{eqnarray}
Hence, the trace distance
\begin{eqnarray}\label{robs1}
    &&||\widetilde{\Phi}(\ket{\psi}_{A^1A^2A^3}\otimes\ket{000}_{A^{1'}A^{2'}A^{3'}})-\Phi(\ket{\psi}_{A^1A^2A^3}\otimes\ket{000}_{A^{1'}A^{2'}A^{3'}})||\nonumber\\
    &&=\frac{1}{8}\sum_{a^1,a^2,a^3\in\{0,1\}} \Bigg|\Bigg|\Big[(\widetilde{Z}_{A^1})^{a^1} (\widetilde{Z}_{A^2})^{a^2} (\widetilde{Z}_{A^3})^{a^3} (1+(-1)^{a^1} \widetilde{X}_{A^1}) (1+(-1)^{a^2} \widetilde{X}_{A^2}) (1+(-1)^{a^3} \widetilde{X}_{A^3})\nonumber\\
    &&-(Z_{A^1})^{a^1} (Z_{A^2})^{a^2} (Z_{A^3})^{a^3} (1+(-1)^{a^1} X_{A^1}) (1+(-1)^{a^2} X_{A^2}) (1+(-1)^{a^3} X_{A^3})\Big] \ket{\psi}_{A^1A^2A^3} \ket{a^1a^2a^3}_{A^{1'}A^{2'}A^{3'}}\Bigg|\Bigg|
\end{eqnarray}
The first term will be
\begin{eqnarray}\label{robs2}
   && \Bigg|\Bigg|\Big[(1+ \widetilde{X}_{A^1}) (1+ \widetilde{X}_{A^2}) (1+ \widetilde{X}_{A^3})
    -(1+ X_{A^1}) (1+ X_{A^2}) (1+ X_{A^3})\Big] \ket{\psi}_{A^1A^2A^3} \ket{000}_{A^{1'}A^{2'}A^{3'}}\Bigg|\Bigg|\nonumber\\
    &&\leq \Big[||(\widetilde{X}_{A^1}-X_{A^1})\ket{\psi}_{A^1A^2A^3}||+||(\widetilde{X}_{A^2}-X_{A^2})\ket{\psi}_{A^1A^2A^3}||+||(\widetilde{X}_{A^3}-X_{A^3})\ket{\psi}_{A^1A^2A^3}||+||(\widetilde{X}_{A^1}\widetilde{X}_{A^2}-X_{A^1} X_{A^2})\ket{\psi}_{A^1A^2A^3}||\nonumber\\
    &&+||(\widetilde{X}_{A^2}\widetilde{X}_{A^3}-X_{A^2} X_{A^3})\ket{\psi}_{A^1A^2A^3}||+||(\widetilde{X}_{A^1}\widetilde{X}_{A^3}-X_{A^1} X_{A^3})\ket{\psi}_{A^1A^2A^3}||+||(\widetilde{X}_{A^1}\widetilde{X}_{A^2}\widetilde{X}_{A^3}-X_{A^1} X_{A^2} X_{A^3})\ket{\psi}_{A^1A^2A^3}||\Big]\nonumber\\
    &&\leq[4\epsilon_{A^1} + 2 \epsilon_{A^2}+2\epsilon_{A^2}||\widetilde{X}_{A^1}\ket{\psi}_{A^1A^2A^3}||+\epsilon_{A^3}(||\widetilde{X}_{A^1}\ket{\psi}_{A^1A^2A^3}||+||\widetilde{X}_{A^2}\ket{\psi}_{A^1A^2A^3}||\nonumber\\
 &&+||\widetilde{X}_{A^1}\widetilde{X}_{A^2}\ket{\psi}_{A^1A^2A^3}||)]\quad(\text{Using Eq.~(\ref{rob}), (\ref{robxx}), (\ref{robxxx})})\nonumber\\
    &&\leq f_1(\epsilon_{A^1},\epsilon_{A^2},\epsilon_{A^3})
\end{eqnarray}
In the same way, the second term will be
\begin{eqnarray}\label{robs3}
   && \Bigg|\Bigg|\Big[\widetilde{Z}_{A^3}(1+ \widetilde{X}_{A^1}) (1+\widetilde{X}_{A^2}) (1-\widetilde{X}_{A^3})
    -Z_{A^3}(1+ X_{A^1}) (1+X_{A^2}) (1-X_{A^3})\Big] \ket{\psi}_{A^1A^2A^3} \ket{001}_{A^{1'}A^{2'}A^{3'}}\Bigg|\Bigg|\nonumber\\
    &&\leq\Big[8\alpha_{A^3}+\epsilon_{A^3}\qty(||\widetilde{Z}_{A^3}\ket{\psi}_{A^1A^2A^3}||+||\widetilde{Z}_{A^3}\widetilde{X}_{A^1}\ket{\psi}_{A^1A^2A^3}||+||\widetilde{Z}_{A^3}\widetilde{X}_{A^2}\ket{\psi}_{A^1A^2A^3}||+||\widetilde{Z}_{A^3}\widetilde{X}_{A^1}\widetilde{X}_{A^2}\ket{\psi}_{A^1A^2A^3}||)+4\epsilon_{A^1} ||\widetilde{Z}_{A^3}\ket{\psi}_{A^1A^2A^3}||\nonumber\\
    &&+2\epsilon_{A^2}\qty( ||\widetilde{Z}_{A^3}\ket{\psi}_{A^1A^2A^3}||+||\widetilde{Z}_{A^3}\widetilde{X}_{A^2}\ket{\psi}_{A^1A^2A^3}||)\Big] \quad (\text{Using Eq.~(\ref{rob})-\ref{robzxxx}})\nonumber\\
    &&\leq f_2(\alpha_{A^3},\epsilon_{A^1},\epsilon_{A^2},\epsilon_{A^3})
\end{eqnarray}
The 3rd term will be
\begin{eqnarray}\label{robs4}
   && \Bigg|\Bigg|\Big[\widetilde{Z}_{A^2}(1+ \widetilde{X}_{A^1}) (1-\widetilde{X}_{A^2}) (1+ \widetilde{X}_{A^3})
    -Z_{A^2}(1+ X_{A^1}) (1-X_{A^2}) (1+X_{A^3})\Big] \ket{\psi}_{A^1A^2A^3}\Bigg|\Bigg|\nonumber\\
    &&\leq\Big[8\alpha_{A^2}+\epsilon_{A^2}\qty(||\widetilde{Z}_{A^2}\ket{\psi}_{A^1A^2A^3}||+||\widetilde{Z}_{A^2}\widetilde{X}_{A^1}\ket{\psi}_{A^1A^2A^3}||+||\widetilde{Z}_{A^2}\widetilde{X}_{A^3}\ket{\psi}_{A^1A^2A^3}||+||\widetilde{Z}_{A^2}\widetilde{X}_{A^1}\widetilde{X}_{A^3}\ket{\psi}_{A^1A^2A^3}||)+4\epsilon_{A^1} ||\widetilde{Z}_{A^2}\ket{\psi}_{A^1A^2A^3}||\nonumber\\
    && 2\epsilon_{A^3}\qty(||\widetilde{Z}_{A^2}\ket{\psi}_{A^1A^2A^3}||+||\widetilde{Z}_{A^2}\widetilde{X}_{A^1}\ket{\psi}_{A^1A^2A^3}||)\Big] \quad (\text{Using Eq.~(\ref{rob})-\ref{robzxxx}})\nonumber\\
    &&\leq f_3(\alpha_{A^2},\epsilon_{A^1},\epsilon_{A^2},\epsilon_{A^3})
\end{eqnarray}
The 4th term
\begin{eqnarray}\label{robs5}
   && \Bigg|\Bigg|\Big[\widetilde{Z}_{A^2}\widetilde{Z}_{A^3}(1+ \widetilde{X}_{A^1}) (1-\widetilde{X}_{A^2}) (1-\widetilde{X}_{A^3})
    -Z_{A^2} Z_{A^3}(1+ X_{A^1}) (1-X_{A^2}) (1-X_{A^3})\Big] \ket{\psi}_{A^1A^2A^3} \ket{011}_{A^{1'}A^{2'}A^{3'}}\Bigg|\Bigg|\nonumber\\
    &&\leq\Big[8\alpha_{A^2}+\epsilon_{A^3}\qty(||\widetilde{Z}_{A^2}\widetilde{Z}_{A^3}\ket{\psi}_{A^1A^2A^3}||+||\widetilde{Z}_{A^2}\widetilde{Z}_{A^3}\widetilde{X}_{A^2}\ket{\psi}_{A^1A^2A^3}||+||\widetilde{Z}_{A^2} \widetilde{Z}_{A^3}\widetilde{X}_{A^1}\ket{\psi}_{A^1A^2A^3}||+||\widetilde{Z}_{A^2}\widetilde{Z}_{A^3}\widetilde{X}_{A^1}\widetilde{X}_{A^2}\ket{\psi}_{A^1A^2A^3}||)\nonumber\\
    && +8\alpha_{A^3}||\widetilde{Z}_{A^2}\ket{\psi}_{A^1A^2A^3}||+4\epsilon_{A^1} ||\widetilde{Z}_{A^2} \widetilde{Z}_{A^3}\ket{\psi}_{A^1A^2A^3}||+2\epsilon_{A^2}\qty(||\widetilde{Z}_{A^2} \widetilde{Z}_{A^3}\ket{\psi}_{A^1A^2A^3}||+||\widetilde{Z}_{A^2} \widetilde{Z}_{A^3}\widetilde{X}_{A^1}\ket{\psi}_{A^1A^2A^3}||)\Big](\text{Using Eq.~(\ref{rob})-\ref{robZZXXX}})\nonumber\\
    &&\leq f_4(\alpha_{A^2},\alpha_{A^3},\epsilon_{A^1},\epsilon_{A^2},\epsilon_{A^3}) 
\end{eqnarray}
The 5th term will be
\begin{eqnarray}\label{robs6}
   && \Bigg|\Bigg|\Big[\widetilde{Z}_{A^1}(1- \widetilde{X}_{A^1}) (1+\widetilde{X}_{A^2}) (1+ \widetilde{X}_{A^3})
    -Z_{A^1}(1-X_{A^1}) (1+X_{A^2}) (1+X_{A^3})\Big] \ket{\psi}_{A^1A^2A^3} \ket{100}_{A^{1'}A^{2'}A^{3'}}\Bigg|\Bigg|\nonumber\\
    &&\leq\Big[8\alpha_{A^1}+\epsilon_{A^3}\qty(||\widetilde{Z}_{A^1}\ket{\psi}_{A^1A^2A^3}||+||\widetilde{Z}_{A^1}\widetilde{X}_{A^1}\ket{\psi}_{A^1A^2A^3}||+||\widetilde{Z}_{A^1}\widetilde{X}_{A^2}\ket{\psi}_{A^1A^2A^3}||+||\widetilde{Z}_{A^1}\widetilde{X}_{A^1}\widetilde{X}_{A^2}\ket{\psi}_{A^1A^2A^3}||)4\epsilon_{A^1} ||\widetilde{Z}_{A^1}\ket{\psi}_{A^1A^2A^3}||\nonumber\\
    && + 2\epsilon_{A^2}\qty(||\widetilde{Z}_{A^1}\ket{\psi}_{A^1A^2A^3}||+||\widetilde{Z}_{A^1}\widetilde{X}_{A^1}\ket{\psi}_{A^1A^2A^3}||)\Big] (\text{Using Eq.~(\ref{rob})-\ref{robzxxx}})\nonumber\\
    &&\leq f_5(\alpha_{A^1},\epsilon_{A^1},\epsilon_{A^2},\epsilon_{A^3}) 
\end{eqnarray}
The 6th term
\begin{eqnarray}\label{robs7}
   && \Bigg|\Bigg|\Big[\widetilde{Z}_{A^1}\widetilde{Z}_{A^3}(1- \widetilde{X}_{A^1}) (1+\widetilde{X}_{A^2}) (1-\widetilde{X}_{A^3})
    -Z_{A^1} Z_{A^3}(1-X_{A^1}) (1+X_{A^2}) (1-X_{A^3})\Big] \ket{\psi}_{A^1A^2A^3} \ket{101}_{A^{1'}A^{2'}A^{3'}}\Bigg|\Bigg|\nonumber\\
    &&\leq\Big[8\alpha_{A^1}+\epsilon_{A^3}\qty(||\widetilde{Z}_{A^1}\widetilde{Z}_{A^3}\ket{\psi}_{A^1A^2A^3}||+||\widetilde{Z}_{A^1}\widetilde{Z}_{A^3}\widetilde{X}_{A^2}\ket{\psi}_{A^1A^2A^3}||+||\widetilde{Z}_{A^1} \widetilde{Z}_{A^3}\widetilde{X}_{A^1}\ket{\psi}_{A^1A^2A^3}||+||\widetilde{Z}_{A^1}\widetilde{Z}_{A^3}\widetilde{X}_{A^1}\widetilde{X}_{A^2}\ket{\psi}_{A^1A^2A^3}||)\nonumber\\
    && +4\epsilon_{A^1}||\widetilde{Z}_{A^1}\widetilde{Z}_{A^3}\ket{\psi}_{A^1A^2A^3}||+8\alpha_{A^3}||\widetilde{Z}_{A^1}\ket{\psi}_{A^1A^2A^3}||+2\epsilon_{A^2}\qty(||\widetilde{Z}_{A^1}\widetilde{Z}_{A^3}\ket{\psi}_{A^1A^2A^3}||+||\widetilde{Z}_{A^1}\widetilde{Z}_{A^3}\widetilde{X}_{A^1}\ket{\psi}_{A^1A^2A^3}||)\Big](\text{Using Eq.~(\ref{rob})-\ref{robZZXXX}})\nonumber\\
    &&\leq f_6(\alpha_{A^1},\alpha_{A^3},\epsilon_{A^1},\epsilon_{A^2},\epsilon_{A^3}) 
\end{eqnarray}
The 7th term
\begin{eqnarray}\label{robs8}
   && \Bigg|\Bigg|\Big[\widetilde{Z}_{A^1}\widetilde{Z}_{A^2}(1- \widetilde{X}_{A^1}) (1-\widetilde{X}_{A^2}) (1+\widetilde{X}_{A^3})
    -Z_{A^1} Z_{A^2}(1-X_{A^1}) (1-X_{A^2}) (1+X_{A^3})\Big] \ket{\psi}_{A^1A^2A^3} \ket{110}_{A^{1'}A^{2'}A^{3'}}\Bigg|\Bigg|\nonumber\\
    &&\leq8\alpha_{A^1}+\epsilon_{A^2}\qty(||\widetilde{Z}_{A^1}\widetilde{Z}_{A^2}\ket{\psi}_{A^1A^2A^3}||+||\widetilde{Z}_{A^1}\widetilde{Z}_{A^2}\widetilde{X}_{A^1}\ket{\psi}_{A^1A^2A^3}||+||\widetilde{Z}_{A^1} \widetilde{Z}_{A^2}\widetilde{X}_{A^3}\ket{\psi}_{A^1A^2A^3}||+||\widetilde{Z}_{A^1}\widetilde{Z}_{A^2}\widetilde{X}_{A^1}\widetilde{X}_{A^3}\ket{\psi}_{A^1A^2A^3}||)\nonumber\\
    &&+8\alpha_{A^2}||\widetilde{Z}_{A^1}\ket{\psi}_{A^1A^2A^3}||+4\epsilon_{A^1}||\widetilde{Z}_{A^1}\widetilde{Z}_{A^2}\ket{\psi}_{A^1A^2A^3}||+2\epsilon_{A^3}\qty(||\widetilde{Z}_{A^1}\widetilde{Z}_{A^2}\ket{\psi}_{A^1A^2A^3}||+||\widetilde{Z}_{A^1}\widetilde{Z}_{A^2}\widetilde{X}_{A^1}\ket{\psi}_{A^1A^2A^3}||)\Big](\text{Using Eq.~(\ref{rob})-\ref{robZZXXX}})\nonumber\\
    &&\leq f_7(\alpha_{A^1},\alpha_{A^2},\epsilon_{A^1},\epsilon_{A^2},\epsilon_{A^3}) 
\end{eqnarray}
The 8th term
\begin{eqnarray}\label{robs9}
   && \Bigg|\Bigg|\Big[\widetilde{Z}_{A^1}\widetilde{Z}_{A^2}\widetilde{Z}_{A^3}(1- \widetilde{X}_{A^1}) (1-\widetilde{X}_{A^2}) (1-\widetilde{X}_{A^3})
    -Z_{A^1} Z_{A^2} Z_{A^3} (1-X_{A^1}) (1-X_{A^2}) (1-X_{A^3})\Big] \ket{\psi}_{A^1A^2A^3} \ket{111}_{A^{1'}A^{2'}A^{3'}}\Bigg|\Bigg|\nonumber\\
    &&\leq\Big[8\alpha_{A^1}+8\alpha_{A^2}||\widetilde{Z}_{A^1}\ket{\psi}_{A^1A^2A^3}||+8\alpha_{A^3}||\widetilde{Z}_{A^1}\widetilde{Z}_{A^2} \ket{\psi}_{A^1A^2A^3}||+4\epsilon_{A^1}||\widetilde{Z}_{A^1}\widetilde{Z}_{A^2} \widetilde{Z}_{A^3}\ket{\psi}_{A^1A^2A^3}||+2\epsilon_{A^2}\Big(||\widetilde{Z}_{A^1}\widetilde{Z}_{A^2} \widetilde{Z}_{A^3}\ket{\psi}_{A^1A^2A^3}||\nonumber\\
    &&+||\widetilde{Z}_{A^1}\widetilde{Z}_{A^2} \widetilde{Z}_{A^3}\widetilde{X}_{A^1}\ket{\psi}_{A^1A^2A^3}||\Big)+\epsilon_{A^3}\Big(||\widetilde{Z}_{A^1}\widetilde{Z}_{A^2} \widetilde{Z}_{A^3}\ket{\psi}_{A^1A^2A^3}||+||\widetilde{Z}_{A^1}\widetilde{Z}_{A^2}\widetilde{Z}_{A^3}\widetilde{X}_{A^2}\ket{\psi}_{A^1A^2A^3}||+||\widetilde{Z}_{A^1} \widetilde{Z}_{A^2}\widetilde{Z}_{A^3}\widetilde{X}_{A^1}\ket{\psi}_{A^1A^2A^3}||\nonumber\\
    &&+||\widetilde{Z}_{A^1}\widetilde{Z}_{A^2}\widetilde{Z}_{A^3}\widetilde{X}_{A^1}\widetilde{X}_{A^2}\ket{\psi}_{A^1A^2A^3}||\Big)\Big] \  (\text{Using Eq.~(\ref{rob})-\ref{robZZZXXX}})\nonumber\\
    &&\leq f_8(\alpha_{A^1},\alpha_{A^2},\alpha_{A^3},\epsilon_{A^1},\epsilon_{A^2},\epsilon_{A^3}) 
\end{eqnarray}
Now taking only the modulus value from Eq.~(\ref{robs2})-(\ref{robs9}) and putting it in Eq.~(\ref{robs1}) we get

\begin{eqnarray}\label{robsF}
    &&\hspace{-1cm}||\widetilde{\Phi}(\ket{\psi}_{A^1A^2A^3}\otimes\ket{000}_{A^{1'}A^{2'}A^{3'}})-\Phi(\ket{\psi}_{A^1A^2A^3}\otimes\ket{000}_{A^{1'}A^{2'}A^{3'}})||\nonumber\\
    &&\leq\frac{1}{8}\Big(f_1(\epsilon_{A^1},\epsilon_{A^2},\epsilon_{A^3})+f_2(\alpha_{A^3},\epsilon_{A^1},\epsilon_{A^2},\epsilon_{A^3})+f_3(\alpha_{A^2},\epsilon_{A^1},\epsilon_{A^2},\epsilon_{A^3})\nonumber\\
    &&+f_4(\alpha_{A^2},\alpha_{A^3},\epsilon_{A^1},\epsilon_{A^2},\epsilon_{A^3})+f_5(\alpha_{A^1},\epsilon_{A^1},\epsilon_{A^2},\epsilon_{A^3})\nonumber\\
    &&+f_6(\alpha_{A^1},\alpha_{A^3},\epsilon_{A^1},\epsilon_{A^2},\epsilon_{A^3})+f_7(\alpha_{A^1},\alpha_{A^2},\epsilon_{A^1},\epsilon_{A^2},\epsilon_{A^3})\nonumber\\
    &&+f_8(\alpha_{A^1},\alpha_{A^2},\alpha_{A^3},\epsilon_{A^1},\epsilon_{A^2},\epsilon_{A^3})\Big)\nonumber\\
    &&\leq F_S(\alpha_{A^1},\alpha_{A^2},\alpha_{A^3},\epsilon_{A^1},\epsilon_{A^2},\epsilon_{A^3})
\end{eqnarray}
Note that
\begin{eqnarray}
    &&\lim_{\{\alpha_{A^1},\alpha_{A^2},\alpha_{A^3},\epsilon_{A^1},\epsilon_{A^2},\epsilon_{A^3}\}\to 0} F_S(\alpha_{A^1},\alpha_{A^2},\alpha_{A^3},\epsilon_{A^1},\epsilon_{A^2},\epsilon_{A^3})=0\nonumber\\
    &&\Rightarrow ||\widetilde{\Phi}(\ket{\psi}_{A^1A^2A^3}\otimes\ket{000}_{A^{1'}A^{2'}A^{3'}})||\approx ||\Phi(\ket{\psi}_{A^1A^2A^3}\otimes\ket{000}_{A^{1'}A^{2'}A^{3'}})||\quad \quad
\end{eqnarray}
\subsection{Robust self-testing of observables}
The procedure will be the same as in the previous section; only here will we calculate the robustness with observables, i.e., $X_{A^k}, Z_{A^k}$ ($\forall k\in[3]$). Hence, the robustness of $X_{A^k}$ is represented as
\begin{eqnarray}\label{robob1}
    &&||\widetilde{\Phi}(\widetilde{X}_{A^k}\ket{\psi}_{A^1A^2A^3}\otimes\ket{000}_{A^{1'}A^{2'}A^{3'}})-\Phi(X_{A^k}\ket{\psi}_{A^1A^2A^3}\otimes\ket{000}_{A^{1'}A^{2'}A^{3'}})||\nonumber\\
    &&=\frac{1}{8}\sum_{a^1,a^2,a^3\in\{0,1\}} \Bigg|\Bigg|\Big[(\widetilde{Z}_{A^1})^{a^1} (\widetilde{Z}_{A^2})^{a^2} (\widetilde{Z}_{A^3})^{a^3} (1+(-1)^{a^1} \widetilde{X}_{A^1}) (1+(-1)^{a^2} \widetilde{X}_{A^2}) (1+(-1)^{a^3} \widetilde{X}_{A^3})\widetilde{X}_{A^k}\ket{\psi}_{A^1A^2A^3}\nonumber\\
    &&-(Z_{A^1})^{a^1} (Z_{A^2})^{a^2} (Z_{A^3})^{a^3} (1+(-1)^{a^1} X_{A^1}) (1+(-1)^{a^2} X_{A^2}) (1+(-1)^{a^3} X_{A^3})X_{A^k}\ket{\psi}_{A^1A^2A^3}\Big]  \ket{a^1a^2a^3}_{A^{1'}A^{2'}A^{3'}}\Bigg|\Bigg|\nonumber\\
    &&\approx\frac{1}{8}\sum_{a^1,a^2,a^3\in\{0,1\}} \Bigg|\Bigg|\Big[(\widetilde{Z}_{A^1})^{a^1} (\widetilde{Z}_{A^2})^{a^2} (\widetilde{Z}_{A^3})^{a^3} (1+(-1)^{a^1} \widetilde{X}_{A^1}) (1+(-1)^{a^2} \widetilde{X}_{A^2}) (1+(-1)^{a^3} \widetilde{X}_{A^3})(\epsilon_{A^k} \openone_d+ X_{A^k})\ket{\psi}_{A^1A^2A^3}\nonumber\\
    &&-(Z_{A^1})^{a^1} (Z_{A^2})^{a^2} (Z_{A^3})^{a^3} (1+(-1)^{a^1} X_{A^1}) (1+(-1)^{a^2} X_{A^2}) (1+(-1)^{a^3} X_{A^3})X_{A^k}\ket{\psi}_{A^1A^2A^3}\Big]  \ket{a^1a^2a^3}_{A^{1'}A^{2'}A^{3'}}\Bigg|\Bigg|\nonumber\\
     &&\leq\frac{\epsilon_{A^k}}{8}\sum_{a^1,a^2,a^3\in\{0,1\}} \Bigg|\Bigg|(\widetilde{Z}_{A^1})^{a^1} (\widetilde{Z}_{A^2})^{a^2} (\widetilde{Z}_{A^3})^{a^3} (1+(-1)^{a^1} \widetilde{X}_{A^1}) (1+(-1)^{a^2} \widetilde{X}_{A^2}) (1+(-1)^{a^3} \widetilde{X}_{A^3})\ket{\psi}_{A^1A^2A^3}\ket{a^1a^2a^3}_{A^{1'}A^{2'}A^{3'}}\Bigg|\Bigg|\nonumber\\
     &&+ \frac{1}{8}\sum_{a^1,a^2,a^3\in\{0,1\}} \Bigg|\Bigg|\Big[(\widetilde{Z}_{A^1})^{a^1} (\widetilde{Z}_{A^2})^{a^2} (\widetilde{Z}_{A^3})^{a^3} (1+(-1)^{a^1} \widetilde{X}_{A^1}) (1+(-1)^{a^2} \widetilde{X}_{A^2}) (1+(-1)^{a^3} \widetilde{X}_{A^3})\nonumber\\
    &&-(Z_{A^1})^{a^1} (Z_{A^2})^{a^2} (Z_{A^3})^{a^3} (1+(-1)^{a^1} X_{A^1}) (1+(-1)^{a^2} X_{A^2}) (1+(-1)^{a^3} X_{A^3})\Big]X_{A^k} \ket{\psi}_{A^1A^2A^3} \ket{a^1a^2a^3}_{A^{1'}A^{2'}A^{3'}}\Bigg|\Bigg|\nonumber\\
     &&\leq\frac{\epsilon_{A^k}}{8}\sum_{a^1,a^2,a^3\in\{0,1\}} \Bigg|\Bigg|(\widetilde{Z}_{A^1})^{a^1} (\widetilde{Z}_{A^2})^{a^2} (\widetilde{Z}_{A^3})^{a^3} (1+(-1)^{a^1} \widetilde{X}_{A^1}) (1+(-1)^{a^2} \widetilde{X}_{A^2}) (1+(-1)^{a^3} \widetilde{X}_{A^3})\ket{\psi}_{A^1A^2A^3}\ket{a^1a^2a^3}_{A^{1'}A^{2'}A^{3'}}\Bigg|\Bigg|\nonumber\\
     &&+ \frac{1}{8}\sum_{a^1,a^2,a^3\in\{0,1\}} \Bigg|\Bigg|\Big[(\widetilde{Z}_{A^1})^{a^1} (\widetilde{Z}_{A^2})^{a^2} (\widetilde{Z}_{A^3})^{a^3} (1+(-1)^{a^1} \widetilde{X}_{A^1}) (1+(-1)^{a^2} \widetilde{X}_{A^2}) (1+(-1)^{a^3} \widetilde{X}_{A^3})\nonumber\\
    &&-(Z_{A^1})^{a^1} (Z_{A^2})^{a^2} (Z_{A^3})^{a^3} (1+(-1)^{a^1} X_{A^1}) (1+(-1)^{a^2} X_{A^2}) (1+(-1)^{a^3} X_{A^3})\Big] \ket{\psi}_{A^1A^2A^3} \ket{a^1a^2a^3}_{A^{1'}A^{2'}A^{3'}}\Bigg|\Bigg|\nonumber\\
    &&\leq\frac{\epsilon_{A^k}}{8}\sum_{a^1,a^2,a^3\in\{0,1\}} \Bigg|\Bigg|(\widetilde{Z}_{A^1})^{a^1} (\widetilde{Z}_{A^2})^{a^2} (\widetilde{Z}_{A^3})^{a^3} (1+(-1)^{a^1} \widetilde{X}_{A^1}) (1+(-1)^{a^2} \widetilde{X}_{A^2}) (1+(-1)^{a^3} \widetilde{X}_{A^3})\ket{\psi}_{A^1A^2A^3}\ket{a^1a^2a^3}_{A^{1'}A^{2'}A^{3'}}\Bigg|\Bigg|\nonumber\\
    &&+\Big|\Big|\widetilde{\Phi}(\ket{\psi}_{A^1A^2A^3}\otimes\ket{000}_{A^{1'}A^{2'}A^{3'}})-\Phi(\ket{\psi}_{A^1A^2A^3}\otimes\ket{000}_{A^{1'}A^{2'}A^{3'}})\Big|\Big| \quad (\text{From Eq.~(\ref{robs1})})\nonumber\\
    &&\leq F(\epsilon_{A^k}) + F_S(\alpha_{A^1},\alpha_{A^2},\alpha_{A^3},\epsilon_{A^1},\epsilon_{A^2},\epsilon_{A^3})\nonumber\\
    &&\leq F_{O_X}(\alpha_{A^1},\alpha_{A^2},\alpha_{A^3},\epsilon_{A^1},\epsilon_{A^2},\epsilon_{A^3}) 
\end{eqnarray}
Note that
\begin{eqnarray}
    &&\lim_{\{\alpha_{A^1},\alpha_{A^2},\alpha_{A^3},\epsilon_{A^1},\epsilon_{A^2},\epsilon_{A^3}\}\to 0}\hspace{-1cm} F_{O_X}(\alpha_{A^1},\alpha_{A^2},\alpha_{A^3},\epsilon_{A^1},\epsilon_{A^2},\epsilon_{A^3})=0\nonumber\\
    &&\Rightarrow ||\widetilde{\Phi}(\widetilde{X}_{A^k}\ket{\psi}_{A^1A^2A^3}\otimes\ket{000}_{A^{1'}A^{2'}A^{3'}})||\approx ||\Phi(X_{A^k}\ket{\psi}_{A^1A^2A^3}\otimes\ket{000}_{A^{1'}A^{2'}A^{3'}})||\quad\quad\quad
\end{eqnarray}
 In the same way, the robust self-testing of $Z_{A^k}$
 \begin{eqnarray}\label{robob2}
    &&||\widetilde{\Phi}(\widetilde{Z}_{A^k}\ket{\psi}_{A^1A^2A^3}\otimes\ket{000}_{A^{1'}A^{2'}A^{3'}})-\Phi(Z_{A^k}\ket{\psi}_{A^1A^2A^3}\otimes\ket{000}_{A^{1'}A^{2'}A^{3'}})||\nonumber\\
    &&=\frac{1}{8}\sum_{a^1,a^2,a^3\in\{0,1\}} \Bigg|\Bigg|\Big[(\widetilde{Z}_{A^1})^{a^1} (\widetilde{Z}_{A^2})^{a^2} (\widetilde{Z}_{A^3})^{a^3} (1+(-1)^{a^1} \widetilde{X}_{A^1}) (1+(-1)^{a^2} \widetilde{X}_{A^2}) (1+(-1)^{a^3} \widetilde{X}_{A^3})\widetilde{Z}_{A^k}\ket{\psi}_{A^1A^2A^3}\nonumber\\
    &&-(Z_{A^1})^{a^1} (Z_{A^2})^{a^2} (Z_{A^3})^{a^3} (1+(-1)^{a^1} X_{A^1}) (1+(-1)^{a^2} X_{A^2}) (1+(-1)^{a^3} X_{A^3})Z_{A^k}\ket{\psi}_{A^1A^2A^3}\Big]  \ket{a^1a^2a^3}_{A^{1'}A^{2'}A^{3'}}\Bigg|\Bigg|\nonumber\\
    &&\approx\frac{1}{8}\sum_{a^1,a^2,a^3\in\{0,1\}} \Bigg|\Bigg|\Big[(\widetilde{Z}_{A^1})^{a^1} (\widetilde{Z}_{A^2})^{a^2} (\widetilde{Z}_{A^3})^{a^3} (1+(-1)^{a^1} \widetilde{X}_{A^1}) (1+(-1)^{a^2} \widetilde{X}_{A^2}) (1+(-1)^{a^3} \widetilde{X}_{A^3})(\alpha_{A^k} \openone_d+ Z_{A^k})\ket{\psi}_{A^1A^2A^3}\nonumber\\
    &&-(Z_{A^1})^{a^1} (Z_{A^2})^{a^2} (Z_{A^3})^{a^3} (1+(-1)^{a^1} X_{A^1}) (1+(-1)^{a^2} X_{A^2}) (1+(-1)^{a^3} X_{A^3})Z_{A^k}\ket{\psi}_{A^1A^2A^3}\Big]  \ket{a^1a^2a^3}_{A^{1'}A^{2'}A^{3'}}\Bigg|\Bigg|\nonumber\\
     &&\leq\frac{\alpha_{A^k}}{8}\sum_{a^1,a^2,a^3\in\{0,1\}} \Bigg|\Bigg|(\widetilde{Z}_{A^1})^{a^1} (\widetilde{Z}_{A^2})^{a^2} (\widetilde{Z}_{A^3})^{a^3} (1+(-1)^{a^1} \widetilde{X}_{A^1}) (1+(-1)^{a^2} \widetilde{X}_{A^2}) (1+(-1)^{a^3} \widetilde{X}_{A^3})\ket{\psi}_{A^1A^2A^3}\ket{a^1a^2a^3}_{A^{1'}A^{2'}A^{3'}}\Bigg|\Bigg|\nonumber\\
     &&+ \frac{1}{8}\sum_{a^1,a^2,a^3\in\{0,1\}} \Bigg|\Bigg|\Big[(\widetilde{Z}_{A^1})^{a^1} (\widetilde{Z}_{A^2})^{a^2} (\widetilde{Z}_{A^3})^{a^3} (1+(-1)^{a^1} \widetilde{X}_{A^1}) (1+(-1)^{a^2} \widetilde{X}_{A^2}) (1+(-1)^{a^3} \widetilde{X}_{A^3})\nonumber\\
    &&-(Z_{A^1})^{a^1} (Z_{A^2})^{a^2} (Z_{A^3})^{a^3} (1+(-1)^{a^1} X_{A^1}) (1+(-1)^{a^2} X_{A^2}) (1+(-1)^{a^3} X_{A^3})\Big]Z_{A^k} \ket{\psi}_{A^1A^2A^3} \ket{a^1a^2a^3}_{A^{1'}A^{2'}A^{3'}}\Bigg|\Bigg|\nonumber\\
     &&\leq\frac{\alpha_{A^k}}{8}\sum_{a^1,a^2,a^3\in\{0,1\}} \Bigg|\Bigg|(\widetilde{Z}_{A^1})^{a^1} (\widetilde{Z}_{A^2})^{a^2} (\widetilde{Z}_{A^3})^{a^3} (1+(-1)^{a^1} \widetilde{X}_{A^1}) (1+(-1)^{a^2} \widetilde{X}_{A^2}) (1+(-1)^{a^3} \widetilde{X}_{A^3})\ket{\psi}_{A^1A^2A^3}\ket{a^1a^2a^3}_{A^{1'}A^{2'}A^{3'}}\Bigg|\Bigg|\nonumber\\
     &&+ \frac{1}{8}\sum_{a^1,a^2,a^3\in\{0,1\}} \Bigg|\Bigg|\Big[(\widetilde{Z}_{A^1})^{a^1} (\widetilde{Z}_{A^2})^{a^2} (\widetilde{Z}_{A^3})^{a^3} (1+(-1)^{a^1} \widetilde{X}_{A^1}) (1+(-1)^{a^2} \widetilde{X}_{A^2}) (1+(-1)^{a^3} \widetilde{X}_{A^3})\nonumber\\
    &&-(Z_{A^1})^{a^1} (Z_{A^2})^{a^2} (Z_{A^3})^{a^3} (1+(-1)^{a^1} X_{A^1}) (1+(-1)^{a^2} X_{A^2}) (1+(-1)^{a^3} X_{A^3})\Big] \ket{\psi}_{A^1A^2A^3} \ket{a^1a^2a^3}_{A^{1'}A^{2'}A^{3'}}\Bigg|\Bigg|\nonumber\\
    &&\leq\frac{\alpha_{A^k}}{8}\sum_{a^1,a^2,a^3\in\{0,1\}} \Bigg|\Bigg|(\widetilde{Z}_{A^1})^{a^1} (\widetilde{Z}_{A^2})^{a^2} (\widetilde{Z}_{A^3})^{a^3} (1+(-1)^{a^1} \widetilde{X}_{A^1}) (1+(-1)^{a^2} \widetilde{X}_{A^2}) (1+(-1)^{a^3} \widetilde{X}_{A^3})\ket{\psi}_{A^1A^2A^3}\ket{a^1a^2a^3}_{A^{1'}A^{2'}A^{3'}}\Bigg|\Bigg|\nonumber\\
     &&+ \frac{1}{8}\sum_{a^1,a^2,a^3\in\{0,1\}} \Bigg|\Bigg|\Big[(\widetilde{Z}_{A^1})^{a^1} (\widetilde{Z}_{A^2})^{a^2} (\widetilde{Z}_{A^3})^{a^3} (1+(-1)^{a^1} \widetilde{X}_{A^1}) (1+(-1)^{a^2} \widetilde{X}_{A^2}) (1+(-1)^{a^3} \widetilde{X}_{A^3})\nonumber\\
    &&-(Z_{A^1})^{a^1} (Z_{A^2})^{a^2} (Z_{A^3})^{a^3} (1+(-1)^{a^1} X_{A^1}) (1+(-1)^{a^2} X_{A^2}) (1+(-1)^{a^3} X_{A^3})\Big] \ket{\psi}_{A^1A^2A^3} \ket{a^1a^2a^3}_{A^{1'}A^{2'}A^{3'}}\Bigg|\Bigg|\nonumber\\
    &&\leq\frac{\alpha_{A^k}}{8}\sum_{a^1,a^2,a^3\in\{0,1\}} \Bigg|\Bigg|(\widetilde{Z}_{A^1})^{a^1} (\widetilde{Z}_{A^2})^{a^2} (\widetilde{Z}_{A^3})^{a^3} (1+(-1)^{a^1} \widetilde{X}_{A^1}) (1+(-1)^{a^2} \widetilde{X}_{A^2}) (1+(-1)^{a^3} \widetilde{X}_{A^3})\ket{\psi}_{A^1A^2A^3}\ket{a^1a^2a^3}_{A^{1'}A^{2'}A^{3'}}\Bigg|\Bigg|\nonumber\\
    &&+\Big|\Big|\widetilde{\Phi}(\ket{\psi}_{A^1A^2A^3}\otimes\ket{000}_{A^{1'}A^{2'}A^{3'}})-\Phi(\ket{\psi}_{A^1A^2A^3}\otimes\ket{000}_{A^{1'}A^{2'}A^{3'}})\Big|\Big| \quad (\text{From Eq.~(\ref{robs1})})\nonumber\\
    &&\leq F(\alpha_{A^k}) + F_S(\alpha_{A^1},\alpha_{A^2},\alpha_{A^3},\epsilon_{A^1},\epsilon_{A^2},\epsilon_{A^3})\nonumber\\
    &&\leq F_{O_Z}(\alpha_{A^1},\alpha_{A^2},\alpha_{A^3},\epsilon_{A^1},\epsilon_{A^2},\epsilon_{A^3}) 
\end{eqnarray}
Note that
{\small\begin{eqnarray}
    \lim_{\{\alpha_{A^1},\alpha_{A^2},\alpha_{A^3},\epsilon_{A^1},\epsilon_{A^2},\epsilon_{A^3}\}\to 0} \hspace{-1cm}F_{O_Z}(\alpha_{A^1},\alpha_{A^2},\alpha_{A^3},\epsilon_{A^1},\epsilon_{A^2},\epsilon_{A^3})=0\Rightarrow ||\widetilde{\Phi}(\widetilde{Z}_{A^k}\ket{\psi}_{A^1A^2A^3}\otimes\ket{000}_{A^{1'}A^{2'}A^{3'}})||\approx ||\Phi(Z_{A^k}\ket{\psi}_{A^1A^2A^3}\otimes\ket{000}_{A^{1'}A^{2'}A^{3'}})||\quad\quad\quad
\end{eqnarray}}
\subsection{Special case:Only the \texorpdfstring{$3$}{3}rd party implements imperfect observables}
Now, if we consider only the third number of parties implement the imperfect observables ($\mathcal{A}^3_{i_3}$), then $ ||(\widetilde{X}_{A^3}-X_{A^3})\ket{\psi}_{A^1A^2A^3}||\leq \delta,||(\widetilde{Z}_{A^3}-Z_{A^3})\ket{\psi}_{A^1A^2A^3}||\leq \delta$, where we consider that the error is the same, i.e., $\alpha_{A^3}=\epsilon_{A^3}=\delta\geq 0$. Again, the error of each observable of the first party is $||(\mathcal{A}^3_{i_3}-A^3_{i_3})\ket{\psi_{A^1A^2A^3}}||\leq \beta\implies \mathcal{A}^3_{i_3}\approx A^3_{i_3}+\beta \ \openone_d, \forall i_3\in[n], \ \beta \geq 0$, which implies $||(\widetilde{X}_{A^3}-X_{A^3})\ket{\psi}_{A^1A^2A^3}||=||(\mathcal{A}^3_1-A^3_1)\ket{\psi}_{A^1A^2A^3}||\leq \beta \Rightarrow \beta\approx \delta$. In the presence of noise, $L^3_{i_1i_2i_3}$ becomes $\widetilde{L}^3_{i_1i_2i_3}$
, and from the main text we know that in the presence of noise,
\begin{eqnarray}\label{noisygaman}
    \Tr[\widetilde{\gamma}_n \ \rho_{A^1A^2A^3}]&=& \frac{1}{2}\sum_{i_3=1}^{n}\sum_{i_1=1}^{n} w_{i_1i_2i_3} \Tr[\widetilde{L}^{3\dag}_{i_1i_2 i_3} \widetilde{L}^3_{i_1i_2i_3} \ \rho_{A^1A^2A^3}]=\frac{1}{2}\sum_{i_3=1}^{n}\sum_{i_1=1}^{n}w_{i_1i_2i_3}  \ \xi_{i_1i_2i_3}^2 =\xi
\end{eqnarray}
Here, $\widetilde{L}^3_{i_1i_2i_3}$ is defined as 
\begin{eqnarray}\label{SOS}
    \widetilde{L}^3_{i_1i_2i_3} &=& \openone_d \otimes A^2_{(i_1+i_3-2)\oplus_n +1} \otimes \mathcal{A}^3_{i_3} - \widetilde{A}^1_{i_1} \otimes \openone_d \otimes \openone_d \nonumber\\
    &=& \openone_d \otimes A^2_{(i_1+i_3-2)\oplus_n +1} \otimes (A^3_{i_3}+\beta) - \widetilde{A}^1_{i_1} \otimes \openone_d \otimes \openone_d\nonumber\\
    &=& L^3_{i_1i_2i_3}+\openone_d\otimes \beta A^2_{(i_1+i_3-2)\oplus_n +1}  \otimes \openone_d\quad (\text{ $L^3_{i_1i_2i_3} = \openone_d\otimes A^2_{(i_1+i_3-2)\oplus_n +1} \otimes  A^3_{i_3} - \widetilde{A}^1_{i_1} \otimes \openone_d \otimes \openone_d$})\quad\quad
\end{eqnarray}
Hence, 
\begin{eqnarray}\label{noisyLij}
    \Tr[\widetilde{L}^{3\dag}_{i_1i_2i_3}\widetilde{L}^3_{i_1i_2i_3} \ \rho_{A^1A^2A^3}]&=&\langle (L^{3\dag}_{i_1i_2i_3}+\openone_d\otimes \beta A^2_{(i_1+i_3-2)\oplus_n +1} \otimes \openone_d)(L^3_{i_1i_2i_3}+\openone_d\otimes \beta A^2_{(i_1+i_3-2)\oplus_n +1} \otimes \openone_d)\rangle\nonumber\\
    &=& \langle L^{3\dag}_{i_1i_2i_3} L^3_{i_1i_2i_3} \rangle+\beta \langle L^{3\dag}_{i_1i_2i_3} A^2_{(i_1+i_3-2)\oplus_n +1} \rangle+\beta \langle A^2_{(i_1+i_3-2)\oplus_n +1}L^3_{i_1i_2i_3} \rangle+\beta^2\nonumber\\
    &=&\beta^2 \quad (\text{Ideal scenario $\langle L_{i_2 i_3}^\dagger L_{i_2 i_3} \rangle=\langle L_{i_2 i_3}^\dagger A^2_{(i_1+i_3-2)\oplus_n +1}\rangle=\langle A^2_{(i_1+i_3-2)\oplus_n +1} L_{i_2 i_3} \rangle=0$})\quad\quad \ \
\end{eqnarray}
Again, using the optimization condition, we have $w_{i_1 i_2 i_3} = w_{i_1} = \sqrt{2 - \langle \{A^1_{i_1}, A^1_{i_1 \ \oplus_n + 1}\} \rangle} = 2\cos\frac{\pi}{2n}$, where $\langle \cdots \rangle_{\rho_{A^1A^2A^3}} = \langle \cdots \rangle$. Substituting this expression for $w_{i_1 i_2 i_3}$ together with Eq.~(\ref{noisyLij}) into Eq.~(\ref{noisygaman}), we obtain the following.
\begin{eqnarray}
    &&\frac{1}{2}\sum_{i_3=1}^{n}\sum_{i_1=1}^{n}2\cos{\frac{\pi}{2n}}  \ \xi_{i_1i_2i_3}^2 =\xi\Rightarrow n^2 \beta^2 \ \cos{\frac{\pi}{2n}}=\xi\nonumber\\
    &&\hspace{4cm}\Rightarrow\beta =\sqrt{\frac{\xi }{n^2 \cos \left(\frac{\pi }{2 n}\right)}}
\end{eqnarray}
This is Eq.~(28) in the main text. For robust self-testing of the state from Eq.~(\ref{robs1}) we get
\begin{eqnarray}\label{noisy1}
    &&||\widetilde{\Phi}(\ket{\psi}_{A^1A^2A^3}\otimes\ket{000}_{A^{1'} A^{2'} A^{3'}})-\Phi(\ket{\psi}_{A^1A^2A^3}\otimes\ket{000}_{A^{1'} A^{2'} A^{3'}})||\nonumber\\
    &&=\frac{1}{8}\sum_{a^1,a^2,a^3\in\{0,1\}} \Bigg|\Bigg|\Big[(Z_{A^1})^{a^1} (Z_{A^2})^{a^2} (\widetilde{Z}_{A^3})^{a^3} (1+(-1)^{a^1} X_{A^1}) (1+(-1)^{a^2} X_{A^2}) (1+(-1)^{a^3} \widetilde{X}_{A^3})\nonumber\\
    &&-(Z_{A^1})^{a^1} (Z_{A^2})^{a^2} (Z_{A^3})^{a^3} (1+(-1)^{a^1} X_{A^1}) (1+(-1)^{a^2} X_{A^2}) (1+(-1)^{a^3} X_{A^3})\Big] \ket{\psi}_{A^1A^2A^3} \ket{a^1a^2a^3}_{A^{1'} A^{2'} A^{3'}}\Bigg|\Bigg|\nonumber\\
    &&=\frac{1}{8}\sum_{a^1,a^2,a^3\in\{0,1\}} \Bigg|\Bigg|(Z_{A^1})^{a^1}(Z_{A^2})^{a^2} \Big[(\widetilde{Z}_{A^3})^{a^3}  (1+(-1)^{a^3} \widetilde{X}_{A^3})-(Z_{A^3})^{a^3}  (1+(-1)^{a^3} X_{A^3})\Big]\nonumber\\
    &&\quad\quad\times(1+(-1)^{a^2} X_{A^2}) (1+(-1)^{a^1} X_{A^1}) \ket{\psi}_{A^1A^2A^3} \ket{a^1a^2a^3}_{A^{1'} A^{2'} A^{3'}}\Bigg|\Bigg|\nonumber\\
    &&\leq\frac{1}{8}\Bigg[\sum_{a^1,a^2,a^3\in\{0,1\}} \Bigg|\Bigg|\qty((\widetilde{Z}_{A^3})^{a^3}-(Z_{A^3})^{a^3}+(-1)^{a^3} ((\widetilde{Z}_{A^3})^{a^3} \widetilde{X}_{A^3}-(Z_{A^3})^{a^3} X_{A^3}))\Big(1+(-1)^{a^2} X_{A^2}\nonumber\\
    &&\hspace{8cm}+(-1)^{a^1} X_{A^1}+(-1)^{a_1+a_2} X_{A^1} X_{A^2}\Big) \ket{\psi}_{A^1A^2A^3} \ket{a^1a^2a^3}_{A^{1'} A^{2'} A^{3'}}\Bigg|\Bigg|\Bigg]\nonumber\\
    &&\leq\frac{1}{8}\Bigg[\sum_{a^3\in\{0,1\}} \Bigg|\Bigg|\qty((\widetilde{Z}_{A^3})^{a^3}-(Z_{A^3})^{a^3}+(-1)^{a^3} ((\widetilde{Z}_{A^3})^{a^3} \widetilde{X}_{A^3}-(Z_{A^3})^{a^3} X_{A^3}))\sum_{a^1,a^2\in\{0,1\}}\Big(1+(-1)^{a^2} X_{A^2}\nonumber\\
    &&\hspace{8cm}+(-1)^{a^1} X_{A^1}
    +(-1)^{a^1+a^2} X_{A^1} X_{A^2}\Big) \ket{\psi}_{A^1A^2A^3} \ket{a^1a^2a^3}_{A^{1'} A^{2'} A^{3'}}\Bigg|\Bigg|\Bigg]\nonumber\\
    &&\leq\frac{1}{8}\Bigg[\sum_{a^3\in\{0,1\}} \Bigg|\Bigg|\qty((\widetilde{Z}_{A^3})^{a^3}-(Z_{A^3})^{a^3}+(-1)^{a^3} ((\widetilde{Z}_{A^3})^{a^3} \widetilde{X}_{A^3}-(Z_{A^3})^{a^3} X_{A^3}))\sum_{a^1a^2\in\{0,1\}} \ket{\psi}_{A^1A^2A^3} \ket{a^1a^2a^3}_{A^{1'} A^{2'} A^{3'}}\Bigg|\Bigg|\nonumber\\
    &&+\sum_{a^3\in\{0,1\}}\Bigg|\Bigg|\qty((\widetilde{Z}_{A^3})^{a^3}-(Z_{A^3})^{a^3}+(-1)^{a^3} ((\widetilde{Z}_{A^3})^{a^3} \widetilde{X}_{A^3}-(Z_{A^3})^{a^3} X_{A^3}))\sum_{a^1a^2\in\{0,1\}} (-1)^{a^2} X_{A^2} \ket{\psi}_{A^1A^2A^3} \ket{a^1a^2a^3}_{A^{1'} A^{2'} A^{3'}}\Bigg|\Bigg|\nonumber\\
    &&+\sum_{a^3\in\{0,1\}} \Bigg|\Bigg|\qty((\widetilde{Z}_{A^3})^{a^3}-(Z_{A^3})^{a^3}+(-1)^{a^1} ((\widetilde{Z}_{A^3})^{a^3} \widetilde{X}_{A^3}-(Z_{A^3})^{a^3} X_{A^3}))\sum_{a^1,a^2\in\{0,1\}}(-1)^{a^3} X_{A^1}\ket{\psi}_{A^1A^2A^3} \ket{a^1a^2a^3}_{A^{1'} A^{2'} A^{3'}}\Bigg|\Bigg|\nonumber\\
    &&+\sum_{a^3\in\{0,1\}} \Bigg|\Bigg|\qty((\widetilde{Z}_{A^3})^{a^3}-(Z_{A^3})^{a^3}+(-1)^{a^3} ((\widetilde{Z}_{A^3})^{a^3} \widetilde{X}_{A^3}-(Z_{A^3})^{a^3} X_{A^3}))\sum_{a^1a^2\in\{0,1\}}(-1)^{a^1+a^2} X_{A^1} X_{A^2}\ket{\psi}_{A^1A^2A^3} \ket{a^1a^2a^3}_{A^{1'} A^{2'} A^{3'}}\Bigg|\Bigg|\Bigg]\nonumber\\
    &&\leq 2\sum_{a^3\in\{0,1\}} \Bigg|\Bigg|\qty((\widetilde{Z}_{A^3})^{a^3}-(Z_{A^3})^{a^3}+(-1)^{a^3} ((\widetilde{Z}_{A^3})^{a^3} \widetilde{X}_{A^3}-(Z_{A^3})^{a^3} X_{A^3})) \ket{\psi}_{A^1A^2A^3} \Bigg|\Bigg|\nonumber\quad\\
    &&\leq 2\qty(||(\widetilde{Z}_{A^3}-Z_{A^3}) \ket{\psi}_{A^1A^2A^3} ||+||(\widetilde{X}_{A^3}-X_{A^3}) \ket{\psi}_{A^1A^2A^3} ||+||(\widetilde{Z}_{A^3}\widetilde{X}_{A^3}-Z_{A^3} X_{A^3})\ket{\psi}_{A^1A^2A^3} ||) \nonumber\\
    &&\leq 2\qty(2\beta+||(\widetilde{Z}_{A^3}\widetilde{X}_{A^3}-Z_{A^3} X_{A^3})\ket{\psi}_{A^1A^2A^3} ||)
\end{eqnarray}
Again
\begin{eqnarray}\label{noisyzaxa}
    ||(\widetilde{Z}_{A^3}\widetilde{X}_{A^3}-Z_{A^3} X_{A^3})\ket{\psi}_{A^1A^2A^3} ||&\approx&||(\widetilde{Z}_{A^3}(\beta\openone_d+X_{A^3})-Z_{A^3} X_{A^3})\ket{\psi}_{A^1A^2A^3} ||\nonumber\\
    &\leq&\beta ||\widetilde{Z}_{A^3}\ket{\psi}_{A^1A^2A^3}||+||(\widetilde{Z}_{A^3}-Z_{A^3})X_{A^3} \ket{\psi}_{A^1A^2A^3}||\nonumber\\
    &\leq& \beta ||(\beta \openone_d + Z_{A^3})\ket{\psi}_{A^1A^2A^3}||+\beta\leq 2\beta +\beta^2
\end{eqnarray}
Now, putting Eq.~(\ref{noisyzaxa}) in Eq.~(\ref{noisy1}) we get
\begin{eqnarray}\label{noisy2}
    ||\widetilde{\Phi}(\ket{\psi}_{A^1A^2A^3}\otimes\ket{000}_{A^{1'} A^{2'} A^{3'}})-\Phi(\ket{\psi}_{A^1A^2A^3}\otimes\ket{000}_{A^{1'} A^{2'} A^{3'}})||\leq8\beta+2\beta^2
\end{eqnarray}
For Robust self-testing of the observable from Eq.~(\ref{robob1}), we get
\begin{eqnarray}
    &&||\widetilde{\Phi}(\widetilde{X}_{A^3}\ket{\psi}_{A^1A^2A^3}\otimes\ket{000}_{A^{1'} A^{2'} A^{3'}})-\Phi(X_{A^3}\ket{\psi}_{A^1A^2A^3}\otimes\ket{000}_{A^{1'} A^{2'} A^{3'}})||\nonumber\\
    &&\leq\frac{\beta}{8}\sum_{a^1,a^2,a^3\in\{0,1\}} \Bigg|\Bigg|(Z_{A^1})^{a^1} (Z_{A^2})^{a^2}  (\widetilde{Z}_{A^3})^{a^3}(1+(-1)^{a^1} X_{A^1}) (1+(-1)^{a^2} X_{A^2}) (1+(-1)^{a^3} \widetilde{X}_{A^3})\ket{\psi}_{A^1A^2A^3}\ket{a^1a^2a^3}_{A^{1'} A^{2'} A^{3'}}\Bigg|\Bigg|\nonumber\\
    &&+\Big|\Big|\widetilde{\Phi}(\ket{\psi}_{A^1A^2A^3}\otimes\ket{000}_{A^{1'} A^{2'} A^{3'}})-\Phi(\ket{\psi}_{A^1A^2A^3}\otimes\ket{000}_{A^{1'} A^{2'} A^{3'}})\Big|\Big|\nonumber\\
    &&\leq\frac{\beta}{8}\sum_{a^1,a^2,a^3\in\{0,1\}} \Bigg|\Bigg|(\widetilde{Z}_{A^3})^{a^3}  (1+(-1)^{a^1} X_{A^1}) (1+(-1)^{a^2} X_{A^2}) (1+(-1)^{a^3} \widetilde{X}_{A^3})\ket{\psi}_{A^1A^2A^3}\ket{a^1a^2a^3}_{A^{1'} A^{2'} A^{3'}}\Bigg|\Bigg|+8\beta +2\beta^2\nonumber\\
    &&\leq\frac{\beta}{8}\sum_{a^3\in\{0,1\}} \Bigg|\Bigg|(\widetilde{Z}_{A^3})^{a^3}  (1+(-1)^{a^3} \widetilde{X}_{A^3}) \sum_{a^1a^2\in\{0,1\}}\qty(1+(-1)^{a^2} X_{A^2}+(-1)^{a^1} X_{A^1}+(-1)^{a_1+a_2}X_{A^1} X_{A^2})\ket{\psi}_{A^1A^2A^3}\ket{a^1a^2a^3}_{A^{1'} A^{2'} A^{3'}}\Bigg|\Bigg|\nonumber\\
    &&+8\beta +2\beta^2\nonumber\\
    &&\leq2\beta\sum_{a^3\in\{0,1\}} \Bigg|\Bigg|(\widetilde{Z}_{A^3})^{a^3}  (1+(-1)^{a^3} \widetilde{X}_{A^3}) \ket{\psi}_{A^1A^2A^3}\Bigg|\Bigg|+8\beta +2\beta^2\nonumber\quad \\
    &&\leq2\beta\qty(1+||\widetilde{X}_{A^3} \ket{\psi}_{A^1A^2A^3}||+||\widetilde{Z}_{A^3} \ket{\psi}_{A^1A^2A^3}||+||\widetilde{Z}_{A^3} \widetilde{X}_{A^3} \ket{\psi}_{A^1A^2A^3}||)+8\beta +2\beta^2\nonumber\\
    &&\leq2\beta\qty(1+||(\beta\openone_d+X_{A^3}) \ket{\psi}_{A^1A^2A^3}||+||(\beta\openone_d+Z_{A^3})\ket{\psi}_{A^1A^2A^3}||+|| (\beta\openone_d+Z_{A^3})(\beta\openone_d+X_{A^3})\ket{\psi}_{A^1A^2A^3}||)+8\beta +2\beta^2\nonumber\\
    &&\leq2\beta\qty(1+2(\beta+1)+(\beta^2+2\beta+1))+8\beta +2\beta^2\nonumber\\
    &&\leq 16\beta+10\beta^2+2\beta^3
\end{eqnarray}
Similarly, 
\begin{eqnarray}
    &&||\widetilde{\Phi}(\widetilde{Z}_{A^3}\ket{\psi}_{A^1A^2A^3}\otimes\ket{000}_{A^{1'} A^{2'} A^{3'}})-\Phi(Z_{A^3}\ket{\psi}_{A^1A^2A^3}\otimes\ket{000}_{A^{1'} A^{2'} A^{3'}})||
    \leq 16\beta+10\beta^2+2\beta^3
\end{eqnarray}
Re-collecting all the results
\begin{eqnarray}
    &&||\widetilde{\Phi}(\ket{\psi}_{A^1A^2A^3}\otimes\ket{000}_{A^{1'} A^{2'} A^{3'}})-\Phi(\ket{\psi}_{A^1A^2A^3}\otimes\ket{000}_{A^{1'} A^{2'} A^{3'}})||\leq 8\beta +2\beta^2\\
    &&||\widetilde{\Phi}(\widetilde{O}\ket{\psi}_{A^1A^2A^3}\otimes\ket{000}_{A^{1'} A^{2'} A^{3'}})-\Phi(O\ket{\psi}_{A^1A^2A^3}\otimes\ket{000}_{A^{1'} A^{2'} A^{3'}})||\leq 16\beta+10\beta^2+2\beta^3 \quad \forall O\in\{X_{A^3},Z_{A^3}\}
\end{eqnarray}
This is Eq.~(27) in main-text. Again, the relation between $\beta$ and observed value $\xi$ is $\beta =\sqrt{\frac{\xi }{n^2 \cos \left(\frac{\pi }{2 n}\right)}}$. To determine the robust self-testing limits for both state and observables, we defined the relative observed violations $r$ as a function of the number of inputs $n$ and the observed value $\xi$.
\begin{eqnarray}
    &&r=\frac{(\widetilde{\Gamma}^3_n)_Q-(\Gamma^3_n)_{BL}}{(\Gamma^3_n)_Q-(\Gamma^3_n)_{BL}}= 1-\frac{\xi}{2 n^2 \cos{\frac{\pi}{2n}}-2n(n-1)}\\
    &&\xi =(1-r) \left(2 n^2 \cos \left(\frac{\pi }{2 n}\right)-2 n(n-1)\right)
\end{eqnarray}
These are Eqs.~(29) and (30) in the main text. Here $(\widetilde{\Gamma}^3_n)_Q=2n^2 \cos{\frac{\pi}{2n}}-\xi, (\Gamma^3_n)_Q=2n^2 \cos{\frac{\pi}{2n}}$ and the bi-local value $(\Gamma^3_n)_{BL}=2n(n-1)$. 
Hence, the trace distance between the observed and ideal state in terms of the relative operator $r$ is
\begin{eqnarray}
    ||\widetilde{\Phi}(\ket{\psi}_{A^1A^2A^3}\otimes\ket{000}_{A^{1'} A^{2'} A^{3'}})-\Phi(\ket{\psi}_{A^1A^2A^3}\otimes\ket{000}_{A^{1'} A^{2'} A^{3'}})||\leq f_s(r)\quad
\end{eqnarray}
 where $f_s(r)=\frac{4 (n-1) (r-1) \sec \left(\frac{\pi }{2 n}\right)}{n}+8 \sqrt{2} \sqrt{\frac{(r-1) \left((n-1) \sec \left(\frac{\pi }{2 n}\right)-n\right)}{n}}-4 r+4$. Now, according to Fuchs-Van de Graaf~\cite{wilde2013quantum}, the approximate relation between the trace distance $f_s(r)$ and robust fidelity $F_s(r)$ is 
 \begin{eqnarray}
     2\qty(1-\sqrt{F_s(r)})\leq f_s(r)\leq 2\sqrt{1-F_s(r)}
 \end{eqnarray}
 This implies the lower bound of fidelity in terms of trace distance, i.e
 \begin{eqnarray}
     F_s(r)\geq \qty(1-\frac{1}{2}f_s(r))^2
 \end{eqnarray}
 In the same way for the observables, we can define the fidelity $F_o(r)$ in terms of trace distance $f_o(r)$, i.e.
  \begin{eqnarray}
     F_o(r)\geq \qty(1-\frac{1}{2}f_o(r))^2
 \end{eqnarray}
 where $f_o(r)=4 \left(\sqrt{2} \left(\frac{(r-1) \left((n-1) \sec \left(\frac{\pi }{2 n}\right)-n\right)}{n}\right)^{3/2}+\frac{5 (r-1) \left((n-1) \sec \left(\frac{\pi }{2 n}\right)-n\right)}{n}+4 \sqrt{2} \sqrt{\frac{(r-1) \left((n-1) \sec \left(\frac{\pi }{2 n}\right)-n\right)}{n}}\right)$.
\subsection{Robustness of genuine randomness in tripartite scenario }\label{Rob RC}
In the experimental scenario, achieving the optimal quantum violation is nearly impossible due to noisy systems. In this scenario, we introduce equal noise parameters for each of the observables of $A^3$, assuming the state is perfect. Thus, Bell's functional takes the form
 \begin{eqnarray}
     \widetilde{\Gamma}^3_n&=&\sum_{i_3=1}^n \qty(\sum_{{i_1}=1}^n(A^1_{i_1}-A^1_{{i_1}\oplus_n+1})\otimes A^2_{({i_1}+i_3-2)\oplus_n +1 })\otimes \mathcal{A}^3_{i_3},\forall n\in odd 
 \end{eqnarray}
 
We can write the noisy observables in the following form
\begin{eqnarray}
    ||(\mathcal{A}^3_{i_3}-A^3_{i_3})\ket{\psi}_{AB}||\leq \beta&&\implies  \mathcal{A}^3_{i_3} \approx A^3_{i_3}+ \beta \openone_d\nonumber\\
    &&\implies \mathcal{A}^3_{i_3}= \frac{ A^3_{i_3}+ \beta \openone_d}{\sqrt{1+\beta^2}} \quad (\text{where $\Tr[A^3_{i_3} \ \rho_{A^1A^2A^3}]=0$})
\end{eqnarray}
where $\mathcal{A}^3_{i_3}$ is a dichotomic observable. For a given measurement settings $A^1_{i_1}$, $A^2_{i_2}$ and $\mathcal{A}^3_{i_3}$, the probability \linebreak $P\left(a^1,a^2,a^3|A^1_{i_1},A^2_{i_2}, \mathcal{A}^3_{i_3}\right)$ is defined as follows.

\begin{eqnarray}
P\left(a^1,a^2,a^3|A^1_{i_1},A^2_{i_2}, \mathcal{A}^3_{i_3}\right)&=&\Tr[ \bigg(\frac{\openone_d+a^1 A^1_{i_1}}{2}\otimes \frac{\openone_d+a^2A^2_{i_2}}{2}\otimes \frac{\openone_d+a^3\mathcal{A}^3_{i_3}}{2}\bigg)\ \rho_{A^1A^2A^3}]\quad (\text{where $a^k\in \pm 1,\forall k\in[3]$})\nonumber\\
&=&\frac{\sqrt{1+\beta^2}+a^3\beta+a^1a^2a^3 \langle A^1_{i_1}\otimes A^2_{i_2}\otimes A^3_{i_3}\rangle}{8\sqrt{1+\beta^2}}, \quad \qty(\text{$\langle A^k_{i_k}\rangle=\langle A^{k_1}_{i_{k_1}}A^{k_2}_{i_{k_2}}\rangle_{k_1\neq k_2}=0,\forall k,k_i\in[3]$})\qquad\label{rabd}
\end{eqnarray}
From Eq.~(\ref{maxraibjck}), we know $\bra{\psi}_{A^1A^2A^3}\left( A^1_{1+(i_2-i_3+\frac{n+1}{2})\oplus_n}\otimes A^2_{i_2} \otimes A^3_{i_3}\right)\ket{\psi}_{A^1A^2A^3}=0$. Substituting $i_1=1+(i_2-i_3+\frac{n+1}{2})\oplus_n$ into Eq.~(\ref{rabd}), we obtain
\begin{eqnarray}
&&P\left(a^1,a^2,a^3 \mid A^1_{1+(i_2-i_3+\frac{n+1}{2})\oplus_n}, A^2_{i_2}, \mathcal{A}^3_{i_3}\right) = \frac{\sqrt{1+\beta^2}+a^3\beta+a^1a^2a^3 \langle A^1_{1+(i_2-i_3+\frac{n+1}{2})\oplus_n}\otimes A^2_{i_2}\otimes A^3_{i_3}\rangle}{8\sqrt{1+\beta^2}}\\
&&P_{max}\left(a^1,a^2,a^3 \mid A^1_{1+(i_2-i_3+\frac{n+1}{2})\oplus_n}, A^2_{i_2}, \mathcal{A}^3_{i_3}\right) = \frac{\sqrt{1+\beta^2}+\beta}{8\sqrt{1+\beta^2}}
\end{eqnarray}
Hence, in the presence of noise, the maximum randomness $\mathcal{R}^m_{\text{max}}$ changes to 
\begin{eqnarray}
\widetilde{\mathcal{R}}^m_{\text{max}} = \log_2 \left(\frac{8\sqrt{1+\beta^2}}{\sqrt{1+\beta^2}+\beta}\right)
\end{eqnarray}
This is Eq.~(33) in the main text.

\twocolumngrid
\bibliographystyle{apsrev4-2}
\bibliography{references}

\end{document}